\documentclass[aps,pra,reprint,superscriptaddress,nofootinbib]{revtex4-1}
\usepackage{afterpage}
\usepackage{amsmath}
\usepackage{amssymb}
\usepackage{amsfonts}
\usepackage{mathrsfs}
\usepackage{graphicx}
\usepackage[caption=false]{subfig}
\usepackage{enumitem}
\usepackage[normalem]{ulem}
\usepackage[percent]{overpic}
\usepackage{bm}
\usepackage{rotating}
\usepackage[usenames, dvipsnames]{color}
\usepackage{hyperref}
\usepackage{cancel}
\usepackage{verbatim}
\usepackage{mathtools}
\usepackage{tensor}
\usepackage{verbatim}
\usepackage{tikz}
\usepackage{booktabs}
\graphicspath{ {images/} }

\usepackage{braket}

\numberwithin{equation}{section}

\newcommand{\ii}{\mathrm{i}}

\renewcommand{\d}{\mathrm{d}}

\newcommand{\be}{\begin{equation}}
\newcommand{\bel}[1]{\begin{equation}\label{#1}}
\newcommand{\ee}{\end{equation}}

\begin{document}
\title{A Discrete Analog of General Covariance - Part 1:\\
Could the world be fundamentally set on a lattice?}
\author{Daniel Grimmer}
\email{daniel.grimmer@philosophy.ox.ac.uk}
\affiliation{Reuben College, University of Oxford, Oxford, OX2 6HW United Kingdom}
\affiliation{Faculty of Philosophy, University of Oxford, Oxford, OX2 6GG United Kingdom}
\affiliation{Barrio RQI, Waterloo, Ontario N2L 3G1, Canada}

\begin{abstract}
A crucial step in the history of General Relativity was Einstein's adoption of the principle of general covariance which demands a coordinate independent formulation for our spacetime theories. General covariance helps us to disentangle a theory's substantive content from its merely representational artifacts. It is an indispensable tool for a modern understanding of spacetime theories, especially regarding their background structures and symmetry. Motivated by quantum gravity, one may wish to extend these notions to quantum spacetime theories (whatever those are). Relatedly, one might want to extend these notions to discrete spacetime theories (i.e., lattice theories). This paper delivers such an extension with surprising consequences.

One's first intuition regarding discrete spacetime theories may be that they introduce a great deal of fixed background structure (i.e., a lattice) and thereby limit our theory's possible symmetries down to those which preserve this fixed structure (i.e., only certain discrete symmetries). By so restricting symmetries, lattice structures appear to be both theory-distinguishing and fundamentally ``baked-into'' our discrete spacetime theories. However, as I will discuss, all of these intuitions are doubly wrong and overhasty. Discrete spacetime theories can and do have continuous translation and rotation symmetries. Moreover, the exact same lattice theory can be given a wide variety of lattice structures and can even be described making reference to no lattice whatsoever. As my discrete analog of general covariance will reveal: lattice structure is rather less like a fixed background structure and rather more like a coordinate system, i.e., merely a representational artifact. Ultimately, I show that the lattice structure supposedly underlying any discrete ``lattice'' theory has the same level of physical import as coordinates do, i.e., none at all. Thus, the world cannot be ``fundamentally set on a square lattice'' (or any other lattice) any more than it could be ``fundamentally set in a certain coordinate system''. Like coordinate systems, lattice structures are just not the sort of thing that can be fundamental; they are both thoroughly merely representational. Spacetime cannot be a lattice (even when it might be representable as such).
\end{abstract}

\maketitle

\section{Introduction}\label{SecIntro}
A crucial step in the history of General Relativity (GR) was Einstein's adoption of the principle of general covariance\footnote{The terms general covariance, diffeomorphism invariance, and background independence are sometimes used interchangeably in the physics literature. In this paper, I understand these terms as distinct as laid out in~\cite{Pooley2015}. In Appendix~\ref{SecGenCov}, I will give explicit examples which separate these three concepts.} which states that the form of our physical laws should be independent of any choice of coordinate systems. At first, Einstein thought this property was unique to GR and that this is what set his theory apart from all of its predecessors. However, in 1917 Kretschmann pointed out that any physical theory can be written in a generally covariant form (i.e., in a coordinate-independent way). See~\cite{Norton1993} for a historical review of this point.

The modern understanding of the principle of general covariance is best summarized by Friedman~\cite{Friedman1983}:
\begin{quote}
\dots the principle of general covariance has no physical content whatever: it specifies no particular physical theory; rather it merely expresses our commitment to a certain style of formulating physical theories.
\end{quote}
However, despite this lack of physical content\footnote{Some argue that this principle does, in fact, have physical content at least when it is applied to isolated subsystems~\cite{TehFreidel}. E.g., in Galileo's thought experiment when only the ship subsystem is boosted relative to the un-boosted shore.}, the conceptual benefits of writing a theory in a coordinate-free way are immense. A generally covariant formulation of a theory has at least two major benefits: 1) it more clearly exposes the theory's geometric background structure, and 2) it thereby helps clarify our understanding of the theory's symmetries (i.e., its structure/solution preserving transformations). It does both of these by disentangling the theory's substantive content from representational artifacts which arise in particular coordinate representations~\cite{Pooley2015,Norton1993,EarmanJohn1989Weas}. Thus, general covariance is an indispensable tool for a modern understanding of spacetime theories. 

Motivated by quantum gravity, one may wish to extend these notions to quantum spacetime theories (whatever those are) for the following reason. It is now widely believed that the key conceptual shift which sets GR apart from its predecessors is not its general covariance as Einstein thought but rather its background independence, i.e., its complete lack of background structure (whatever this means exactly~\cite{Norton1993,Pitts2006,Pooley2010,ReadThesis,Pooley2015,Teitel2019,Belot2011}). Moreover, it is often claimed~\cite{rovelli_2001,rovelli_2004,SmolinLee2006TCfB,SmolinLee2008Ttwp,deHaroSebastian2017Daeg,DeHaroSebastian2017TIoD} that any successor theory to GR (e.g., quantum gravity) should follow GR's lead and be similarly background independent (whatever this ends up meaning in a quantum setting~\cite{ReadThesis}). As I discussed above, for classical spacetime theories our best tool for exposing background structures is general covariance. As such, it would be great if we could extend this tool towards quantum spacetime theories.

This paper develops a parallel extension, not towards quantum spacetime theories but towards discrete spacetime theories (i.e., lattice theories\footnote{Given the results of this paper, calling these ``lattice theories'' is a bit of a misnomer. This would be analogous to referring to continuum spacetime theories as ``coordinate theories''. As I will discuss, in both cases the coordinate systems/lattice structure are merely representational artifacts and so do not deserve ``first billing'' so to speak. All lattice theories are best thought of as lattice-representable theories. Similarly, the term ``discrete spacetime theories'' ought to be here read as ``discretely-representable spacetime theories''. As I will discuss, the defining feature of such theories is that they have a finite density of degrees of freedom, see the work of Achim Kempf~\cite{UnsharpKempf,Kempf2003,Kempf2004a,Kempf2004b,Kempf2006}.}) with surprising consequences. These extensions are not unrelated: quantization literally means discretization. Many approaches to quantum gravity assume some kind of discrete spacetime: causal sets, cellular automata, loop quantum gravity, spin foams, etc. Indeed, we have good reason to believe that a full non-perturbative theory of quantum gravity will have something like a finite density of degrees of freedom\footnote{Firstly, basic thought experiments in quantum gravity suggest the existence of something like a minimum possible length scale at approximately the Planck length. Measurements which resolve things at this scale, or attempts to store information at this scale both seem to lead to the creation of black holes. Relatedly, the Bekenstein bound suggests that only a finite amount of information can be stored in any given volume (with this bound scaling with the region's area) before a black hole forms.}. While the lattice theories considered here are notably simpler than those used in modern approaches to quantum gravity, it is likely that the lessons learned here can be applied there as well.

\subsection{Central Claims}\label{SecCentralClaims}
\begin{figure}[t!]
\includegraphics[width=0.4\textwidth]{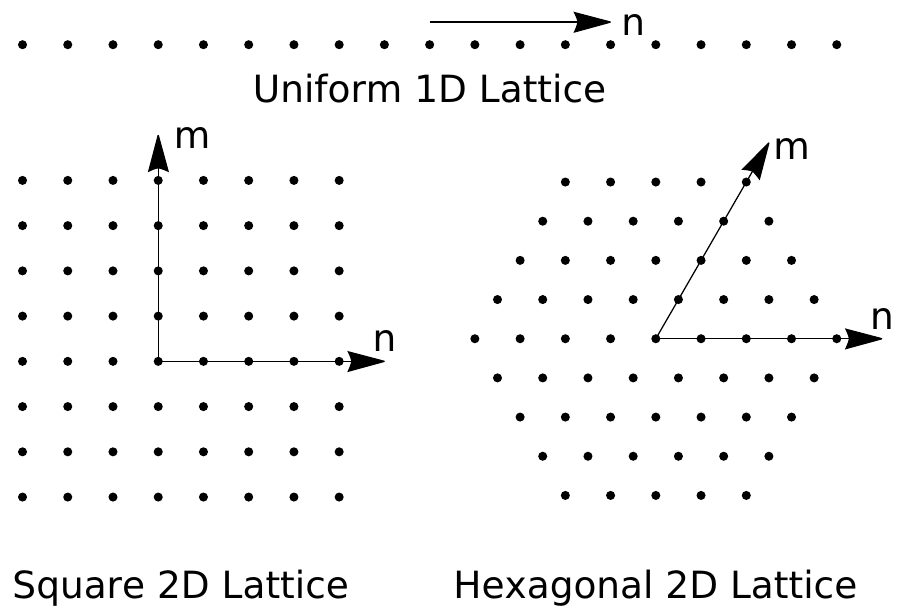}
\caption{Three of the lattice structures considered throughout this paper. The arrows indicate the indexing conventions for the lattice sites.}\label{FigLat}
\end{figure}

Intuitively, it is in the realm of metaphysical possibility that the world could be fundamentally set on a lattice. This should be distinguished from the possibility of a world with a fundamentally continuous spacetime manifold and dynamics which causes the world's matter content to arrange itself lattice-wise, at least approximately, (e.g., like the crystals studied by solid state physics). Here I am considering the possibility that the spacetime \textit{itself} is a lattice. Indeed, as discussed above, quantum gravity seems to point us roughly in this direction. However, independent of applications to quantum gravity, it is interesting to try and take the possibility of a fundamentally lattice-based world seriously.

Let's consider the following empirical situation and follow our first intuitions interpreting it. Suppose that by empirically investigating the world on the smallest scales we discover microscopic symmetry restriction. Namely, we find that only quarter turns or perhaps one-sixth turns (but not continuous rotations) preserve the dynamics. 

One might intuitively interpret these effects as ``lattice artifacts'' which reflect the discrete symmetries of the underlying lattice structure. For instance, Fig.~\ref{FigLat} shows a square lattice and a hexagonal lattice. Intuitively, a theory set on a square lattice can only have the symmetries of that lattice (i.e., discrete shifts, quarter rotations, and certain mirror reflections). Similarly for a hexagonal lattice but with one-sixth rotations. Even with an unstructured lattice it is intuitive that our theories could at most have discrete symmetries (i.e., permutations of lattice sites). Thus, assuming that the world has a fundamental underlying lattice structure would explain the restricted rotation symmetries we found at a microscopic scale. 

Suppose we buy into this fundamental-underlying-lattice hypothesis, one might then hope to discover which kind of lattice structure the world has (e.g., square, hexagonal, irregular, etc.) by investigating the theory's dynamical symmetries and finding the matching lattice structure. (Such an exercise is carried out in Sec.~\ref{SecHeat1}.) Further, one might hope to discover what kind of interactions there are on this lattice: nearest neighbor, next-to-nearest neighbor, infinite-range, etc. Intuitively, one could discover these sorts of things through detailed microscopic investigations.

Suppose that after substantial empirical investigation we find such ``lattice artifacts'' and moreover we have great predictive success when modeling the world as being set on (for instance) a square lattice with next-to-nearest neighbor interactions. Would this in any way prove that the world is fundamentally set on such a lattice? No, all this would prove is that the world can be \textit{faithfully represented} on such a lattice with such interactions, at least empirically. Anything can be faithfully represented in any number of ways, this is just mathematics. Some extra-empirical work must be done to know if we should take the lattice structure appearing in this representation seriously. That is, we must ask which parts of the theory are substantive and which parts are merely representational? The discrete analog of general covariance delivered in this paper answers this question: lattice structures are coordinate-like representational artifacts and so ultimately have no physical content.

To flesh out the contrary received position however, let us proceed without this discrete analog for the moment. We can ask: beyond merely appearing in our hypothetical empirically successful theory, what reason do we have to take the lattice structures which appear in this theory seriously? Well, intuitively the lattice structures appear to play a very substantial role in these theories, not merely a representational one. One likely has the following three interconnected first intuitions regarding the role that the lattice and lattice structure play in discrete spacetime theories:
\begin{itemize}
\item[1.] They restrict our possible symmetries. Taking the lattice structure to be a part of the theory's fixed background structure, our possible symmetries are limited to those which preserve this fixed structure. As discussed above, intuitively a theory set on a square lattice can only have the symmetries of that lattice. Similarly for a hexagonal lattice, or even an unstructured lattice.
\item[2.] Differing lattice structures distinguishes our theories. Two theories with different lattice structures (e.g., square, hexagonal, irregular, etc.) cannot be identical. As suggested above they have different fixed background structures and as therefore have different symmetries.
\item[3.] The lattice is fundamentally ``baked-into'' the theory. Firstly, it is what the fundamental fields are defined over: they map lattice sites (and possibly times) into some value space. Secondly, the bare lattice is what the lattice structure structures. Thirdly, it is what limits us to discrete permutation symmetries in advance of further limitations from the lattice structure.
\end{itemize}
These intuitions will be fleshed out and made more concrete in Sec.~\ref{SecHeat1}. However, as this paper demonstrates, each of the above intuitions are doubly wrong and overhasty.

What goes wrong with the above intuitions is that we attempted to directly transplant our notions of background structure and symmetry from continuous to discrete spacetime theories. This is an incautious way to proceed and is apt to lead us astray. Recall that, as discussed above, our notions of background structure and symmetry are best understood in light of general covariance. It is only once we understand what is substantial and what is merely representational in our theories that we have any hope of properly understanding them. Therefore, we ought to instead first transplant a notion of general covariance into our discrete spacetime theories and then see what conclusions we are led to regarding the role that the lattice and lattice structure play in our discrete spacetime theories. This paper does just that. 

As my discrete analog of general covariance will reveal: lattice structure is rather less like a fixed background structure and rather more like a coordinate system, i.e., merely a representational artifact. Indeed, this paper develops a rich analogy between the lattice structures which appear in our discrete spacetime theories and the coordinate systems which appear in our continuum spacetime theories. Three lessons learned throughout this paper\footnote{These lessons are also visible in some corners of the physics literature, particularly in the work of Achim Kempf~\cite{UnsharpKempf,Kempf2000b,Kempf2003,Kempf2004a,Kempf2004b,Kempf2006,Martin2008,Kempf_2010,Kempf2013,Pye2015,Kempf2018} among others~\cite{PyeThesis,Pye2022,BEH_2020}. For an overview see~\cite{Kempf2018}.} point us in this direction. Each of these lessons negates one of the above discussed first intuitions.

Firstly, as I will show, taking any lattice structure seriously as a fixed background structure systematically under predicts the symmetries that discrete theories can and do have. Indeed, as I will show neither the bare lattice itself nor its lattice structure in any way restrict a theory's possible symmetries. In fact, there is no conceptual barrier to having a theory with continuous translation and rotation symmetries formulated on a discrete lattice. As I will discuss, this is analogous to the familiar fact that there is no conceptual barrier to having a continuum theory with rotational symmetry formulated on a Cartesian coordinate system.

Secondly, as I will show, discrete theories which are initially presented to us with very different lattice structures (i.e., square vs. hexagonal) may nonetheless turn out to be substantially equivalent theories or as overlapping parts of some larger theory. Moreover, given any discrete theory with some lattice structure we can always re-describe it using a different lattice structure. As I will discuss, this is analogous to the familiar fact that our continuum theories can be described in different coordinates, and moreover we can switch between these coordinate systems freely.

Thirdly, as I will show, in addition to being able to switch between lattice structures, we can also reformulate any discrete theory in such a way that it has no lattice structure whatsoever. Indeed, we can always do away with the lattice altogether. As I will discuss, this is analogous to the familiar fact that any continuum theory can be written in a generally covariant (i.e., coordinate-free) way.

These three lessons combine to give us a rich analogy between lattice structures and coordinate systems. It is from this rich analogy that the central claims of this paper follow. Namely, from this analogy it follows that the lattice structure supposedly underlying any discrete ``lattice'' theory has the same level of physical import as coordinates do, i.e., none at all. Thus, the world cannot be ``fundamentally set on a square lattice'' (or any other lattice) any more than it could be ``fundamentally set in a certain coordinate system''. Like coordinate systems, lattice structures are just not the sort of thing that can be fundamental; they are both thoroughly merely representational. Spacetime cannot be a lattice (even when it might be representable as such). Specifically, I claim that properly understood, there are no such things as lattice-fundamental theories, rather there are only lattice-representable theories.

This paper is largely inspired by the brilliant work of mathematical physicist Achim Kempf~\cite{Kempf_1997,UnsharpKempf,Kempf2000b,Kempf2003,Kempf2004a,Kempf2004b,Kempf2006,Martin2008,Kempf_2010,Kempf2013,Pye2015,Kempf2018} among others~\cite{PyeThesis,Pye2022,BEH_2020}. A key feature present both here and in Kempf's work is the sampling property of bandlimited function revealed by the Nyquist-Shannon sampling theory~\cite{GARCIA200263,SamplingTutorial,UnserM2000SyaS}. I review sampling theory in more detail in Sec.~\ref{SecSamplingTheory}, but let me overview here. Bandlimited functions are those with have a limited extent in Fourier space (i.e., compact support). Bandlimited functions have the following sampling property: they can be exactly reconstructed knowing only the values that they take on any sufficiently dense sample lattice. What ``sufficiently dense'' means is fixed in terms of the size of the function's support in Fourier space.

Nyquist-Shannon sampling theory was first discovered in the context of information processing as a way of converting between analog and digital signals (i.e., between continuous and discrete information). Sampling theory found its first application in fundamental spacetime physics with Kempf's \cite{Kempf_1997,UnsharpKempf}, ultimately leading to his thesis that ``Spacetime could be simultaneously continuous and discrete, in the same way that information can be'' \cite{Kempf_2010}. Kempf's thoughts on these topics is the primary inspiration for this paper and deserves wider appreciation by the philosophy of physics community. For an overview of Kempf's works on this topic see~\cite{Kempf2018}.

My thesis here is in broad agreement with Kempf's with one crucial alteration. I stress that the sampling property of bandlimited functions indicates that bandlimited physics can be simultaneously \textit{represented as} continuous and discrete, (i.e., on a continuous or discrete spacetime). However, I further argue that when one investigates these two representations one finds substantial issue with taking the discrete representation as fundamental. These issues stem from the rich analogy between the lattice structures and coordinate systems developed throughout this paper (and extended to a Lorentzian context in \cite{DiscreteGenCovPart2}).

\subsection{Outline of the Paper}
In Sec.~\ref{SecSevenHeat}, I will introduce seven discrete heat equations in an interpretation-neutral way and solve their dynamics. Then, in Sec.~\ref{SecHeat1}, I will make a first attempt at interpreting these theories. I will (ultimately wrongly) identify their underlying manifold, locality properties, and symmetries. Among other issues, a central problem with this first attempt is that it takes the lattice over time to be the underlying spacetime manifold and thereby unequivocally cannot support continuous translation and rotation symmetries. This systematically under predicts the symmetries that these theories can and do have.

In Sec.~\ref{SecHeat2}, I will provide a second attempt at interpreting these theories which fixes this issue (albeit in a slightly unsatisfying way). In particular, in this second attempt I deny that the lattice over time is the underlying spacetime manifold. Instead, I ``internalize'' it into the theory's value space. Fruitfully, this second interpretation does allow for continuous translation and rotation symmetries. However, the key move here of ``internalization'' has several unsatisfying consequences. For instance, the continuous translation and rotation symmetries we find are here classified as internal (i.e, associated with the value space) whereas intuitively they ought to be external (i.e, associated with the manifold).

We thus will need a third attempt at interpreting these theories which externalizes these symmetries. Sec.~\ref{SecExtPart1} - Sec.~\ref{SecExtPart2} lay the groundwork for this third interpretation. In particular, they describe a principled way of 1) inventing a continuous spacetime manifold for our formerly discrete theories to live on and 2) embedding our theory's states/dynamics onto this manifold as a new dynamical field. In the middle of this, in Sec.~\ref{SecSamplingTheory}, I will provide an informal overview of the primary mathematical tools used in this paper in the latter half of this paper. Namely, I review the basics of Nyquist-Shannon sampling theory and bandlimited functions. 

With this groundwork complete, in Sec.~\ref{SecHeat3} and Sec.~\ref{SecHeat3Extra} I will provide a third attempt at interpreting these seven theories which fixes all issues arising in the previous two interpretations. For instance, like in my second attempt, this third interpretation can support continuous translation and rotation symmetries. However, unlike the second attempt it realizes them as external symmetries (i.e., associated with the underlying manifold, not the theory's value space).

In Sec.~\ref{SecDisGenCov}, I will review the lessons learned in these three attempts at interpretation. As I will discuss, the lessons learned combine to give us a rich analogy between lattice structures and coordinate systems. As I will discuss, there are actually two ways of fleshing out this analogy: one internal and one external. This section spells out these analogies in detail, each of which gives us a discrete analog of general covariance. I find reason to prefer the external notion, but either is likely to be fruitful for further investigation/use. Sec.~\ref{PerfectRotation} provides us with a perfectly rotation invariant lattice theory.

Finally, in Sec. \ref{SecConclusion} I will summarize the results of this paper, discuss its implications, and provide an outlook of future work. For comments on what this means for quantum gravity, see \cite{DiscreteGenCovPart2} where I extend this work to a Lorentzian context. Here I will mostly discuss the impact these results have on the dynamical vs geometrical spacetime debate~\cite{EarmanJohn1989Weas,TwiceOver,BelotGordon2000GaM,Menon2019,BrownPooley1999,Nonentity,HuggettNick2006TRAo,StevensSyman2014Tdat,DoratoMauro2007RTbS,Norton2008,Pooley2013} especially in regards to a complaint against the dynamical approach raised by Norton~\cite{Norton2008}.

\section{Seven Discrete Heat Equations}\label{SecSevenHeat}
In this section I will introduce seven discrete heat equations (H1-H7) in an interpretation-neutral way and solve their dynamics. These theories are all describable as being set on a lattice in space. I consider the following three cases for the lattice in space: a uniform 1D lattice, a square 2D lattice, and a hexagonal 2D lattice, see Fig.~\ref{FigLat}.

As harmful as these choices seem to be to continuous translation and rotation invariance, I will show they ultimately pose no barrier to our theories having these symmetries. As I will argue, these choices of lattice are ultimately merely choices of representation which have absolutely nothing to do with the thing being represented. In particular, there is no need for our representational structure to have the same symmetries as the thing being represented. There is no issue with using Cartesian coordinates to describe a rotationally invariant state/dynamics. I claim that analogously there is no issue with using a lattice to describe a state/dynamics with continuous translation and rotation invariance. (In \cite{DiscreteGenCovPart2} I demonstrate this claim for Lorentz invariance as well.)

\subsection{Introducing H1-H7}
To begin, let us consider the continuum heat equation in one and two dimensions: 
\begin{align}\label{HeatEq00}
&\text{\bf Heat Equation 00 (H00):}\\
\nonumber
&\partial_t \psi(t,x) = \bar{\alpha}\, \partial_x^2\,\psi(t,x)\\
\label{HeatEq0}
&\text{\bf Heat Equation 0 (H0):}\\
\nonumber
&\partial_t \psi(t,x,y) = \frac{\bar{\alpha}}{2}\, (\partial_x^2+\partial_y^2)\,\psi(t,x,y) 
\end{align}
with some diffusion rate $\bar{\alpha}\geq0$. For a generally covariant (i.e., coordinate-free) view of these theories, see Sec.~\ref{SecFullGenCov}.

For our first three discrete heat equations, let us consider the theories with only nearest-neighbor (N.N.) interactions on the above discussed lattices which best approximate H0 and H00. Namely:
\begin{align}
\label{H1Long}
&\text{\bf 1D N.N. Heat Equation (H1):}\\
\nonumber
&\frac{\d }{\d t}\phi_n(t)
=\alpha\,[\phi_{n-1}(t)-2\phi_n(t)+\phi_{n+1}(t)]\\
\label{H4Long}
&\text{\bf Square N.N. Heat Equation (H4):}\\
\nonumber
&\frac{\d }{\d t}\phi_{n,m}(t)
=\frac{\alpha}{2}\big[\phi_{n-1,m}(t) -2\phi_{n,m}(t)+\phi_{n+1,m}(t)\\
\nonumber
&\qquad\qquad\qquad+\phi_{n,m-1}(t) -2\phi_{n,m}(t)+\phi_{n,m+1}(t)\big]\\
\label{H5Long}
&\text{\bf Hexagonal N.N. Heat Equation (H5):}\\
\nonumber
&\frac{\d }{\d t}\phi_{n,m}(t)
=\frac{\alpha}{3}\big[\, \phi_{n-1,m}(t) -2\phi_{n,m}(t)+\phi_{n+1,m}(t)\\
\nonumber
&\qquad\qquad\qquad+\phi_{n,m-1}(t) -2\phi_{n,m}(t)+\phi_{n,m+1}(t)\\
\nonumber
&\qquad\qquad\qquad+\phi_{n+1,m-1}(t)-2\phi_{n,m}(t)+\phi_{n-1,m+1}(t)\big]
\end{align}
with $n\in\mathbb{Z}$ and $m\in\mathbb{Z}$ indexing the lattice sites at each time $t\in\mathbb{R}$. See Fig.~\ref{FigLat} for the indexing convention. Here $\alpha\in\mathbb{R}$ is a dimensionless number playing the role of the dispersion rate. The terms in square brackets in the above expressions are the best possible approximations of the second derivative on each lattice which make use of only nearest neighbor interactions. These theories are named H1, H4, and H5 in anticipation of their further treatment later in this section.

In the above three theories time is still treated as continuous, while space is discrete. This is an incidental property of these examples. In \cite{DiscreteGenCovPart2} I extend the conclusions of this paper to Lorentzian theories. For instance, in \cite{DiscreteGenCovPart2} I consider the following discrete Klein Gordon equation among six others:
\begin{align}\label{KG1Long}
&\text{\bf 1D N.N. Klein Gordon Eq. (KG1):}\\
\nonumber
&[\phi_{j-1,n}-2\phi_{j,n}+\phi_{j+1,n}]\\
\nonumber
&\quad=[\phi_{j,n-1}-2\phi_{j,n}+\phi_{j,n+1}]
-\mu^2 \phi_{j,n}
\end{align}
with $j\in\mathbb{Z}$ indexing time, $n\in\mathbb{Z}$ indexing space and $\mu\in\mathbb{R}$ a dimensionless number playing the role of the field's mass. While KG1 is not Lorentz invariant, some lattice theories are. See \cite{DiscreteGenCovPart2} for details.

Returning to H1, H4, and H5, this section has promised to introduce these theories in an interpretation neutral way. As such, some of the above discussion needs to be hedged. In particular, in introducing these theories I have made casual comparison was made between parts of these theories' dynamical equations and various approximations of the second derivative. While, as I will discuss, such comparisons can be made, to do so immediately is unearned. It comes dangerously close to imagining the lattices shown in Fig.~\ref{FigLat} as being embedded in a continuous manifold. This may be something we want to do later (see Sec.~\ref{SecExtPart1}) but it is a non-trivial interpretational move which ought not be done so casually. 

Crucially, in this paper I will begin by analyzing these theories as discrete-native theories, that is as self-sufficient theories in their own right. We must not begin by thinking of them as various discretizations or bandlimitations of the continuum theories. While, as I will discuss, these discrete theories have some notable relationships to various continuum theories it is important to resist any temptation to see these continuum theories as ``where they came from''. Rather, let us pretend these theories ``came from nowhere'' and let us see what sense we can make of them. Contrast this with the following.

It is common in physics for theories to be set on a continuum manifold with dynamics which ultimately cause the theory's matter content to arrange itself lattice-wise, at least approximately. This is the case, for instance, regarding crystal structures considered by solid state physics. These crystal theories are not the sorts of lattice theories considered in this paper. For such crystal theories it may be convenient to approximate and recast their (fundamentally continuous) dynamics in terms of nearest-neighbor equations like Eq.~\eqref{H1Long} - Eq.~\eqref{H5Long}. However, it would be wrong to start trying to read-off the commitments (either material or spacetime) of these crystal theories from these approximated (and therefore non-fundamental) expressions. It would be wrong for the same reason it is wrong to expand QFT in a perturbative series and then take the resulting virtual particles seriously. One should be careful when trying to extract a theory's physical commitments, not to take its approximated forms too seriously. 

I am here taking Eq.~\eqref{H1Long} - Eq.~\eqref{H5Long} to be non-approximate and hence potentially worth taking seriously as expressing the real contents of these theories. As formulated above these theories seem to suggest that the spacetime manifold itself (not just the matter content) has a lattice structure. This is markedly different than the crystal theories discussed above. (However, as I will discuss throughout the paper, the above lattice-focused formulations of H1, H4 and H5 turn out to be problematic, with other reformulations being more fundamental.)

Another bit of hedging: in introducing the above three theories I casually associated them with the lattice structures shown in Fig.~\ref{FigLat}: square, hexagonal, etc. Making such associations ab initio is unwarranted. While we may eventually associate these theories with those lattice structures we cannot do so immediately. Such an association would need to be made following careful consideration of the dynamics. (Such an exercise is carried out in Sec.~\ref{SecHeat1}.) Beginning here in an interpretation-neutral way these theories ought to be seen as being defined over a completely unstructured lattice.

I will reflect this concern in my notation as follows. The labels for the lattice sites are presently too structured (e.g., $n\in\mathbb{Z}$ and $(n,m)\in\mathbb{Z}^2$). Instead we ought to think of the lattice sites as having labels $\ell\in L$ for some set $L$. Crucially, at this point the set of labels for the lattice sites, $L$, is just that, an unstructured set.

Up to isomorphism (here, generic bijections, i.e. generic relabelings), sets are uniquely specified by their cardinality. The set of labels for the lattice sites is here countable, $\ell\in L\cong\mathbb{Z}\cong\mathbb{Z}^2\cong\mathbb{Z}^3$. Reframed this way, the above discussed theories each consider the same discrete variables $\phi_\ell(t)\in\mathbb{R}$. In particular, H1 considers variables $\phi_\ell(t)$ which under some convenient relabeling of the lattice sites, $\ell\in L \mapsto n\in\mathbb{Z}$, satisfies Eq.~\eqref{H1Long}. Similarly, H4 and H5 consider variables $\phi_\ell(t)$ which under some convenient relabeling of the lattice sites, $\ell\in L \mapsto (n,m)\in\mathbb{Z}^2$, satisfy Eq.~\eqref{H4Long} and Eq.~\eqref{H5Long} respectively.

It's important to stress that the mere existence of these convenient relabelings by itself has no interpretative force. The fact that our labels $n\in\mathbb{Z}$ and $(n,m)\in\mathbb{Z}^2$ in some sense form a uniform 1D lattice and a square 2D lattice in no way forces us to think of $L$ as being structured in this way (indeed, we might later like to think of $L$ as a hexagonal 2D lattice). In particular, the fact that these labels are in a sense equidistant from each other does not force us to think of the lattice sites as being equidistant from each other. Nor are we forced to think that ``the distance between lattice sites'' to be meaningful at all. Dynamical considerations may later push us in this direction, but the mere convenience of this labeling should not.

I have above introduced three out of seven discrete heat equations. In order to introduce the other four theories it is convenient (but not necessary) to first reformulate things. In particular, let us reorganize the $\phi_\ell(t)$ variables into a vector, namely,
\begin{align}\label{PhiDef}
\bm{\Phi}(t)\coloneqq \sum_{\ell\in L} \phi_\ell(t) \, \bm{b}_\ell. 
\end{align}
where $\bm{b}_\ell$ is a linearly independent basis vector for each $\ell\in L$ and $\bm{\Phi}(t)$ is a vector in the vector space \mbox{$\mathbb{R}^L\coloneqq\text{span}(\{\bm{b}_\ell\}_{\ell\in L})$}. For later reference, it should be noted that $\phi_\ell(t)$ is also a vector in a vector space: namely, $F_L$ the space of functions $f:L\to\mathbb{R}$. Note that Eq.~\eqref{PhiDef} is an vector space isomorphism between these vector spaces, $\mathbb{R}^L\cong F_L$. Everything which follows concerning $\bm{\Phi}(t)\in\mathbb{R}^L$ has an isomorphic description in terms of $\phi_\ell(t)\in F_L$.

Recall that for H1 the lattice sites $\ell\in L$ have a convenient relabeling in terms of an integer index, \mbox{$\ell\in L\mapsto n\in\mathbb{Z}$}. We can use this relabeling to map $\mathbb{R}^L$ onto the vector space $\mathbb{R}^\mathbb{Z}\coloneqq\text{span}(\{\bm{e}_n\}_{n\in \mathbb{Z}})$ by taking $\bm{b}_\ell\mapsto\bm{e}_n$ where 
\begin{align}
\bm{e}_n = (\dots,0,0,1,0,0,\dots)^\intercal\in\mathbb{R}^\mathbb{Z}
\end{align}
with the 1 in the $n^\text{th}$ position. Under this restructuring of H1 we have, 
\begin{align}\label{PhiVec1}
\bm{\Phi}(t)&= \sum_{n\in\mathbb{Z}} \phi_{n}(t) \, \bm{e}_n\\
\nonumber
&=(\dots,\phi_{-1}(t),\phi_0(t),\phi_1(t),\phi_2(t),\dots)^\intercal. 
\end{align}
In these terms the dynamics of H1 is given by,
\begin{align}
\label{DH1}
&\text{\bf Heat Equation 1 (H1):}\\
\nonumber
&\frac{\d }{\d t}\bm{\Phi}(t)=\alpha \, \Delta_{(1)}^2 \bm{\Phi}(t) 
\end{align}
where $\Delta_{(1)}^2$ is the following bi-infinite Toeplitz matrix:
\begin{align}\label{Delta12}
\Delta_{(1)}^2=\{\Delta^+,\Delta^-\}
&=\text{Toeplitz}(1,\,-2,\,1)\\
\nonumber
\Delta^+&=\text{Toeplitz}(0,-1,\,1)\\
\nonumber
\Delta^-&=\text{Toeplitz}(-1,\,1,\,0)
\end{align}
where the curly brackets indicate the anticommutator, $\{A,B\}= \frac{1}{2}(A\,B + B\,A)$. Recall that Toeplitz matrices are so called diagonal-constant matrices with \mbox{$[A]_{i,j}=[A]_{i+1,j+1}$}. Thus, the values in the above expression give the matrix's values on either side of the main diagonal.

Although above I warned about thinking in terms of derivative approximations prematurely, a few comments are here warranted. Note that $\Delta^+$ is associated with the forward derivative approximation, $\Delta^-$ is be associated with the backwards derivative approximation, and $\Delta_{(1)}^2$ is associated with the nearest neighbor second derivative approximation,
\begin{align}
\nonumber
\Delta_{(1)}^2/\epsilon^2:\ \partial_x^2 f(x)
&\approx\frac{f(x+\epsilon)-2 f(x)+f(x-\epsilon)}{\epsilon^2}.
\end{align}
As stressed above, we ought to be cautious not to lean too heavily on these relationships when interpreting these discrete theories. 

In addition to H1, I will also consider two more theories with ``improved derivative approximations''. Namely,
\begin{align}
\label{DH2}
&\text{\bf Heat Equation 2 (H2):}\\
\nonumber
&\frac{\d }{\d t}\bm{\Phi}(t)=\alpha \, \Delta_{(2)}^2 \bm{\Phi}(t)\\
\label{DH3}
&\text{\bf Heat Equation 3 (H3):}\\
\nonumber
&\frac{\d }{\d t}\bm{\Phi}(t)=\alpha \, D^2 \bm{\Phi}(t).
\end{align}
where
\begin{align}\label{BigToeplitz}
\Delta_{(2)}^2&=\text{Toeplitz}(\frac{-1}{12},\,\frac{4}{3},\,\frac{-5}{2},\,\frac{4}{3},\,\frac{-1}{12})\\
\nonumber
D&=\text{Toeplitz}(\dots,\!\frac{-1}{5},\!\frac{1}{4},\!\frac{-1}{3},\!\frac{1}{2},\!-1,\!0,\!1,\!\frac{-1}{2},\!\frac{1}{3},\!\frac{-1}{4},\!\frac{1}{5},\!\dots)\\
\nonumber
D^2&=\text{Toeplitz}(\dots,\!\frac{-2}{16},\!\frac{2}{9},\!\frac{-2}{4},\!\frac{2}{1},\!\frac{-2\pi^2}{6},\!\frac{2}{1},\!\frac{-2}{4},\!\frac{2}{9},\!\frac{-2}{16},\!\dots).
\end{align}
Note that $\Delta_{(2)}^2$ is related to the next-to-nearest-neighbor approximation to the second derivative. Obviously, the longer range we make our derivative approximations the more accurate they can be. (For reasons I will discuss in Sec.~\ref{SecIntuitiveLocality}, this is actually a bit mysterious.) The infinite-range operator $D$ (and its square $D^2$) in some sense are the best discrete approximations to the derivative (and second derivative) possible. The defining property of $D$ is that it is diagonal in the (discrete) Fourier basis with spectrum,
\begin{align}\label{LambdaD}
\lambda_D(k)=-\ii\,\underline{k} 
\end{align}
where $\underline{k}=k$ for $k\in[-\pi,\pi]$ repeating itself cyclically with period $2\pi$ outside of this region. This is in tight connection with the continuum derivative operator $\partial_x$ which is diagonal in the (continuum) Fourier basis with spectrum \mbox{$\lambda_{\partial_x}(k)=-\ii\, k$} for $k\in[-\infty,\infty]$. 

Alternatively, one can construct $D^2$ in the following way: generalize $\Delta_{(1)}^2$ and $\Delta_{(2)}^2$ to $\Delta_{(n)}^2$ namely the best second derivative approximation which considers up to $n^\text{th}$ neighbors to either side. Taking the limit $n\to\infty$ gives $D^2=\lim_{n\to\infty}\Delta_{(n)}^2$. Other aspects of $D$ will be discussed in Sec.~\ref{SecSamplingTheory} (including its related derivative approximation Eq.~\eqref{ExactDerivative}) but enough has been said for now.

While these connections to derivative approximations allow us to export some intuitions from the continuum theories into these discrete theories, we must resist this (at least for now). In particular, I should stress again that we should not be thinking of any of H1, H2 and H3 as coming from the continuum theory under some approximation of the derivative.

Let's next reformulate H4 and H5 in terms of \mbox{$\bm{\Phi}(t)\in\mathbb{R}^L$}. In these cases we have a convenient relabeling of the lattice sites in terms of two integer indices, $\ell\mapsto (n,m)$. We can use this relabeling to grant the vector space a tensor product structure as $\mathbb{R}^L\mapsto\mathbb{R}^\mathbb{Z}\otimes\mathbb{R}^\mathbb{Z}$ by taking by taking \mbox{$\bm{b}_\ell\mapsto\bm{e}_n\otimes\bm{e}_m$}. Under this restructuring we have,
\begin{align}\label{PhiVec2}
\bm{\Phi}(t)= \sum_{n,m\in\mathbb{Z}} \phi_{n,m}(t) \, \bm{e}_n\otimes\bm{e}_m. 
\end{align}
In these terms the dynamics of H4 given above (namely, Eq.~\eqref{H4Long}) is now given by,
\begin{align}\label{DH4}
&\text{\bf Heat Equation 4 (H4):}\\
\nonumber
&\frac{\d}{\d t}\bm{\Phi}(t) =\frac{\alpha}{2} \, (\Delta^2_{(1),\text{n}}+ \Delta^2_{(1),\text{m}}) \, \bm{\Phi}(t),
\end{align}
where the notation $A_\text{n}\coloneqq A\otimes\openone$ and $A_\text{m}\coloneqq\openone\otimes A$ mean $A$ acts only on the first or second tensor space respectively. Thus, $\Delta^2_{(1),\text{n}}$ is just $\Delta^2_{(1)}$ applied to the first index, $n$. Likewise $\Delta^2_{(1),\text{m}}$ is just $\Delta^2_{(1)}$ applied to the second index, $m$.

A similar treatment of the dynamics of H5 (namely, Eq.~\eqref{H5Long}) gives us,
\begin{align}\label{DH5}
&\text{\bf Heat Equation 5 (H5):}\\
\nonumber
&\frac{\d}{\d t}\bm{\Phi}(t) =\frac{\alpha}{3} \, \Big[\Delta^2_{(1),\text{n}}+ \Delta^2_{(1),\text{m}}\\
\nonumber
&\qquad\qquad\quad \ +\big\{\Delta^+_\text{m}-\Delta^+_\text{n},\Delta^-_\text{m}-\Delta^-_\text{n}\big\}\Big] \, \bm{\Phi}(t),
\end{align}
While the third term looks complicated, it is just the analog of $\Delta^2_{(1),\text{n}}$ and $\Delta^2_{(1),\text{m}}$ but in the $m-n$ direction. See Eq.~\eqref{Delta12}.

Finally, in addition to H4 and H5 I consider the following two theories:
\begin{align}
\label{DH6}
&\text{\bf Heat Equation 6 (H6):}\\
\nonumber
&\frac{\d}{\d t}\bm{\Phi}(t) =\frac{\alpha}{2} \, (D^2_\text{n}+ D^2_\text{m}) \, \bm{\Phi}(t)\\
\label{DH7}
&\text{\bf Heat Equation 7 (H7):}\\
\nonumber
&\frac{\d}{\d t}\bm{\Phi}(t) =\frac{\alpha}{3}\left(D^2_\text{n}+D^2_\text{m}+(D_\text{m}-D_\text{n})^2\right) \, \bm{\Phi}(t)
\end{align}
which resemble H4 and H5 but which make use of an infinite range coupling between lattice sites. Having introduced these seven theories, let us next solve their dynamics.

\subsection{Solving Their Dynamics}
Conveniently, each of H1-H7 admit planewave solutions. Moreover, in each case these planewave solutions form a complete basis of solutions.

Considering first H1-H3 we have solutions,
\begin{align}
\phi_n(t;k)=\phi_n(k) \,e^{-\Gamma(k)\,t}\text{ where }\phi_n(k)\coloneqq e^{-\ii k \, n}
\end{align}
with $k\in\mathbb{R}$ and $\Gamma(k)$ some wavenumber-dependent decay rate. It should be noted however, that outside of the range $k\in[-\pi,\pi]$ these planewaves $\phi_n(k)$ repeat themselves with period $2\pi$ due to Euler's identity, $\exp(2\pi\ii)=1$. In terms of $\mathbb{R}^L\cong\mathbb{R}^\mathbb{Z}$ these planewaves are:
\begin{align}
\bm{\Phi}(k)=\sum_{n\in\mathbb{Z}} \phi_n(k) \, \bm{e}_n. 
\end{align}
From this planewave basis we can recover the $\bm{e}_n$ basis as:
\begin{align}
\bm{e}_n
&=\frac{1}{2\pi}\int_{-\pi}^\pi e^{\ii\, k\, n}\, \bm{\Phi}(k)\,\d k. 
\end{align}
For H1-H3 the wavenumber-dependent decay rate, $\Gamma(k)$, can be straight-forwardly calculated from the theory's dynamics:
\begin{align}
\text{H1}:& \quad \!\! \Gamma(k)= \alpha \, (2-2\text{cos}(k))\\
\nonumber
\text{H2}:& \quad \!\! \Gamma(k)= \frac{\alpha}{6}\,(\text{cos}(2\,k)-16\,\text{cos}(k)+15)\\
\nonumber
\text{H3}:& \quad \!\! \Gamma(k)= \alpha \, \underline{k}^2
\end{align}
where $\underline{k}=k$ for $k\in[-\pi,\pi]$ repeating itself cyclically with period $2\pi$ outside of this region.
Note that $\Gamma(k)$ for H3 follows Eq.~\eqref{LambdaD}, essentially from the definition of $D$.

\begin{figure}[t]
\includegraphics[width=0.4\textwidth]{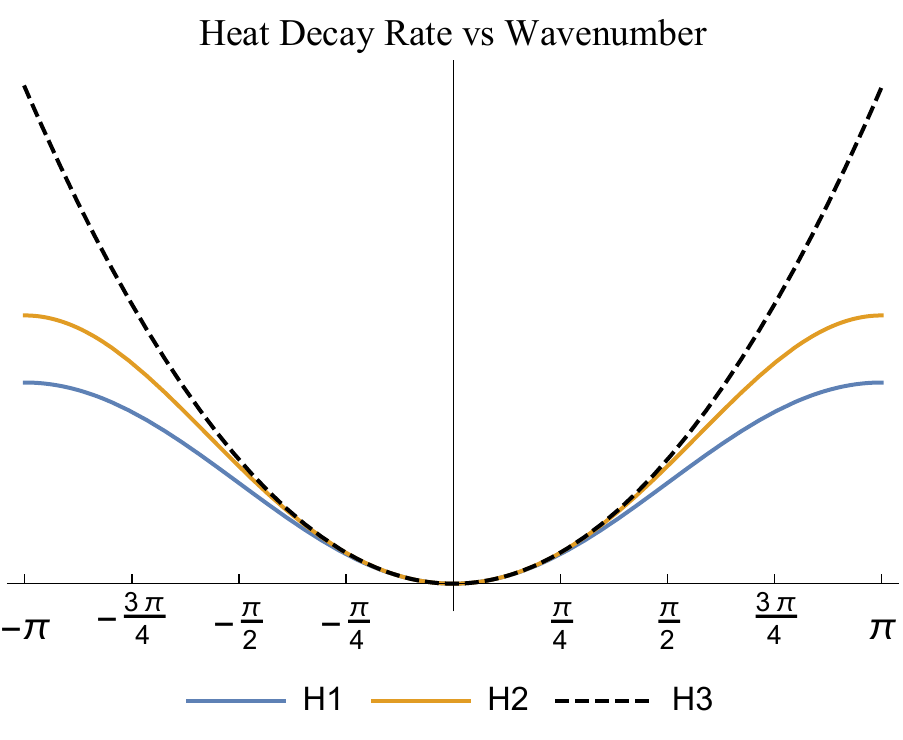}
\caption{The decay rates for the planewave solutions to the discrete heat equations are plotted as a function of wavenumber for H1, H2 and H3 (bottom to top). It should be noted that these decay rates repeats themselves cyclically with period $2\pi$ outside of this region.}\label{FigHeatDecay}
\end{figure}

Fig.~\ref{FigHeatDecay} shows these decay rates as a function of wavenumber restricted to $k\in[-\pi,\pi]$. Thus, what sets these theories apart is the rate at which high frequency planewaves decay. Let's investigate how these theories behave for planewaves with wavelengths which span many lattice sites, that is with $\vert k\vert \ll \pi$. 
\begin{align}
\text{H1}:& \quad \!\! \Gamma(k)= \alpha \, \left(k^2-\frac{k^4}{12}+\mathcal{O}(k^6)\right)\\
\nonumber
\text{H2}:& \quad \!\! \Gamma(k)= \alpha \left(k^2-\frac{k^6}{90}+\mathcal{O}(k^8)\right)\\
\nonumber
\text{H3}:& \quad \!\! \Gamma(k)= \alpha \, k^2.
\end{align}
Note that the decay rate $\Gamma(k)$ for H3 exactly matches the decay rate of the continuum theory not only in this regime but for all $k\in[-\pi,\pi]$. In the $\vert k\vert \ll \pi$ regime, H2 gives a better approximation of the continuum theory than H1 does, This is due to its longer range coupling giving a better approximation of the derivative. 

If we consider only solutions with all or most of their wavenumber support in Fourier space with $k \ll \pi$, we have an approximate one-to-one correspondence between the solutions to these theories. This is roughly why each of these theories have the same continuum limit, namely H00 defined above. In terms of the rate at which these theories converge to the continuum theory in the continuum limit, one can expect H3 to outpace H2 which outpaces H1. (As I will discuss in Sec.~\ref{SecIntuitiveLocality}, this is in a way counter-intuitive: why does the most non-local discrete theory give the best approximation of our perfectly local continuum theory?)

However, while interesting in their own right, these relationships with the continuum theory are not directly helpful in helping us understand H1-H3 in their own terms as discrete-native theories.

Moving on to H4-H7, their planewave solutions are, 
\begin{align}
&\phi_{n,m}(t;k_1,k_2)=\phi_{n,m}(k_1,k_2) \,e^{-\Gamma(k_1,k_2)\,t}\\
\nonumber
&\text{where }\phi_{n,m}(k_1,k_2)\coloneqq e^{-\ii k_1 n-\ii k_2 m}
\end{align}
with $k_1,k_2\in\mathbb{R}^2$. Again, it should be noted however, that outside of the range \mbox{$k_1,k_2\in[-\pi,\pi]$} these planewaves $\phi_{n,m}(k_1,k_2)$ repeat themselves with period $2\pi$ due to Euler's identity, $\exp(2\pi\ii)=1$. In terms of $\mathbb{R}^L\cong\mathbb{R}^\mathbb{Z}\otimes\mathbb{R}^\mathbb{Z}$ these planewaves are:
\begin{align}
\bm{\Phi}(k_1,k_2)=\sum_{n,m\in\mathbb{Z}} \phi_{n,m}(k_1,k_2) \, \bm{e}_n\otimes\bm{e}_m. 
\end{align}
From this planewave basis we can recover the $\bm{e}_n\otimes\bm{e}_m$ basis as:
\begin{align}
\nonumber
\bm{e}_n\otimes\bm{e}_m=\frac{1}{(2\pi)^2}\int\!\!\!\!\int_{-\pi}^\pi e^{\ii\, k_1 n+\ii\, k_2 m}\,\bm{\Phi}(k_1,k_2)\,\d k_1 \d k_2. 
\end{align}
The wavenumber-dependent decay rate $\Gamma(k_1,k_2)$ for each theory is given by:
\begin{align}
\text{H4}:& \ \Gamma(k_1,k_2)= \alpha \left(2\!-\!\text{cos}(k_1)\!-\!\text{cos}(k_2)\right)\\
\nonumber
\text{H5}:& \ \Gamma(k_1,k_2)= \frac{2\alpha}{3} \left(3\!-\!\text{cos}(k_1)\!-\!\text{cos}(k_2)\!-\!\text{cos}(k_2-k_1)\right)\\
\nonumber
\text{H6}:& \ \Gamma(k_1,k_2)= \frac{\alpha}{2} \, (\underline{k_1}^2+\underline{k_2}^2)\\
\nonumber
\text{H7}:& \ \Gamma(k_1,k_2)=
\frac{\alpha}{3}\left(\underline{k_1}^2+\underline{k_2}^2+(\underline{k_2}-\underline{k_1})^2\right).
\end{align}
Note that $\Gamma(k_1,k_2)$ for H6 and H7 follow from Eq.~\eqref{LambdaD}, essentially from the definition of $D$.

Unlike H1-H3, these theories do not all agree with each other in the small $k_1,k_2$ regime. H4 and H6 agree that for $\vert k_1\vert,\vert k_2\vert \ll \pi$ we have \mbox{$\Gamma(k_1,k_2)\approx \frac{\alpha}{2} (k_1^2+k_2^2)$}. Moreover, H5 and H7 agree with each other in this regime, but not with H4 and H6. Do we have two different results in the continuum limit here?

Closer examination reveals that we do not. The key to realizing this is to note that under the transformation, 
\begin{align}\label{SkewH7H6}
k_1\mapsto k_1,\quad
k_2\mapsto \frac{1}{2}\, k_1+\frac{\sqrt{3}}{2}\,k_2.
\end{align}
we have $\Gamma(k_1,k_2)$ for H7 mapping exactly onto $\Gamma(k_1,k_2)$ for H6.
The inverse of this map is
\begin{align}\label{SkewH6H7}
k_1\mapsto k_1,\quad
k_2\mapsto \frac{2k_2-k_1}{\sqrt{3}}.
\end{align}
Technically, when acting on the planewaves $\bm{\Phi}(k_1,k_2)$ these transformations are only each other's inverses when we have \mbox{$k_1,k_2\in[-\pi,\pi]$} both before and after the transformation. This is due to the $2\pi$ periodicity of these planwaves. A substantial portion of Fourier space has this property. As I will soon discuss, this means we have an exact one-to-one correspondence between a substantial portion of H6 and H7's solutions (much ado will be made about this later.) Applying this transformation to H5 does not map it onto H4, but it does bring their $\vert k_1\vert,\vert k_2\vert\ll\pi$ behavior into agreement.

Thus, if we consider only solutions with all or most of their wavenumber support with $\vert k_1\vert, \vert k_2\vert \ll \pi$ (or the appropriately transformed regime for H5 and H7) we have an approximate one-to-one correspondence between the solutions to these theories. Within this regime we can identify their common continuum limit, H0 defined above. Repeating our analysis of the convergence rates of H1-H3 here, we expect H6 and H7 to converge in the continuum limit faster than H4 and H5.

\vspace{0.25cm}

This paper will make three attempts at interpreting these seven discrete theories. Allow me to identify in advance three important points of comparison between these interpretations. 

The first important point of comparison is what sense they make of these different convergence rates in the continuum limit. As discussed above, in terms of this convergence rate we expect \mbox{$\text{H3}>\text{H2}>\text{H1}$} and similarly \mbox{$\text{H6}, \text{H7} > \text{H4}, \text{H5}$} with higher rated theories converging more quickly. As I will discuss in detail in Sec.~\ref{SecIntuitiveLocality}, this is in tension with our intuitive sense of locality for these theories: judging locality by the number of lattice sites coupled together we have
\mbox{$\text{H1} > \text{H2} > \text{H3}$} with higher rated theories being more local and similarly
\mbox{$\text{H4},\text{H5} > \text{H6},\text{H7}$}. How is it that our most non-local discrete theories are somehow the nearest to our perfectly local continuum theory?

A second important point of comparison between these three interpretations will be what sense they make of the above-noted exact one-to-one correspondence between a substantial portion of H6 and H7's solutions. (More will be said about this in Sec.~\ref{SecHeat2}.) It is important to note that the mere existence of such a one-to-one correspondence does not automatically mean that these theories are identical or even equivalent; All it means technically is that their solution spaces have the same cardinality. As I will discuss, some of the coming interpretations recognize H6 and H7 as being substantially identical whereas others do not.

A third important point of comparison between the coming interpretations will be what sense they make of these theories having continuous symmetries. For instance, the decay rate for H6 appears to be in some sense rotation invariant (at least in Fourier space and staying inside of the region \mbox{$k_1,k_2\in[-\pi,\pi]$}). In a sense, H7 might have these symmetries too: given the above-noted one-to-one correspondence between a substantial portion of the solutions of H6 and H7, there may be some (skewed) sense in which H7 is rotation invariant as well. All of this will be made precise later on. As I will discuss, some interpretations consider H6 and H7 to have a rotational symmetry whereas others do not.

Having introduced these theories and solved their dynamics in an interpretation-neutral way. We can now make a first (ultimately misled) attempt at interpreting them.

\section{A First Attempt at Interpreting H1-H7}\label{SecHeat1}
 Now that we have introduced these seven discrete theories and solved their dynamics, let's get on to interpreting them. Let us begin by following our first intuitions and analyze these seven discrete theories concerning their underlying manifold, locality properties and symmetries. Ultimately however, as I will discuss later, much of the following is misled and will need to be revisited and revised later. Luckily, retracing where we went wrong here will be instructive later.

Let's start by taking the initial formulation of the above theories in terms of $\phi_\ell(t)$ seriously, i.e. Eq.~\eqref{H1Long}, Eq.~\eqref{H4Long} and Eq.~\eqref{H5Long}. Taken literally as written, what are these theories about? Intuitively these theories are about a field $\phi_\ell(t)$ which maps lattice sites ($\ell\in L\cong\mathbb{Z}\cong\mathbb{Z}^2$) and times ($t\in\mathbb{R}$) into temperatures ($\phi_\ell(t)\in\mathbb{R}$). That is a field $\phi:Q\to \mathcal{V}$ with a discrete manifold $Q=L\times\mathbb{R}$ and value space $\mathcal{V}=\mathbb{R}$. Thus, taking $\phi:Q\to \mathcal{V}$ seriously as a fundamental field leads us to thinking of $Q=L\times\mathbb{R}$ as the theory's underlying manifold and $\mathcal{V}=\mathbb{R}$ as the theory's value space. It is important to note that here $Q$ is the entire manifold, it is not being thought of as embedded in some larger manifold. (However, a view like this will be considered in Sec.~\ref{SecExtPart1}.)

%For later reference it should be noted that the space of all functions $f:L\to\mathbb{R}$ is closed under addition and scalar multiplication and is therefore a vector space. Let us call this vector space $F_L$. Note that $F_L$ has a countably infinite dimension, $F_L\cong \mathbb{R}^\mathbb{Z}$. Indeed, $F_L\cong\mathbb{R}^L$ with Eq.~\eqref{PhiDef} being a vector space isomorphism between them. But let's continue here thinking in terms of $\phi_\ell(t)\in F_L$ rather than $\bm{\Phi}(t)\in\mathbb{R}^L$.

Let's see what consequences this interpretive stance has for these theories' locality and symmetry.

\subsection{Intuitive Locality}\label{SecIntuitiveLocality}
Firstly, let's develop a sense of comparative locality for H1, H2, and H3 taking $Q$ to be the underlying manifold. In a highly intuitive sense, theory H1 is the most local in that it couples together the fewest lattice sites: the instantaneous rate of change of $\phi_n(t)$ only depends on itself, $\phi_{n-1}(t)$, and $\phi_{n+1}(t)$. It is this sense of locality which justifies us calling these sites its ``nearest neighbors''. After this, H2 is the next most local in the same sense: it couples only next-to-nearest neighbors. Finally, in this sense H3 is the least local, it has an infinite range coupling: the instantaneous rate of change of $\phi_n(t)$ depends on the current value at every lattice site. Thus at least on this intuitive notion of locality, $\text{H1} > \text{H2} > \text{H3}$ with higher rated theories being more local. Similarly, assessing H4-H7 on this intuitive notion of locality gives the ratings, 
$\text{H4},\text{H5} > \text{H6},\text{H7}$.

There is some tension however with these intuitive locality ratings and the rate we expect each theory to converge at in the continuum limit. For H1-H3 our intuitive locality ratings are $\text{H1} > \text{H2} > \text{H3}$ but as discussed in the previous section we expect convergence rates in the continuum limit to be $\text{H3} > \text{H2} > \text{H1}$. Similar tension exists for H4-H7. How is it that our most non-local discrete theory is somehow the nearest to our perfectly local continuum theory? 

In one sense there is no mystery here, when we make our derivative approximation longer range (and intuitively more non-local) they can clearly get more accurate. But the question remains how exactly does a sequence of increasingly non-local operations on the lattice \mbox{(i.e, $\Delta^2_{(1)}\to\Delta^2_{(2)}\to\dots\to D^2$)} give us an increasingly good approximation of a perfectly local operation (i.e, $\partial_x^2$) on the continuum?

Not much can be said on this first interpretation to relieve this tension. However, this tension will be dissolved and resolved in our second and third interpretations respectively. In particular, as I will discuss in Sec.~\ref{SecInternalLocality} and Sec.~\ref{SecBandlimitedLocality}, these later interpretations do this by negating or reversing all of the above intuitive locality judgments.

\subsection{Intuitive Symmetries}
With this manifold $Q=L\times\mathbb{R}$ and value space $\mathcal{V}=\mathbb{R}$ picked out, what can we expect of these theories' symmetries? For any spacetime theory there are roughly three kinds of symmetries: 1) external symmetries associated with automorphisms of the manifold, here $\text{Auto}(Q)$, 2) internal symmetries associated with automorphisms of the value space, here $\text{Auto}(\mathcal{V})$, and gauge symmetries which result from allowing these internal symmetries to vary smoothly across the manifold. See Appendix~\ref{SecGenCov} for further discussion. But what are the relevant notions of automorphism here?

Answering this question for $\text{Auto}(Q)$ will require us to distinguish what structures are ``built into'' $Q$ and what are ``built on top of'' $Q$. The analogous distinction in the continuum case is that we generally take the manifold's differentiable structure to be built into it while the Minkowski metric, for instance, is something additional built on top of the manifold. In this paper, I am officially agnostic on where we draw this line in the discrete case. However, for didactic purposes I will here be as conservative as possible giving $Q$ as little structure as is sensible. Note that the less structure we associate with $Q$ the wider the class of relevant automorphisms $\text{Auto}(Q)$ will be. Thus, I am taking $\text{Auto}(Q)$ to be as large as it can reasonably be. 

Here the minimal structure we can reasonably associate with $Q=L\times\mathbb{R}$ is that of a set times a differentiable manifold. As such the largest $\text{Auto}(Q)$ could reasonably be is permutations of the lattice sites together with time reparemetrizations, \mbox{$\text{Auto}(Q)=\text{Perm}(L)\times\text{Diff}(\mathbb{R})$}. These act on $\phi_\ell(t)$ as 
\begin{align}
s_\text{External}:\quad&\phi_\ell(t)\mapsto \phi_{P(\ell)}(\tau(t))
\end{align}
for some smooth monotone $\tau(t)$ and some permutation $P:L\to L$.

In addition to $\text{Auto}(Q)$ we might also have internal symmetries $\text{Auto}(\mathcal{V})$ and gauge symmetries. While in general there may be abundant internal or gauge symmetries, for the present cases there are not many. In particular, for all of the above-mentioned theories we only have $\mathcal{V}=\mathbb{R}$. As mentioned following Eq.~\eqref{PhiDef}, our (potentially off-shell) discrete fields are themselves vectors $\phi_\ell(t)\in F_L$. Namely, they are closed under addition and scalar multiplication and hence form a vector space. This addition and scalar multiplication is carried out lattice-site-by-lattice-site at each time. Thus, the field's value space $\mathcal{V}=\mathbb{R}$ is also structured like a vector space. 

The value space $\mathcal{V}=\mathbb{R}$ may additionally have more structure than this. However, as above, for didactic purposes I will here minimize assumed structure in order to maximize possible symmetries. We can even drop the zero vector from our consideration taking $\mathcal{V}=\mathbb{R}$ to be an affine vector space. Therefore, I will take \mbox{$\text{Auto}(\mathcal{V})=\text{Aff}(\mathbb{R})$} such that our internal symmetries are linear-affine rescalings of $\phi_\ell(t)$, namely \mbox{$\phi_\ell(t)\mapsto c_1\phi_\ell(t)+c_2$}. We can then find the theory's gauge symmetries by letting $c_1,c_2\in\mathbb{R}$ vary smoothly across $Q$. That is, 
\begin{align}
s_\text{Gauge}:\quad &\phi_\ell(t)\mapsto c_{\ell,1}(t) \, \phi_{\ell}(t)+c_{\ell,2}(t)
\end{align}
for some smooth functions $c_{\ell,1}(t),\,c_{\ell,2}(t)\in\mathbb{R}$, 

Thus, in total, for H1-H7 the widest scope of symmetry transformations available to us (at least on this interpretation) are:
\begin{align}\label{PermutationLong}
s:\ \phi_\ell(t)\!\mapsto\! c_{P(\ell),1}(\tau(t)) \, \phi_{P(\ell)}(\tau(t))\!+c_{P(\ell),2}(\tau(t)) 
\end{align}

For later reference it will be convenient to translate these potential symmetry transformations in terms of the vector, $\bm{\Phi}(t)\in\mathbb{R}^L$, as 
\begin{align}\label{Permutation}
s:\quad\bm{\Phi}(t)\mapsto C_1(\tau(t))\,P\,\bm{\Phi}(\tau(t))+\bm{c}_2(\tau(t)),
\end{align}
for some permutation matrix, $P$, a diagonal matrix $C_1(t)$ and a vector $\bm{c}_2(t)$. Here $P$ and $\tau(t)$ captures the theory's possible external symmetries: the possibility of permuting lattice sites and reparametrizing time. The diagonal matrix $C_1(t)$ and the vector $\bm{c}_2(t)$ capture the theory's possible gauge symmetries: the possibility of linear-affine rescalings of $\phi_\ell(t)$ which vary smoothly across $Q$.

I will next discuss which transformations of this form preserve the dynamics of H1-H7. It should be clear from the outset however that (at least on this interpretation) these theories cannot have continuous spacial translation and rotation symmetries. Indeed, I have been charitable considering the lattice sites structured only as a set times a differentiable manifold (perhaps artificially) increasing the size of $\text{Auto}(Q)$. Given this, it would be highly surprising if we found H1-H7 to have symmetries outside of this set. (Such a surprise is coming in the Sec.~\ref{SecHeat2}.) 

Indeed, as I will show in Sec.~\ref{SecHeat2}, this first interpretation of these theories systematically under predicts the symmetries that discrete spacetime theories can and do have. Fixing this issue will lead us to develop the two discrete analogs of general covariance promised by this paper. We here under-predict symmetries because we are taking these theories' lattice structures too seriously. Properly understood, they are merely a coordinate-like representational artifact which do not limit our symmetries. Before that however, let's see the symmetries these theories have on this interpretation.

\subsubsection*{Symmetries of H1-H7: First Attempt}
What then are the symmetries of H1-H7 according to this interpretation? A technical investigation of the symmetries of H1-H7 on this interpretation is carried out in Appendix~\ref{AppB}, but the results are the following. For H1-H3 the dynamical symmetries of the form Eq.~\eqref{PermutationLong} are:
\begin{flushleft}\begin{enumerate}
 \item[1)] discrete shifts which map lattice site $n\mapsto n-d_1$ for some integer $d_1\in\mathbb{Z}$,
 \item[2)] negation symmetry which maps lattice site \mbox{$n\mapsto -n$},
 \item[3)] constant time shifts which map $\phi_\ell(t)\mapsto \phi_\ell(t-\tau)$ for some real $\tau\in\mathbb{R}$,
 \item[4)] global linear rescaling which maps \mbox{$\phi_\ell(t)\mapsto c_1\phi_\ell(t)$} for some $c_1\in\mathbb{R}$,
 \item[5)] local affine rescaling which maps \mbox{$\phi_\ell(t)\mapsto \phi_\ell(t) + c_{2,\ell}(t)$} for some $c_{2,\ell}(t)$ which is also a solution to the dynamics.
\end{enumerate}\end{flushleft}
These are the symmetries of a uniform 1D lattice \mbox{$z_n=n\in\mathbb{R}$} (plus time shifts and linear-affine rescalings). The above negation symmetry corresponds to mirror reflection. Previously I had warned against prematurely interpreting the lattice sites underlying H1-H3 as being organized into a uniform grid. Now, however, having investigated this theory's dynamical symmetries we have some motivation to do so.

What about H4 and H6? For H4 and H6 the dynamical symmetries of the form Eq.~\eqref{PermutationLong} are:
\begin{flushleft}\begin{enumerate}
 \item[1)] discrete shifts which map lattice site \mbox{$(n,m)\mapsto (n-d_2,m-d_3)$} for some integers $d_2,d_3\in\mathbb{Z}$,
 \item[2)] two negation symmetries which map lattice site $(n,m)\mapsto (-n,m)$ and $(n,m)\mapsto (n,-m)$ respectively,
 \item[3)] a 4-fold symmetry which maps lattice site \mbox{$(n,m)\mapsto (m,-n)$},
 \item[4)] constant time shifts which map $\phi_\ell(t)\mapsto \phi_\ell(t-\tau)$ for some real $\tau\in\mathbb{R}$,
 \item[5)] global linear rescaling which maps \mbox{$\phi_\ell(t)\mapsto c_1\phi_\ell(t)$} for some $c_1\in\mathbb{R}$,
 \item[6)] local affine rescaling which maps \mbox{$\phi_\ell(t)\mapsto \phi_\ell(t) + c_{2,\ell}(t)$} for some $c_{2,\ell}(t)$ which is also a solution to the dynamics.
\end{enumerate}\end{flushleft}
These are the symmetries of a square 2D lattice \mbox{$z_{n,m}=(n,m)\in\mathbb{R}^2$} (plus time shifts and linear-affine rescalings). The above 4-fold symmetry corresponds to quarter rotation. Previously I had warned against prematurely interpreting the lattice sites underlying H4-H7 as being organized into a square lattice. Now, however, having investigated these theories' dynamical symmetries we have some motivation to do so at least for H4 and H6.

What about H5 and H7? For H5 and H7 the dynamical symmetries of the form Eq.~\eqref{PermutationLong} are:
\begin{flushleft}\begin{enumerate}
 \item[1)] discrete shifts which map lattice site \mbox{$(n,m)\mapsto (n-d_2,m-d_3)$} for some integers $d_2,d_3\in\mathbb{Z}$,
 \item[2)] an exchange symmetry which maps lattice site $(n,m)\mapsto (m,n)$,
 \item[3)] a 6-fold symmetry which maps lattice site \mbox{$(n,m)\mapsto (-m,n+m)$}. (Roughly, this permutes the three terms in Eq.~\eqref{DH5} for H5 and Eq.~\eqref{DH7} for H7.),
 \item[4)] constant time shifts which map $\phi_\ell(t)\mapsto \phi_\ell(t-\tau)$ for some real $\tau\in\mathbb{R}$,
 \item[5)] global linear rescaling which maps \mbox{$\phi_\ell(t)\mapsto c_1\phi_\ell(t)$} for some $c_1\in\mathbb{R}$,
 \item[6)] local affine rescaling which maps \mbox{$\phi_\ell(t)\mapsto \phi_\ell(t) + c_{2,\ell}(t)$} for some $c_{2,\ell}(t)$ which is also a solution to the dynamics.
\end{enumerate}\end{flushleft}
These are the symmetries of a hexagonal 2D lattice \mbox{$z_{n,m}=(n+m/2,\sqrt{3}m/2)\in\mathbb{R}^2$} (plus time shifts and linear-affine rescalings). The above 6-fold symmetry corresponds to one-sixth rotation. Previously I had warned against prematurely interpreting the lattice sites underlying H4-H7 as being organized into a square lattice. Indeed, as we can now see for H5 and H7 this would have been faulty.

Thus, by investigating these theories' dynamical symmetries we were able to find what sort of lattice structure the assumed-to-be unstructured lattice $L$ actually has for each theory (e.g. a uniform 1D lattice, a square 2D lattice, and a hexagonal 2D lattice).

\vspace{0.25cm}

Finally, in this interpretation what sense can be made of H6 and H7 having a nice one-to-one correspondence between a substantial portion of their solutions as was discussed at the end of Sec.~\ref{SecSevenHeat}? While this correspondence between solutions certainly exists, little sense can be made of it here in support of the equivalence of these theories. As the above discussion has revealed, this interpretation associates very different symmetries to H6 and H7 and correspondingly very different lattice structures. While there is nothing technically wrong per se with this assessment our later interpretations will make better sense of this correspondence.

\vspace{0.25cm}

To summarize, this interpretation has the benefit of being highly intuitive. Taking the fields given to us, \mbox{$\phi:Q\to\mathbb{R}$}, seriously we identified the underlying manifold as $Q=L\times\mathbb{R}$. From this we got some intuitive notions of locality. Moreover, by finding these theories' dynamical symmetries we were able to grant some more structure to their assumed-to-be unstructured lattice sites (e.g. a uniform 1D lattice, square 2D lattice and a hexagonal 2D lattice). By and large, the interpretation seems to validate all of the first intuitions laid out in Sec.~\ref{SecCentralClaims}. On this interpretation, the lattice seems to play a substantive role in the theory: it seems to restrict our symmetries, it seems to distinguish our theories from one another, and it seems to be essentially ``baked-into'' the formalism. (As I will discuss in the next section, none of this is right.)

However, there are three major issues with this interpretation which will become clear in light of our later interpretations. Firstly, our locality assessments are in tension with the rates at which these theories converge to the (perfectly local) continuum theory in the continuum limit. Secondly, despite the niceness of the one-to-one correspondence between substantial portions of the solutions to H6 and H7, this interpretation regards them as significantly different theories: with different lattice structures and (here consequently) different symmetries. The final issue (which will become clear in the next section) is that this interpretation drastically under predicts the kinds of symmetries which H1-H7 can and do have. In fact, each of H1-H7 have a hidden continuous translation symmetry. Moreover, H6 and H7 have a hidden continuous rotational symmetry.

As I will discuss, the root of all of these issues is taking the theory's lattice structure too seriously. As I will argue, when properly understood, they are merely a coordinate-like representational artifact. Indeed, as I will show in the next section they do not limit our theory's symmetries. Moreover, theories appearing initially with different lattice structures may nonetheless be equivalent. Finally, these theories can always be reformulated to refer to no lattice structure at all. Making these three points will establish a rich analogy between the lattice structures appearing in our discrete spacetime theories and the coordinate systems appearing in our continuum theories. Ultimately, spelling out this analogy in detail in Sec.~\ref{SecDisGenCov} will give us a discrete analog of general covariance.

\section{A Second Attempt at Interpreting H1-H7}\label{SecHeat2}
In the previous section, I claimed that H1-H7 have hidden continuous translation and rotation symmetries. But how can this be? How can discrete spacetime theories (i.e. lattice theories) have such continuous symmetries? As I discussed in the previous section, if we take our underlying manifold to be $Q=L\times\mathbb{R}$ then these theories clearly cannot support continuous translation and rotation symmetries.

In order to avoid this conclusion we must deny the premise, $Q$ must not be the underlying manifold. What led us to believe $Q$ was the underlying manifold? We arrived at this conclusion by focusing on $\phi_\ell(t)\in F_L$ and thereby taking the real scalar field $\phi:Q\to\mathcal{V}$ to be fundamental. $Q$ is the underlying manifold because it is where our fundamental field maps from. In order to avoid this conclusion we must deny the premise, the field $\phi:Q\to\mathcal{V}$ must not be fundamental. 

But if $\phi:Q\to\mathcal{V}$ is not fundamental then what is? Fortunately, our above discussion has already provided us with another field which we might take as fundamental. Namely, the vector field $\bm{\Phi}(t)$ defined in Eq.~\eqref{PhiDef}. These vectors \mbox{$\bm{\Phi}(t)\in\mathbb{R}^L$} are in a one-to-one correspondence with the discrete fields $\phi_\ell(t)\in F_L$. Moreover, as noted following Eq.~\eqref{PhiDef}, these vector spaces are isomorphic $\mathbb{R}^L\cong F_L$ with Eq.~\eqref{PhiDef} being a vector space isomorphism between them.

On this second interpretation I will be taking the formulations of H1-H7 in terms of $\bm{\Phi}(t)\in\mathbb{R}^L$ seriously: namely Eq.~\eqref{DH1}, Eq.~\eqref{DH2}, Eq.~\eqref{DH3}, and Eqs.~\eqref{DH4}-\eqref{DH7}. Taken literally as written, what are these theories about? These theories are intuitively about a field $\bm{\Phi}(t)$ which maps times ($t\in\mathbb{R}$) into infinite dimensional vectors (\mbox{$\bm{\Phi}(t)\in\mathbb{R}^L$}). That is a field $\bm{\Phi}:\mathcal{M}\to \mathcal{V}$ with manifold $\mathcal{M}=\mathbb{R}$ of times and value space $\mathcal{V}=\mathbb{R}^L$. Taking $\bm{\Phi}:\mathcal{M}\to \mathcal{V}$ seriously as a fundamental field leads us to thinking of $\mathcal{M}=\mathbb{R}$ (not $Q=L\times\mathbb{R}$) as the theory's underlying manifold. Indeed, in this interpretation H1-H7 are continuum spacetime theories of the sort we are used to interpreting (albeit ones with an abnormally high-dimensional value space).

Notice that in this interpretation the lattice sites, $L$, are no longer a part of our manifold. They have been ``internalized'' into the value space $\mathcal{V}=\mathbb{R}^L$. In particular, in defining this vector space we have associated with each lattice site $\ell\in L$ a basis vector $\bm{b}_\ell$. See Eq.~\eqref{PhiDef}. However, as I will discuss, these particular basis vectors play no special role in these theories. Indeed, looking back at the dynamics for each of H1-H7 written in terms of $\bm{\Phi}(t)$, one can see that in each case it can be made basis-independent. 

Let's see what consequences this interpretive stance has for these theories' locality and symmetry. To preview: this second interpretation either dissolves or resolves all of our issues with the first interpretation. The tension between locality and convergence in the continuum limit is dissolved. H6 and H7 are seen to be equivalent in a stronger sense. And, perhaps most importantly, this interpretation reveals H1-H7's hidden continuous translation and rotation symmetries. However, as I will discuss, this interpretation has some issues of its own which will ultimately require us to make a third attempt at interpreting these theories in Sec.~\ref{SecHeat3}.

\subsection{Internalized Locality}\label{SecInternalLocality}
Before discussing the effect this internalization move has on the theories' possible symmetries, let's think briefly about what it does to our sense of locality. I claimed above that this interpretation dissolves the tension between convergence in the continuum limit and the intuitive sense of locality developed in Sec.~\ref{SecIntuitiveLocality}. It does this by dissolving the possibility of any notion of locality stemming from the lattice sites. 

In this interpretation the lattice sites have been internalized, they are no longer part of the spacetime manifold and therefore we no longer have a right to extract intuitions about locality from them. In this interpretation, the manifold consists only of times, $t\in\mathbb{R}$. Consequently our only notion of locality is locality in time. The dynamics of each of H1-H7 are local in time and are therefore local in every possible sense. There is no longer any tension concerning how the differences in locality line up with the differences in continuum convergence rate; there simply are no differences in locality anymore.

If this seems unsatisfying to you I agree. One might feel that internalization's ban on extracting notions of locality from the lattice sites is far too extreme. Intuitively, more strongly coupled lattice sites ought to be in some sense closer together. Moreover, one might rightly hope for an interpretation which not only dissolves the tension between a theory's locality and its convergence rates in the continuum limit, but rather resolves it by bringing them into harmony. Indeed, if we have no notion of locality between lattice sites it is difficult to see how we get a notion of locality in the continuum limit.

These are all valid complaints which will be addressed in Sec.~\ref{SecHeat3} as I make a third attempt at interpreting H1-H7. 

\subsection{Internalized Symmetries}
But how does this internalization move affect a theory's capacity for symmetry? How can we now have continuous translation and rotation symmetries? At first glance, this may appear to have made things worse. If our manifold is now just times $t\in\mathbb{R}$ then our only possible external symmetries are time-reparametrizations (i.e., not continuous translations and rotations). However, while there are certainly less possible external symmetries, we are now open to a wider range of internal (and gauge) symmetries. It is among these internal symmetries that we will find H1-H7's hidden continuous translation and rotation transformations. As I will argue these symmetries can reasonably be given these names despite being internal symmetries. (In Sec.~\ref{SecHeat3} I will present a third attempt at interpreting these theories which ``externalizes'' these symmetries, making them genuinely spacial translations and rotations.)

With our focus now on $\bm{\Phi}:\mathcal{M}\to \mathcal{V}$, let us consider its possibilities for symmetries. As discussed above, associated with the manifold we have only time reparametrizations, $\text{Auto}(\mathcal{M})=\text{Diff}(\mathbb{R})$. However, associated with the value space (i.e., an infinite dimensional vector space) we now have the full range of invertible linear-affine transformations over $\mathbb{R}^L$, namely $\text{Auto}(\mathcal{V})=\text{Aff}(\mathbb{R}^L)$. Allowing these internal symmetries to vary smoothly across the manifold, we may also have gauge symmetries.

Thus, taken together the possibly symmetries for our theories under this interpretation are,
\begin{align}\label{GaugeVR}
s:\quad\bm{\Phi}(t)\mapsto \Lambda(\tau(t))\,\bm{\Phi}(\tau(t))+\bm{c}(\tau(t))
\end{align}
for some smooth monotone $\tau(t)$, any smoothly-varying invertible linear transformation $\Lambda(t)\in\text{GL}(\mathbb{R}^L)$, and any smoothly-varying vector $\bm{c}(t)\in \mathbb{R}^L$.

Contrast this with the symmetries available to us on our first interpretation, namely Eq.~\eqref{Permutation}. We can role this new class of possible symmetries back onto our first interpretation as follows: Note that because $\mathbb{R}^L \cong F_L$ we have $\text{Aff}(\mathbb{R}^L)\cong \text{Aff}(F_L)$. So translated, the above transformations are $\text{Aff}(F_L)$ smoothly varying over time, \mbox{$\text{Diff}(\mathbb{R})$}. This is much larger than what we previously considered: namely, \mbox{$\text{Auto}(\mathcal{V})=\text{Aff}(\mathbb{R})$} varying smoothly over \mbox{$\text{Auto}(Q)=\text{Perm}(L)\times\text{Diff}(\mathbb{R})$}. Indeed, the present interpretation has a significantly wider class of symmetries than before. 

Moving back to our second interpretation, our previous class of transformations (i.e, \mbox{$\text{Auto}(\mathcal{V})=\text{Aff}(\mathbb{R})$} varying smoothly over \mbox{$\text{Auto}(Q)=\text{Perm}(L)\times\text{Diff}(\mathbb{R})$}) corresponds to only a subset of our present consideration:  $\text{Aff}(\mathbb{R}^L)$ varying smoothly over \mbox{$\text{Diff}(\mathbb{R})$}. The difference is that before we could only apply a permutation matrix $P$ followed by a diagonal matrix $C_1(t)$ whereas now we are allowed a general linear transformation $\Lambda(t)$. 

Note that permutation and diagonal matrices are basis-dependent notions. Our first interpretation took the lattice sites $\ell\in L$ seriously as a part of the manifold $Q=L\times\mathbb{R}$ and this is reflected in its conception of symmetries. Converted into $\mathbb{R}^L$ this first conception of these theories' possible symmetries gives special treatment to the basis associated with the lattice sites, namely $\{\bm{b}_\ell\}_{\ell\in L}$. In particular, on our first interpretation, our possible symmetries are those of the form Eq.~\eqref{GaugeVR} which \textit{additionally} preserve this basis (up to rescaling, and reordering).

This basis receives no special treatment on this second interpretation. While it is true that $\{\bm{b}_\ell\}_{\ell\in L}$ were used in the initial construction of $\bm{\Phi}(t)$, after this they no longer play any special role. We are always free to redescribe $\bm{\Phi}(t)$ in a different basis if we wish. Indeed, here any change of basis transformation is of the form Eq.~\eqref{GaugeVR} and hence a symmetry.

With this attachment to the basis $\{\bm{b}_\ell\}_{\ell\in L}$ dropped, we have a wider class of symmetries. Indeed, everything which was previously considered a symmetry will be here as well and possibly more. Perhaps among this more general class of symmetries we may find continuous translation and rotation symmetries. Let's see.

\subsubsection*{Symmetries of H1-H7: Second Attempt}
Which of the above transformations are symmetries for H1-H7? A non-exhaustive investigation of the symmetries of H1-H7 on this interpretation is carried out in Appendix~\ref{AppB}, but the results are the following. For H1-H3 the dynamical symmetries of the form Eq.~\eqref{GaugeVR} are:
\begin{flushleft}\begin{enumerate}
 \item[1)] action by $T^\epsilon$ sending $\bm{\Phi}(t)\mapsto T^\epsilon \bm{\Phi}(t)$ where $T^\epsilon$ is defined below.
 \item[2)] a negation symmetry which maps basis vectors as $\bm{e}_n\mapsto \bm{e}_{-n}$,
 \item[3)] constant time shifts $T^\tau_\text{t}$ which map $T^\tau_\text{t}:\bm{\Phi}(t)\mapsto \bm{\Phi}(t-\tau)$ for some real $\tau\in\mathbb{R}$,
 \item[4)] a local Fourier rescaling symmetry which maps \mbox{$\bm{\Phi}(k)\mapsto \tilde{f}(t;k)\,\bm{\Phi}(k)$} for some non-zero complex function $\tilde{f}(t;k)\in\mathbb{C}$ with $t\in\mathbb{R}$ and $k\in[-\pi,\pi]$,
 \item[5)] local affine rescaling which maps \mbox{$\bm{\Phi}(t)\mapsto \bm{\Phi}(t) + \bm{c}_2(t)$} for some $\bm{c}_2(t)$ which is also a solution of the dynamics.
\end{enumerate}\end{flushleft}
These are exactly the same symmetries that we found on the previous interpretation with two differences: Firstly, global rescaling $\phi_\ell\mapsto c_1 \phi_\ell$ has been refined to a local Fourier rescaling. (Note that $\text{Aff}(\mathbb{R}^L)$ is very wide indeed: the discrete Fourier transform itself is in $\text{Aff}(\mathbb{R}^L)$ and so is in the class of potential symmetries considered here.)

Secondly, discrete shifts have been replaced with action by 
\begin{align}\label{TDef}
T^\epsilon\coloneqq\text{exp}(-\epsilon D)
\end{align}
with $\epsilon\in\mathbb{R}$. A straight-forward calculation shows that $T^\epsilon$ acts on the planewave basis $\bm{\Phi}(k)$ with $k\in[-\pi,\pi]$ as
\begin{align}
\nonumber
T^\epsilon:\bm{\Phi}(k)\mapsto \exp(\ii k \epsilon)\,\bm{\Phi}(k) 
\end{align}
and the basis $\bm{e}_m$ as
\begin{align}
T^\epsilon: \bm{e}_m \mapsto \sum_{b\in\mathbb{Z}} S_m(b+\epsilon) \, \bm{e}_b
\end{align}
where 
\begin{align}\label{SincDef}
S(y)=\frac{\sin(\pi y)}{\pi y}, \quad\text{and}\quad
S_m(y)=S(y-m), 
\end{align}
are the normalized and shifted sinc functions. Note that $S_n(m)=\delta_{nm}$ for integers $n$ and $m$.

As I will now discuss, $T^\epsilon$ can be thought of as a continuous translation operator for three reasons despite it being here classified as an internal symmetry. Note that none of these reasons depend on $T^\epsilon$ being a symmetry of the dynamics.

First note that $T^\epsilon$ is a generalization of the discrete shift operation in the sense that taking $\epsilon=d_1\in\mathbb{Z}$ reduces action by $T^\epsilon$ to the map $T^{d_1}:\bm{e}_m\mapsto \bm{e}_{m-d_1}$ on basis vectors and relatedly the map $m\mapsto m-d_1$ on lattice sites. 

Second note that $T^\epsilon$ is additive in the sense that $T^{\epsilon_1}\,T^{\epsilon_2}=T^{\epsilon_1+\epsilon_2}$. In the language or representation theory $T^\epsilon$ is a representation of the translation group on the vector space $\mathbb{R}^\mathbb{Z}\cong\mathbb{R}^L$. In particular, this means \mbox{$T^{1/2}\,T^{1/2}=T^1$}: there is something we can do twice to move one space forward. The same is true for all fractions adding to one. 

Third, recall from the discussion following Eq.~\eqref{LambdaD} that $D$ is closely related to the continuum derivative operator $\partial_x$, exactly matching its spectrum for $k\in[-\pi,\pi]$. Recall also that the derivative is the generator of translation, i.e. $h(x-\epsilon)=\text{exp}(-\epsilon\, \partial_x) h(x)$. In this sense also \mbox{$T^\epsilon=\text{exp}(-\epsilon D)$} is a translation operator. More will be said about $T^\epsilon$ in Sec.~\ref{SecSamplingTheory}.

Thus we have our first big lesson: despite the fact that H1-H3 can be represented on a lattice, they nonetheless have a continuous translation symmetry. This continuous translation symmetry was hidden from us on our first interpretation because we there took the lattice to be hard-wired in as a part of the manifold. Here, we do not take the lattice structure so seriously. We have internalized it into the value space where it then disappears from view as just one basis among many.

Next let's consider H4-H7. In the previous interpretation the symmetries of H4 and H6 matched each other, both being associated with a square 2D lattice. Moreover, the symmetries of H5 and H7 matched each other, both being associated with a hexagonal 2D lattice. Here, however, these pairings are broken up and a new matching pair is formed between H6 and H7. More will be said about this momentarily.

Let's consider H4 first. For H4 the dynamical symmetries of the form Eq.~\eqref{GaugeVR} include:
\begin{flushleft}\begin{enumerate}
 \item[1)] action by $T^\epsilon_\text{n}$ sending $\bm{\Phi}(t)\mapsto T^\epsilon_\text{n} \bm{\Phi}(t)$ where \mbox{$T^\epsilon_\text{n}=T^\epsilon\otimes\openone$}. Similarly for $T^\epsilon_\text{m}=\openone\otimes T^\epsilon$,
 \item[2)] two negation symmetries which map basis vectors as \mbox{$\bm{e}_{n}\otimes\bm{e}_{m}\mapsto \bm{e}_{-n}\otimes\bm{e}_{m}$}, and \mbox{$\bm{e}_{n}\otimes\bm{e}_{m}\mapsto \bm{e}_{n}\otimes\bm{e}_{-m}$} respectively,
 \item[3)] a 4-fold symmetry which maps basis vectors as \mbox{$\bm{e}_{n}\otimes\bm{e}_{m}\mapsto \bm{e}_{m}\otimes\bm{e}_{-n}$},
 \item[4)] constant time shifts $T^\tau_\text{t}$ which map $T^\tau_\text{t}:\bm{\Phi}(t)\mapsto \bm{\Phi}(t-\tau)$ for some real $\tau\in\mathbb{R}$,
 \item[5)] a local Fourier rescaling symmetry which maps \mbox{$\bm{\Phi}(k_1,k_2)\mapsto \tilde{f}(t;k_1,k_2)\,\bm{\Phi}(k_1,k_2)$} for some non-zero complex function $\tilde{f}(t;k_1,k_2)\in\mathbb{C}$ with $t\in\mathbb{R}$ and $k_1,k_2\in[-\pi,\pi]$,
 \item[6)] local affine rescaling which maps \mbox{$\bm{\Phi}(t)\mapsto \bm{\Phi}(t) + \bm{c}_2(t)$} for some $\bm{c}_2(t)$ which is also a solution of the dynamics.
\end{enumerate}\end{flushleft}
These are exactly the symmetries which we found on our first interpretation (plus local Fourier rescaling) but with action by $T^\epsilon_\text{n}$ and $T^\epsilon_\text{m}$ replacing the discrete shifts. The same discussion following Eq.~\eqref{TDef} applies here, justifying us calling these continuous translation operations. Thus, H4 has two continuous translation symmetries despite initially being represented on a lattice, Eq.~\eqref{H4Long}.

Let's next consider H5. For H5 the dynamical symmetries of the form Eq.~\eqref{GaugeVR} include:
\begin{flushleft}\begin{enumerate}
 \item[1)] action by $T^\epsilon_\text{n}$ sending $\bm{\Phi}(t)\mapsto T^\epsilon_\text{n} \bm{\Phi}(t)$ where \mbox{$T^\epsilon_\text{n}=T^\epsilon\otimes\openone$}. Similarly for $T^\epsilon_\text{m}=\openone\otimes T^\epsilon$,
 \item[2)] an exchange symmetry which maps basis vectors as \mbox{$\bm{e}_{n}\otimes\bm{e}_{m}\mapsto \bm{e}_{m}\otimes\bm{e}_{n}$},
 \item[3)] a 6-fold symmetry which maps basis vectors as \mbox{$\bm{e}_{n}\otimes\bm{e}_{m}\mapsto \bm{e}_{-m}\otimes\bm{e}_{n+m}$}. (Roughly, this permutes the three terms in Eq.~\eqref{DH5}),
 \item[4)] constant time shifts $T^\tau_\text{t}$ which map $T^\tau_\text{t}:\bm{\Phi}(t)\mapsto \bm{\Phi}(t-\tau)$ for some real $\tau\in\mathbb{R}$,
 \item[5)] a local Fourier rescaling symmetry which maps \mbox{$\bm{\Phi}(k_1,k_2)\mapsto \tilde{f}(t;k_1,k_2)\,\bm{\Phi}(k_1,k_2)$} for some non-zero complex function $\tilde{f}(t;k_1,k_2)\in\mathbb{C}$ with $t\in\mathbb{R}$ and $k_1,k_2\in[-\pi,\pi]$,
 \item[6)] local affine rescaling which maps \mbox{$\bm{\Phi}(t)\mapsto \bm{\Phi}(t) + \bm{c}_2(t)$} for some $\bm{c}_2(t)$ which is also a solution of the dynamics.
\end{enumerate}\end{flushleft}
These are exactly the symmetries which we found on our first interpretation (plus local Fourier rescaling) but with action by $T^\epsilon_\text{n}$ and $T^\epsilon_\text{m}$ replacing the discrete shifts. The same discussion following Eq.~\eqref{TDef} applies here, justifying us calling these continuous translation operations. Thus, H5 has two continuous translation symmetries despite initially being represented on a lattice, Eq.~\eqref{H5Long}.

Before moving on to analyze the symmetries of H6 and H7, let's first see what this interpretation has to say about them being equivalent to one another. As noted at the end of Sec.~\ref{SecSevenHeat}, H6 and H7 have a nice one-to-one correspondence between a substantial portion of their solutions. Allow me to spell this out in detail now. 

Before this, however, it is worth briefly noting a rather weak sense in which each of H4-H7 are equivalent to each other. As noted following Eq.~\eqref{SkewH7H6} and Eq.~\eqref{SkewH6H7} there is an approximate one-to-one correspondence between each of these theories in the $\vert k_1\vert,\vert k_2\vert\ll\pi$ regime as they approach their common continuum limit, H0. By contrast, as I will show, H6 and H7 have an \textit{exact} one-to-one correspondence over \textit{the whole of} $\sqrt{k_1^2+k_2^2}<\pi$ (and indeed more, but ultimately not all of $k_1,k_2\in[-\pi,\pi]$).

This one-to-one correspondence is mediated by the transformations Eq.~\eqref{SkewH7H6} and Eq.~\eqref{SkewH6H7}. Let's first rewrite these in terms of $\bm{\Phi}(t)\in\mathbb{R}^L\cong\mathbb{R}^\mathbb{Z}\otimes\mathbb{R}^\mathbb{Z}$ as follows. Consider first the transformation which maps the decay rate for H7 onto the one for H6 (namely, Eq.~\eqref{SkewH7H6}). Consider its action on planewave basis $\bm{\Phi}(k_1,k_2)$ with \mbox{$k_1,k_2\in[-\pi,\pi]$}, namely
\begin{align}
\Lambda_{\text{H7}\to\text{H6}}:\bm{\Phi}(k_1,k_2)\mapsto \bm{\Phi}\left(k_1,\frac{1}{2}k_1+\frac{\sqrt{3}}{2}k_2\right). \end{align}
A straight-forward calculation shows this acts on the \mbox{$\bm{e}_n\otimes\bm{e}_m$} basis as:
\begin{align}\label{SkewH6H7Basis}
&\Lambda_{\text{H7}\to\text{H6}}:\bm{e}_n\otimes\bm{e}_m\mapsto \\
\nonumber
&\qquad\qquad\sum_{b_1,b_2\in\mathbb{Z}}S_n(b_1+b_2/2) \,S_m(\sqrt{3}\, b_2/2)\, \bm{e}_{b_1}\otimes\bm{e}_{b_2}. 
\end{align}
Consider also the transformation which maps the decay rate for H6 onto the one for H7 (namely, Eq.~\eqref{SkewH6H7}). Consider its action on planewave basis $\bm{\Phi}(k_1,k_2)$ with \mbox{$k_1,k_2\in[-\pi,\pi]$}, namely
\begin{align}
\Lambda_{\text{H6}\to\text{H7}}:\bm{\Phi}(k_1,k_2)\mapsto \bm{\Phi}\left(k_1,\frac{2k_2-k_1}{\sqrt{3}}\right) 
\end{align}
A straight-forward calculation shows this acts on the \mbox{$\bm{e}_n\otimes\bm{e}_m$} basis as:
\begin{align}
&\Lambda_{\text{H6}\to\text{H7}}:\bm{e}_n\otimes\bm{e}_m\mapsto \\
\nonumber
&\qquad\qquad\sum_{b_1,b_2\in\mathbb{Z}}S_n(b_1-b_2/\sqrt{3}) \,S_m(2\, b_2/\sqrt{3})\, \bm{e}_{b_1}\otimes\bm{e}_{b_2}. 
\end{align}

It should be noted, however, that despite the fact that Eq.~\eqref{SkewH7H6} and Eq.~\eqref{SkewH6H7} are each other's inverses, $\Lambda_{\text{H7}\to\text{H6}}$ and $\Lambda_{\text{H6}\to\text{H7}}$ are not each other's inverses (at least not on the whole of $\mathbb{R}^L\cong\mathbb{R}^\mathbb{Z}\otimes\mathbb{R}^\mathbb{Z}$). This is due to the $2\pi$ periodic identification of the planewaves $\bm{\Phi}(k_1,k_2)$. Indeed, when viewed as acting on $\mathbb{R}^L$, the transformation $\Lambda_{\text{H7}\to\text{H6}}$ is not even invertible. In parallel to this, recall that, as discussed following Eq.~\eqref{SkewH7H6} and Eq.~\eqref{SkewH6H7}, these transformations don't map the decay rate for H6 and H7 onto each other everywhere. They only do so when we have \mbox{$k_1,k_2\in[-\pi,\pi]$} both before and after these transformations.

For these reasons we need to consider the following two vector subspaces of $\mathbb{R}^L\cong\mathbb{R}^\mathbb{Z}\otimes\mathbb{R}^\mathbb{Z}$: 
\begin{align}
\mathbb{R}^L_\text{H7}\coloneqq\text{span}(&\bm{\Phi}(k_1,k_2)\vert\,\text{with }k_1,k_2\in[-\pi,\pi]\\
\nonumber
&\text{ before and after applying Eq.~\eqref{SkewH7H6}})\\
\mathbb{R}^L_\text{H6}\coloneqq\text{span}(&\bm{\Phi}(k_1,k_2)\vert\,\text{with }k_1,k_2\in[-\pi,\pi]\\
\nonumber
&\text{ before and after applying Eq.~\eqref{SkewH6H7}}).
\end{align}
For later reference it should be noted that the rotation invariant subspace,
\begin{align}
\label{RLrotinv}
\mathbb{R}^L_\text{rot.inv}\coloneqq\text{span}&\left(\bm{\Phi}(k_1,k_2)\Big\vert\,\sqrt{k_1^2+k_2^2}<\pi\right).
\end{align}
is a subspace of $\mathbb{R}^L_\text{H6}$, that is $\mathbb{R}^L_\text{rot.inv}\subset \mathbb{R}^L_\text{H6}$.

Restricted to $\mathbb{R}^L_\text{H6}$ and $\mathbb{R}^L_\text{H7}$ these transformations are invertible and indeed are each other's inverses. Thus, $\Lambda_{\text{H6}\to\text{H7}}$ and $\Lambda_{\text{H7}\to\text{H6}}$ give us not only a (partial) one-to-one correspondence between H6 and H7 but a solution-preserving vector-space isomorphism between them in the following sense. Restricting our attention to $\mathbb{R}^L_\text{H6}$ and $\mathbb{R}^L_\text{H7}$ these linear maps are each other's inverses and therefore constitute a vector-space isomorphism. Moreover, this isomorphism maps solutions of H6 in $\mathbb{R}^L_\text{H6}$ to solutions of H7 in $\mathbb{R}^L_\text{H7}$ and vice versa.

The fact that this is solution-preserving vector-space isomorphism rather than merely a one-to-one correspondence has substantial consequences for these theories' symmetries. Namely, this forces H6 and H7 to have the same symmetries (at least restricted to $\mathbb{R}^L_\text{H6}$ and $\mathbb{R}^L_\text{H7}$). This is because these transformations are both of the form Eq.~\eqref{GaugeVR} (but notably not of the form Eq.~\eqref{Permutation}). Thus for any symmetry transformation for H6 (within $\mathbb{R}^L_\text{H6}$) there is a corresponding symmetry transformation for H7 (within $\mathbb{R}^L_\text{H7}$) and vice versa. 

This is in strong contrast to the results of our previous analysis in Sec.~\ref{SecHeat1}. There H6 was seen to have symmetries associated with a square 2D lattice and H7 was seen to have the symmetries associated with a hexagonal 2D lattice. By contrast, in the present interpretation H6 and H7 are thoroughly equivalent: We have a solution-preserving vector-space isomorphism between a substantial portion of their solutions, $\mathbb{R}^L_\text{H6}$ and $\mathbb{R}^L_\text{H7}$. Thus, on this interpretation (for a substantial portion of their solutions) the only difference between H6 and H7 is a change of basis.

Thus we have our second big lesson: despite the fact that H6 and H7 can be represented with very different lattice structures (i.e., a square lattice versus a hexagonal lattice) they have nonetheless turned out to be substantially equivalent to one another. This substantial equivalence was hidden from us on our first interpretation because we there took the lattice too seriously. As I will now discuss, this reduced their continuous rotation symmetries down to quarter rotations and one-sixth rotations respectively and thereby made them inequivalent. Here, we do not take the lattice structure so seriously. We have here internalized it into the value space where it subsequently disappears from view as just one basis among many.

In addition to switching between lattice structures, in this interpretation we can also do away with them altogether. In this interpretation, a lattice structure is associated with a choice of basis for $\mathcal{V}=\mathbb{R}^L$. A choice of basis (like a choice of coordinates) is ultimately a merely representational choice, reflecting no physics. We can always choose, if we like, to work with in a basis-free formulation of these theories. That is, ultimately, a lattice-free formulation of these theories. Thus we have our third big lesson: given a discrete spacetime theory with some lattice structure we can always reformulate it in such a way that it has no lattice structure whatsoever.

In the rest of this subsection I will only discuss the symmetries H6, analogous conclusions are true for H7 after applying $\Lambda_{\text{H6}\to\text{H7}}$. For H6 the dynamical symmetries of the form Eq.~\eqref{GaugeVR} include:
\begin{flushleft}\begin{enumerate}
 \item[1)] action by $T^\epsilon_\text{n}$ sending $\bm{\Phi}(t)\mapsto T^\epsilon_\text{n} \bm{\Phi}(t)$. Similarly for $T^\epsilon_\text{m}=\openone\otimes T^\epsilon$,
 \item[2)] two negation symmetries which map basis vectors as \mbox{$\bm{e}_{n}\otimes\bm{e}_{m}\mapsto \bm{e}_{-n}\otimes\bm{e}_{m}$}, and \mbox{$\bm{e}_{n}\otimes\bm{e}_{m}\mapsto \bm{e}_{n}\otimes\bm{e}_{-m}$} respectively,
 \item[3)] action by $R^\theta$ sending $\bm{\Phi}(t)\mapsto R^\theta \bm{\Phi}(t)$ with $R^\theta$ defined below. (This being a symmetry requires some qualification as I will discuss below.),
 \item[4)] constant time shifts $T^\tau_\text{t}$ which map $T^\tau_\text{t}:\bm{\Phi}(t)\mapsto \bm{\Phi}(t-\tau)$ for some real $\tau\in\mathbb{R}$,
 \item[5)] a local Fourier rescaling symmetry which maps \mbox{$\bm{\Phi}(k_1,k_2)\mapsto \tilde{f}(t;k_1,k_2)\,\bm{\Phi}(k_1,k_2)$} for some non-zero complex function $\tilde{f}(t;k_1,k_2)\in\mathbb{C}$ with $t\in\mathbb{R}$ and $k_1,k_2\in[-\pi,\pi]$,
 \item[6)] local affine rescaling which maps \mbox{$\bm{\Phi}(t)\mapsto \bm{\Phi}(t) + \bm{c}_2(t)$} for some $\bm{c}_2(t)$ which is also a solution of the dynamics.
\end{enumerate}\end{flushleft}
As with H4 and H5, we have here gained local Fourier rescaling and action by $T^\epsilon_\text{n}$ and $T^\epsilon_\text{m}$ has replaced the discrete shifts from before. The same discussion following Eq.~\eqref{TDef} applies here, justifying us calling these continuous translation operations. Thus H6 (and\footnote{Note that both $T^\epsilon_\text{n}$ and $T^\epsilon_\text{m}$ map $\mathbb{R}^L_\text{H6}$ into itself. As such H7 has these translation symmetries as well, at least over $\mathbb{R}^L_\text{H7}$. Indeed, further investigation shows that H7 has two continuous translation symmetries over all of $\mathbb{R}^L\cong\mathbb{R}^\mathbb{Z}\otimes\mathbb{R}^\mathbb{Z}$.} H7) have two continuous translation symmetries despite initially being represented on a lattice.

Additionally, we here have the quarter rotation symmetry from our first interpretation replaced with action by $R^\theta$, which as I will argue is essentially a continuous rotation transformation. Before that, it is worth noting a rather weak sense in which each of H4-H7 are rotation invariant. Each of these theories is approximately rotation invariant in the $\vert k_1\vert,\vert k_2\vert\ll\pi$ regime as they approach the continuum limit. By contrast, as I will show, H6 is \textit{exactly} rotation invariant over \textit{the whole of} $\sqrt{k_1^2+k_2^2}<\pi$. That is, restricting our attention to \mbox{$\mathbb{R}^L_\text{rot.inv.}$} defined in Eq.~\eqref{RLrotinv} the transformation $R^\theta$ will map solutions to solutions in an invertible way and will hence be a symmetry here.

This alleged continuous rotation transformation $R^\theta$ is given by
\begin{align}\label{RthetaDef}
R^\theta \coloneqq \exp(-\theta (N_\text{n} D_\text{m}-N_\text{m} D_\text{n}))
\end{align}
with $\theta\in\mathbb{R}$ and where $N$ is the ``position operator'' which acts on the basis $\bm{e}_j$ as $N\bm{e}_j=j\,\bm{e}_j$. Thus \mbox{$N_\text{n}= N\otimes\openone$} and \mbox{$N_\text{m}=\openone\otimes N$} return the first and second index respectively when acting on $\bm{e}_n\otimes \bm{e}_m$.

A straight-forward calculation shows that $R^\theta$ acts on the planewave basis $\bm{\Phi}(k_1,k_2)$ with \mbox{$k_1,k_2\in[-\pi,\pi]$} as
\begin{align}
R^\theta:\,&\bm{\Phi}(k_1,k_2) \mapsto\\
\nonumber
&\bm{\Phi}(\cos(\theta)k_1-\sin(\theta)k_2,\sin(\theta)k_1+\cos(\theta)k_2).
\end{align}
and acts on the basis $\bm{e}_n\otimes \bm{e}_m$ as
\begin{align}
&R^\theta: \bm{e}_n\otimes\bm{e}_m \mapsto\sum_{b_1,b_2\in\mathbb{Z}} R_{nm}^{b_1 b_2}(\theta)\,
\bm{e}_{b_1}\otimes\bm{e}_{b_2}\\
\nonumber
&R_{nm}^{b_1 b_2}\!(\theta)\!=\!S_n(\cos(\theta) b_1 \!-\! \sin(\theta) b_2) S_m(\sin(\theta) b_1\!+\! \cos(\theta)b_2).
\end{align}

It should be noted that $R^\theta$ is not invertible (at least not on the whole of $\mathbb{R}^L\cong\mathbb{R}^\mathbb{Z}\otimes\mathbb{R}^\mathbb{Z}$). As I will soon discuss, this is due to the $2\pi$ periodic identification of the planewaves $\bm{\Phi}(k_1,k_2)$. However, $R^\theta$ is invertible over $\mathbb{R}^L_\text{rot.inv.}$. In parallel to this, note that $\Gamma(k_1,k_2)$ for H6 does not map onto itself under rotation everywhere. It does so when we have \mbox{$k_1,k_2\in[-\pi,\pi]$} both before and after the rotation. This is guaranteed if $\sqrt{k_1^2+k_2^2}<\pi$. 

To see why $R^\theta$ is not invertible over all of $\mathbb{R}^L$ note that $R^{\pi/4}$ maps two different planewaves to the same place: Firstly note,
\begin{align}
R^{\pi/4}\bm{\Phi}(\pi,\pi)&=\bm{\Phi}(0,\sqrt{2}\pi)\\
\nonumber
&=\bm{\Phi}(0,\sqrt{2}\pi-2\pi)
\end{align}
since the planewaves repeat themselves with period $2\pi$. Secondly note, 
\begin{align}
\nonumber
R^{\pi/4}\bm{\Phi}(\pi-\sqrt{2}\pi,\pi-\sqrt{2}\pi)&=\bm{\Phi}(0,\sqrt{2}\pi-2\pi).
\end{align}
Such issues do not arise when $\sqrt{k_1^2+k_2^2}<\pi$. Thus, when we restrict our attention to $\mathbb{R}^L_\text{rot.inv}$ then $R^\theta$ is invertible. Within $\mathbb{R}^L_\text{rot.inv}$ this transformation is of the form Eq.~\eqref{GaugeVR} and maps solutions to H6 onto solution to H6 in an invertible way, and is hence a symmetry. Similarly for H7\footnote{Restricting H7 to the image of $\mathbb{R}^L_\text{rot.inv}$ under $\Lambda_{\text{H6}\to\text{H7}}$ we have $\Lambda_{\text{H6}\to\text{H7}}R^\theta \Lambda_{\text{H7}\to\text{H6}}$ being of the form Eq.~\eqref{GaugeVR} and mapping solutions to solutions in an invertible way.}. 

As I will now discuss, $R^\theta$ can be thought of as a continuous rotation operator for three reasons despite it being here an internal symmetry. First note that $R^\theta$ is a generalization of quarter rotation operation in the sense that taking $\theta=\pi/2$ reduces action by $R^\theta$ to the map \mbox{$R^{\pi/2}:\bm{e}_{n}\otimes\bm{e}_{m}\mapsto \bm{e}_{m}\otimes\bm{e}_{-n}$} on basis vectors and relatedly the map $(n,m)\mapsto (m,-n)$ on lattice sites.

Second, note that restricted to $\mathbb{R}^L_\text{rot.inv.}$, $R^\theta$ is cyclically additive in the sense that $R^{\theta_1}\,R^{\theta_2}=R^{\theta_1+\theta_2}$ with $R^{2\pi}=\openone$. In the language of representation theory, $R^\theta$ is a representation of the rotation group on the vector space $\mathbb{R}^L_\text{rot.inv.}$. In particular, this means \mbox{$R^{\pi/4}\,R^{\pi/4}=R^{\pi/2}$}. There is something we can do twice to make a quarter rotation. Similarly for all fractional rotations. Moreover, note that together with the above discussed translations, these constitute a representation of the 2D Euclidean group over $\mathbb{R}^L_\text{rot.inv.}$. 

Third, recall from the discussion following Eq.~\eqref{LambdaD} that $D$ is closely related to the continuum derivative operator $\partial_x$, exactly matching its spectrum for $k\in[-\pi,\pi]$. Recall also that rotations are generated through the derivative as \mbox{$h(R^\theta(x,y))= \exp(-\theta (x \partial_y-y \partial_x))h(x,y)$}. In this sense also $R^\theta$ is a rotation operator. More will be said about $R^\theta$ in Sec.~\ref{SecSamplingTheory}.

This adds to our first big lesson: despite the fact that H6 and H7 can be represented on a square and hexagonal lattice respectively, they nonetheless both have a continuous rotation symmetry. This, in addition to their continuous translation symmetries. These continuous translation and rotations symmetries were hidden from us on our first interpretation because we there took the lattice representations too seriously. Here, we do not take the lattice structure so seriously. Instead, we have internalized it into the value space where it then disappears from view as just one basis among many.

\vspace{0.25cm}
To summarize: this second attempt at interpreting H1-H7 has fixed all of the issues with our previous interpretation. Firstly, there is no longer any tension between these theories differing locality properties and the rates at which they converge to the (perfectly local) continuum theory in the continuum limit. (There are no longer any differences in locality.) Secondly, the fact that we have a nice one-to-one correspondence between the solutions to H6 and H7 is now more satisfyingly reflected in their matching symmetries. Finally, this interpretation has exposed the fact that H1-H7 have hidden continuous translation and rotation symmetries.

By and large, the interpretation invalidates all of the first intuitions laid out in Sec.~\ref{SecCentralClaims}. As this interpretation has revealed, the lattice seems to play a merely representational role in the theory: it does not restrict our symmetries. Moreover, theories initially appearing with different lattice structures may nonetheless turn out to be substantially equivalent. The process for switching between lattice structures is here a change of basis in the value space. Indeed, we have a third lesson: there is no sense in which these lattice structures are essentially ``baked-into'' these theories; on this interpretation our theories make no reference to any lattice structure if we work in a basis-independent way. No basis is dynamically favored. 

As I will discuss in Sec.~\ref{SecDisGenCov} these three lessons lay the foundation for a rich analogy between the lattice structures which appear in our discrete spacetime theories and the coordinate systems which appear in our continuum spacetime theories. This ultimately gives rise to a discrete analog of general covariance.

These are substantial lessons, but ultimately this interpretation has a few issues of its own. Firstly, the way that the tension is dissolved between locality and convergence in the continuum limit is unsatisfying. Intuitively, we ought to be able to extract intuitions about locality from the lattice sites.

Moreover, while this interpretation has indeed exposed H1-H7's hidden continuous translation and rotation symmetries, the way that it has classified them seems wrong. They are here classified as internal symmetries (i.e., symmetries of the value space) whereas intuitively they should be external symmetries (i.e., symmetries of the manifold).

The root of these issues is taking the theory's lattice structure to be internalized into the theory's value space. Our third attempt at interpreting these theories will fix this by externalizing these symmetries.

\section{Externalizing H1-H7 - Part 1}\label{SecExtPart1}
 In the previous section it was revealed that H1-H7 have hidden continuous symmetry which intuitively correspond to spacial translations and rotation. In our first attempt at interpreting H1-H7 the possibility of such symmetries were outright denied, see Sec.~\ref{SecHeat1}. In our second attempt, these hidden symmetries were exposed, but they were classified (unintuitively) as internal symmetries, see Sec.~\ref{SecHeat2}. This was due to an ``internalization'' move made in our second interpretation. This move also had the unfortunate consequence of undercutting our ability to use the lattice sites to reason about locality.

In the next three sections I will show how we can externalize these symmetries by in a principled way 1) inventing a continuous manifold for our formerly discrete theories to live on and 2) embedding our theory's states/dynamics onto this manifold as a new dynamical field.

\subsection{A Principled Choice of Spacetime Manifold}
If we are going to externalize these symmetries then we need to have a big enough manifold on which to do the job. What spacetime manifold $\mathcal{M}$ might be up to the task? The first thing we must do is pick out which of our theory's symmetries we would like to externalize (there may be some symmetries we want to keep internal). For H1-H7 we want to externalize the following symmetries: continuous translations, mirror reflections, as well as discrete rotations for H4 and H5 and continuous rotations for H6 and H7. For each theory we can collect these dynamical symmetries together in a group $G^\text{dym}_\text{to-be-ext}$. Clearly, our choice of spacetime manifold $\mathcal{M}$ needs to be big enough to have $G^\text{dym}_\text{to-be-ext}$ as a subgroup of $\text{Diff}(\mathcal{M})$. Let us call this the symmetry-fitting constraint\footnote{As I will soon discuss, there may be transformations which we want to externalize even when they are not symmetries.}. 

Of course, symmetry-fitting alone doesn't uniquely specify which manifold we ought to use. If $\mathcal{M}$ works, then so does any larger $\mathcal{M}'$ with $\mathcal{M}$ as a sub-manifold. For standard Occamistic reasons, it is natural to go with the smallest manifold which gets the job done. The larger the gap between the groups $G^\text{dym}_\text{to-be-ext}$ and $\text{Diff}(\mathcal{M})$ the more fixed spacetime structures will need to be introduced later on (see Sec.~\ref{SecFullGenCov}).

In principle, we are free to pick any large-enough manifold which we like to embed H1-H7 onto. However, perhaps surprisingly, if we make natural choices about how the translation operations we have already identified fit onto the new spacetime manifold then our choice of $\mathcal{M}$ is actually fixed up to diffeomorphism. In particular, I demand the following: Certain translation operations on $\mathbb{R}^L$ are to correspond (perhaps in a complicated way) to parallel transport on the new spacetime manifold $\mathcal{M}$. For H1-H3 these to-be-externalized translation operations are $T^{\tau}_\text{t}$ and $T^{\epsilon_1}$. For H4-H7 these to-be-externalized translation operations are $T^{\tau}_\text{t}$, $T^{\epsilon_1}_\text{n}$, and 
$T^{\epsilon_2}_\text{m}$. Let us call this the translation-matching constraint (this will be spelled out technically later). As I will soon show, this constraint fixes the new spacetime manifold $\mathcal{M}$ up to diffeomorphism.

Before fleshing this out, however, it's worth reflecting on two questions focusing on H4-H7: What exactly makes $T^{\tau}_\text{t}$, $T^{\epsilon_1}_\text{n}$, and $T^{\epsilon_2}_\text{m}$ translation operations? Moreover, what motivation do we have to externalize these particular translation operations? To answer: firstly, these can be thought of as translation operations for the reasons discussed following Eq.~\eqref{TDef}. Note these reasons are unrelated to the fact that these are dynamical symmetries of H4-H7. Suppose the dynamics of H4 given by Eq.~\eqref{H4Long} was modified to have explicit dependence on the index $n$. In this case, $T^{\epsilon_1}_\text{n}$ would no longer be a dynamical symmetry but it would still be a translation operation (and moreover, one worth externalizing).

Secondly, why externalize these translation operations in particular? As I will now discuss, any motivation for externalizing these particular translation operations must come from the dynamics. Forgoing any dynamical considerations, all we can say about H1-H7 is that they concern vectors $\bm{\Phi}(t)\in\mathbb{R}^L$ (or alternatively functions \mbox{$\phi_\ell(t):L\to\mathbb{R}$} in $F_L$). Recall that pre-dynamics the set of labels for lattice sites $L$ is uncountable but otherwise unstructured; We might index it using any number of indices we like, \mbox{$L\cong\mathbb{Z}\cong\mathbb{Z}^2\cong\dots\cong\mathbb{Z}^{17}\cong\dots$}. Using each of these re-indexings we can grant $\mathbb{R}^L$ different tensor product structures, \mbox{$\mathbb{R}^L\cong\mathbb{R}^Z\cong\mathbb{R}^Z\otimes\mathbb{R}^Z\cong\dots$}. In each tensor factor we can define a translation operation $T^\epsilon$ as in Eq.~\eqref{TDef}. Thus, the vector space $\mathbb{R}^L$ supports representations of: the 1D translation group, the 2D translation group, $\dots$, the 17D translation group, etc. Given all these possibilities, why externalize $T^{\tau}_\text{t}$, $T^{\epsilon_1}_\text{n}$, and $T^{\epsilon_2}_\text{m}$ in particular? The answer must come from the dynamics.

One good reason to consider $T^{\tau}_\text{t}$, $T^{\epsilon_1}_\text{n}$, and $T^{\epsilon_2}_\text{m}$ worthy of externalization is that they are dynamical symmetries of H4-H7. However, as mentioned above, something might be a translation operation (and moreover, one worthy of externalizing) even if it's not a dynamical symmetry. Unfortunately, I don't presently have a perfect rule for how to identify such cases. The best I can offer is to check whether its associated derivative (or some function thereof) appears in the dynamical equations.

Regardless, its sufficiently clear for H4-H7 that $T^{\tau}_\text{t}$, $T^{\epsilon_1}_\text{n}$, and 
$T^{\epsilon_2}_\text{m}$ are among the translation operations worthy of externalizing. Our translation-matching constraint suggests that these ought to correspond (perhaps in a complicated way) to parallel transport on the spacetime manifold $\mathcal{M}$. I claim that (at least in this case) this translation-matching constraint fixes the new spacetime manifold $\mathcal{M}$ up to diffeomorphism.

To show this I will first pick out at each point $p\in\mathcal{M}$ on the manifold three independent directions in the tangent space at $p$. To realize the translation-matching constraint, I demand roughly that that differential translation by $T^{\tau}_\text{t}$, $T^{\epsilon_1}_\text{n}$, and $T^{\epsilon_2}_\text{m}$ is then to be carried out on the manifold as parallel transport in these three directions. Note that this already implies that the dimension of $\mathcal{M}$ is at least three.

Indeed, in this case we have good reason to take $\mathcal{M}$ to be exactly three dimensional. To see this, note that 
\begin{align}\label{RLCover}
\mathbb{R}^L&\cong\mathbb{R}^\mathbb{Z}\otimes\mathbb{R}^\mathbb{Z}\\
\nonumber
&=\text{span}(T^{\epsilon_1}_\text{n} T^{\epsilon_2}_\text{m} \, \bm{e}_0\!\otimes\bm{e}_0\vert (\epsilon_1,\epsilon_2)\in\mathbb{R}^2)
\end{align}
That is, beginning from $\bm{e}_0\otimes\bm{e}_0$ these translations cover all of $\mathbb{R}^L\cong\mathbb{R}^\mathbb{Z}\otimes\mathbb{R}^\mathbb{Z}$. That is, they cover all of our second interpretation's state space at a time. Note that $T^{\tau}_\text{t}$ covers time translation. Thus, the three translation operations $T^{\tau}_\text{t}$, $T^{\epsilon_1}_\text{n}$, and $T^{\epsilon_2}_\text{m}$ are thus already enough to cover every kinematic possibility. We need no further dimensions and so by Occam's razor we may then take the spacetime manifold $\mathcal{M}$ to be three dimensional. 

But how does the translation-matching constraint (which I have yet to state technically) fix $\mathcal{M}$ up to diffeomorphism? As I will now discuss, (at least in this case) we can fix $\mathcal{M}$ up to diffeomorphism in a natural way by appealing to the group theoretic properties of $T^{\tau}_\text{t}$, $T^{\epsilon_1}_\text{n}$, and $T^{\epsilon_2}_\text{m}$. Let $G_\text{trans.}$ be the group formed by all compositions of $T^{\tau}_\text{t}$, $T^{\epsilon_1}_\text{n}$, and $T^{\epsilon_2}_\text{m}$. Note that the group $G_\text{trans.}$ is a Lie group, and hence is also a differentiable manifold.

Recall that so far the translation-matching constraint associated to each of our differential translations, a direction in the tangent space of each point $p\in\mathcal{M}$. Since $G_\text{trans.}$ contains nothing but these translations, this gives us a one-to-one correspondence between the algebra $g_\text{trans.}$ of $G_\text{trans.}$ and (at least a subspace of) the of the tangent space of each point on the manifold. Let's take these tangent vectors to vary smoothly across the manifold. Let $\mathcal{M}_\text{trans.}(p)$ be the submanifold of points on the spacetime manifold reachable from $p\in\mathcal{M}$ by following these tangent vectors. The algebra $g_\text{trans.}$ is related to the group $G_\text{trans.}$ in roughly the same way that tangent space it related to $\mathcal{M}_\text{trans.}(p)$, namely by repeated application of the exponential map. Hence, I take translation-matching to demand that $\mathcal{M}_\text{trans.}(p)\cong G_\text{trans.}$ as differentiable manifolds for every $p\in\mathcal{M}$. 

As discussed above, we have reason to take $\mathcal{M}$ to be three dimensional. Hence (assuming that $\mathcal{M}$ is connected) any three dimensional submanifold (e.g. $\mathcal{M}_\text{trans.}(p)$) must in fact be the entire manifold. Therefore we have $\mathcal{M}\cong G_\text{trans.}$ as differentiable manifolds.

But what can we say about this translation group $G_\text{trans.}$ viewed as a manifold? Notice that these three continuous translation operations comprising $G_\text{trans.}$ all commute with each other. As such for any combination of them $T_\text{generic}\in G_\text{trans.}$ we always have a unique factorization of the form,
\begin{align}\label{Tgeneric}
T_\text{generic} = T^{\tau}_\text{t} \ T^{\epsilon_1}_\text{n} \ T^{\epsilon_2}_\text{m}
\end{align}
Moreover, this factorization represents all of $G_\text{trans.}$ without redundancy. Indeed, we can smoothly parameterize all of $G_\text{trans.}$ using parameters \mbox{$(\tau,\epsilon_1,\epsilon_2)\in\mathbb{R}^3$} in a one-to-one way. That is, $G_\text{trans.}\cong\mathbb{R}^3$ as differentiable manifolds. Therefore, we have $\mathcal{M}\cong\mathbb{R}^3$ as well. Note this means that we have access to a global coordinate system for $\mathcal{M}$.

Thus, from the translation-matching constraint alone we are forced to take $\mathcal{M}\cong\mathbb{R}^3$ for H4-H7. For H1-H3 the same translation-matching constraint forces us to take $\mathcal{M}\cong\mathbb{R}^2$. Note that in both cases these manifolds satisfy the symmetry-fitting constraint: $\text{Diff}(\mathcal{M})$ is large enough to contain $G^\text{dyn}_\text{to-be-ext}$.

Before going on to discuss how we might embed our vector field $\bm{\Phi}(t)$ onto this manifold, some more must be said about time on this new spacetime manifold $\mathcal{M}$. Let's focus on H4-H7. As mentioned above, we can smoothly parameterize of $G_\text{trans.}$ via $(\tau,\epsilon_1,\epsilon_2)\in\mathbb{R}^3$. Given translation-matching, this gives us a global coordinate system $(\tau,\epsilon_1,\epsilon_2)\in\mathbb{R}^3$ for $\mathcal{M}$. In particular, the $\tau$ coordinate is straight-forwardly associated with $T^\tau_\text{t}$ and so corresponds to the notion of time appearing in our first and second interpretations. As such, we can sensibly decompose the spacetime manifold $\mathcal{M}$ into a collection of submanifolds $\mathcal{M}(t)$ with fixed time coordinate $\tau=t$. Time in this coordinate system is identified with the time appearing in our first interpretation in $Q=L\times\mathbb{R}$ and in our second interpretation in $\mathcal{M}=\mathbb{R}$.

\subsection{A Principled Choice of Embedding}
Now that we have a spacetime manifold selected, we need to somehow embed $\bm{\Phi}(t)$ (or equivalently $\phi_\ell(t)$) into it. Here I will begin with $\bm{\Phi}(t)$ with the approach from $\phi_\ell(t)$ being addressed in Sec.~\ref{SecExtPart2}. The goal in either case is to construct in a principled way from either $\bm{\Phi}(t)$ or $\phi_\ell(t)$ a new field $\phi:\mathcal{M}\to\mathbb{R}$ defined over this manifold. 

As discussed above, we can sensibly decompose $\mathcal{M}$ into a collection of time-slices $\mathcal{M}(t)$. Let $\phi(t)$ be the new field $\phi$ restricted to one of these time-slices as \mbox{$\phi(t):\mathcal{M}(t)\to\mathbb{R}$}. Given this decomposition, our goal to construct in a principled way from $\bm{\Phi}(t)$ at each time a time-slice of the new field \mbox{$\phi(t):\mathcal{M}(t)\to\mathbb{R}$}.

Before doing this, however, recall that our move from the first to the second interpretation was mediated by means of a vector-space isomorphism $F_L\cong\mathbb{R}^L$ namely Eq.~\eqref{PhiDef}. Here too, our reinterpretation will be mediated by a vector-space isomorphism.

Let $F(\mathcal{M}(t))$ be the set of all real scalar functions \mbox{$f(t):\mathcal{M}(t)\to\mathbb{R}$}. Notice that this is a vector space as it is closed under addition and scalar multiplication. Our goal here is to in a principle way construct an isomorphism between \mbox{$\mathbb{R}^L$} and some vector subspace of \mbox{$F(\mathcal{M}(t))$}. Indeed, $F(\mathcal{M}(t))$ is much larger than $\mathbb{R}^L$ (having an uncountably infinite dimension). As such we will only ever map $\mathbb{R}^L$ onto some (possibly time-dependent) subset of it, $F(t)\subset F(\mathcal{M}(t))$. In order for us to have a vector space isomorphism between $\mathbb{R}^L$ and $F(t)$ the following conditions must be satisfied. $F(t)$ must be: 1) closed under addition and scalar multiplication and hence a vector space, and 2) countably infinite dimensional and hence isomorphic to $\mathbb{R}^L$, i.e., \mbox{$F(t)\cong\mathbb{R}^L\cong F_L$} as vector spaces. With the target subspace characterized, the embedding of $\bm{\Phi}(t)$ onto $\mathcal{M}(t)$ is then accomplished by picking a (possibly time-dependent) vector-space isomorphism $E(t):\mathbb{R}^L\to F(t)$. The embedded field is then gives to us as $\phi(t)\coloneqq E(t)(\bm{\Phi}(t))$.

It appears we have a great deal of freedom here. Suppose that, in line with the translation-matching constraint discussed above, we take $\mathcal{M}\cong\mathbb{R}^2$ for H1-H3 and $\mathcal{M}\cong\mathbb{R}^3$ for H4-H7. In this case, we still have great freedom in picking both the target vector space $F(t)\subset F(\mathcal{M}(t))$ and the isomorphism $E(t):\mathbb{R}^L\to F(t)$. However, as I will now discuss, translation-matching also drastically limits these choices as well. Allow me to demonstrate with H4-H7.

First, allow me to explicitly define a coordinate system for $\mathcal{M}\cong G_\text{trans.}\cong\mathbb{R}^3$. As discussed above we already have an explicit smooth parametrization of $G_\text{trans.}$ via \mbox{$(\tau,\epsilon_1,\epsilon_2)\in\mathbb{R}^3$}, see Eq.~\eqref{Tgeneric}. In addition to this, concretely realizing our translation-matching constraint requires us to fix a smooth map from $G_\text{trans.}$ to $\mathcal{M}$. Combined these two maps give us a global coordinate system $(\tau,\epsilon_1,\epsilon_2)\in\mathbb{R}^3$ for $\mathcal{M}$. We can then rescale these to give new coordinates $(t,x,y)\in\mathbb{R}^3$ by adding in a length scale $a>0$ with $t=\tau$, $x=\epsilon_1 a$, and $y=\epsilon_2 a$ for some fixed length scale $a$.

In total this picks out a diffeomorphism \mbox{$d_\text{trans.}:\mathbb{R}^3\to\mathcal{M}$} which assigns global coordinates $(t,x,y)\in\mathbb{R}^3$ to $\mathbb{cal}$. Note it is exactly in these coordinates that we have decomposed $\mathcal{M}$ into time-slices. For much of the next sections I will work with $\phi$ in these coordinates. Namely I will work with the pull back, $\phi\circ d_\text{trans.}:\mathbb{R}^3\to\mathbb{R}$. One may worry that this choice of coordinates is arbitrary. Indeed, if one changes the smooth map from $G_\text{trans.}$ to $\mathcal{M}$ referenced above we end up with a different coordinate system $d_\text{trans.}$ related to the original by some diffeomorphism. While this is true, no matter how one realizes the translation-matching constraint there will be some coordinate system which would result from the above construction. In what follows, nothing depends on how the translation-matching constraint is realized.

A word of warning. The coordinates used above might end up lying on the manifold in a curvilinear way. Crucially, the naive distance and metric structure associated with these coordinates do not automatically have any spatiotemporal significance. At present, our manifold $\mathcal{M}$ still has no metric. When a metric arises later, it will be due to dynamical considerations, not from any choice of coordinates. A coordinate-independent view of these theories will be given in Sec.~\ref{SecFullGenCov}.

With this said, by construction we have the following correspondences in these coordinate:
\begin{flushleft}\begin{enumerate}
\item Action by $T_\text{n}^\epsilon$ on $\mathbb{R}^L\cong\mathbb{R}^\mathbb{Z}\otimes\mathbb{R}^\mathbb{Z}$ acts on the manifold $\mathcal{M}(t)$ in these coordinates as $(t,x,y)\mapsto(t,x-\epsilon\,a,y)$,
\item Action by $T_\text{m}^\epsilon$ on $\mathbb{R}^L\cong\mathbb{R}^\mathbb{Z}\otimes\mathbb{R}^\mathbb{Z}$ acts on the manifold $\mathcal{M}(t)$ in these coordinates as $(t,x,y)\mapsto(t,x,y-\epsilon\,a)$.
\end{enumerate}\end{flushleft}
In particular, matching up differential translations on $\mathbb{R}^L\cong\mathbb{R}^\mathbb{Z}\otimes\mathbb{R}^\mathbb{Z}$ with those on $\mathcal{M}(t)$ requires
\begin{align}
E(t)(D_\text{n}\bm{\Phi})&=a\,\partial_x E(t)(\bm{\Phi})\\
\nonumber
E(t)(D_\text{m}\bm{\Phi})&=a\,\partial_y E(t)(\bm{\Phi})
\end{align}
in these coordinates for all $\bm{\Phi}\in\mathbb{R}^L$.

Evaluating these conditions in the planewave basis for $\mathbb{R}^L\cong\mathbb{R}^\mathbb{Z}\otimes\mathbb{R}^\mathbb{Z}$ (namely $\bm{\Phi}(k_1,k_2)$ for \mbox{$k_1,k_2\in[-\pi,\pi]$}) we see that $E(t)(\bm{\Phi}(k_1,k_2))$ must be simultaneously an eigenvector of $\partial_x$ and $\partial_y$ with eigenvalues $-\ii k_1/a$ and $-\ii k_2/a$ respectively. This uniquely picks out the continuum planewaves:
\begin{align}\label{PlaneWaveCont}
\text{H1-H3:}\quad&\phi(x;k)
\coloneqq e^{ -\ii k x}\\
\nonumber
\text{H4-H7:}\quad&\phi(x,y;k_1,k_2)
\coloneqq e^{-\ii k_1 x-\ii k_2 y}.
\end{align}
In particular, we are forced to take
\begin{align}\label{Eftilde}
E(t):\,&\bm{\Phi}(k_1,k_2)\\
\nonumber
&\mapsto\tilde{f}(t;k_1,k_2)\, \phi(x,y;k_1/a,k_2/a)
\end{align}
for some complex function $\tilde{f}(t;k_1,k_2)\in\mathbb{C}$. That is, each discrete planewave must map onto something proportional to the corresponding continuous planewaves (rescaled by $a$).

Note that up to a local rescaling of Fourier space (which recall is a symmetry of the dynamics) our embedding is uniquely fixed by translation-matching. Even without this consideration however, our translation-matching constraint still fixes the vector space $F(t)\subset F(\mathcal{M}(t))$ we can embed \mbox{$\bm{\Phi}(t)\in\mathbb{R}^L$} onto as follows. We must have \mbox{$\tilde{f}(t;k_1,k_2)\neq0$} for all \mbox{$k_1,k_2\in[-\pi,\pi]$} and all $t\in\mathbb{R}$ otherwise $E(t)$ will not be invertible an hence not an isomorphism. Therefore $F(t)$ is fixed as,
\begin{align}
&F(t)=\text{span}(E(t)(\bm{\Phi}(k_1,k_2))\vert k_1,k_2\in[-\pi,\pi])\\
\nonumber
&=\text{span}(\tilde{f}(t;k_1,k_2)\,\phi(x,y;k_1/a,k_2/a)\vert k_1,k_2\in[-\pi,\pi])\\
\nonumber
&=\text{span}(\phi(x,y;k_1/a,k_2/a)\vert k_1,k_2\in[-\pi,\pi])\\
\nonumber
&=\text{span}(\phi(x,y;k_1,k_2)\vert k_1,k_2\in[-\pi/a,\pi/a])
\end{align}
independent of time. In light of this, let us define
\begin{align}
&\text{H1-H3:}\\
\nonumber
&F^K\coloneqq\text{span}(\phi(x;k)\vert k\in[-K,K])\\
&\text{H4-H7:}\\
\nonumber
&F^K\coloneqq\text{span}(\phi(x,y;k_1,k_2)\vert k_1,k_2\in[-K,K])
\end{align}
where $K=\pi/a$. These are the spaces of bandlimited functions with bandwidth $k\in[-K,K]$ and \mbox{$k_1,k_2\in[-K,K]$} respectively. Thus, demanding translation-matching we are forced to map $\bm{\Phi}\in\mathbb{R}^L$ onto some bandlimited $\phi(t,x,y)\in F(t)=F^K$ (at least in the coordinates $d_\text{trans.}$).

The next section will discuss in detail some remarkable properties of bandlimited functions, namely their sampling property. Before that however, let's find an interpretation for the $\tilde{f}(t;k_1,k_2)$ function appearing in Eq.~\eqref{Eftilde}. First note that applying $E(t)$ to the basis vector $\bm{e}_0\otimes\bm{e}_0$ we have
\begin{align}\label{Embedf}
E(t)(\bm{e}_0\otimes\bm{e}_0)=f\!\left(t;\frac{x}{a},\frac{y}{a}\right)
\end{align}
where $f(t;x,y)$ is the inverse Fourier transform of $\tilde{f}(t;k_1,k_2)$. Next note that by applying $T_\text{n}^\epsilon$ and $T_\text{m}^\epsilon$ with integer arguments we can get from $\bm{e}_0\otimes\bm{e}_0$ to any other basis vector, $\bm{e}_n\otimes\bm{e}_m$. In these coordinates this means,
\begin{align}
&E(t)(\bm{e}_n\otimes\bm{e}_m)=f\left(t;\frac{x}{a}-n,\frac{y}{a}-m\right)
\end{align}
Thus, we can understand a choice of $\tilde{f}$ as picking a profile $f(t;x,y)\in F^\pi$. A translated and rescaled copy of this profile is then associated with each basis vector.

To review: Our translation-matching considerations have greatly constrained our choice of both the spacetime manifold $\mathcal{M}$ and the embedding of $\bm{\Phi}(t)$ onto a subspace of $F(\mathcal{M}(t))$. In particular, $\mathcal{M}$ was forced to be diffeomorphic to $\mathbb{R}^2$ for H1-H3 and to $\mathbb{R}^3$ for H4-H7. Moreover, the vector space $F(t)\subset F(\mathcal{M}(t))$ which we map $\bm{\Phi}(t)\in\mathbb{R}^L$ into is forced to be the space of bandlimited functions with some bandwidth (i.e., bandlimited in coordinates $d_\text{trans.}$ with $k_1,k_2\in[-K,K]$ on each time-slice). 

Beyond this, our only freedom left is in picking a profile $f$ to associate with each basis vector $\bm{e}_n\otimes\bm{e}_m$. In Sec.~\ref{SecExtPart2} I will motivate a principled choice for $f$. Before that however, let's discuss bandlimited function in some detail.

\section{Brief Review of Bandlimited Functions and Nyquist-Shannon Sampling Theory}\label{SecSamplingTheory}
 The previous section has given us reason to care about bandlimited functions. A bandlimited function is one whose Fourier transform has compact support. The bandwidth of such a function is the extent of its support in Fourier space. As I will now discuss, such functions have a remarkable sampling property: they can be exactly reconstructed knowing only their values at a (sufficiently dense) set of sample points. The study of such functions constitutes Nyquist-Shannon sampling theory. For a selection of introductory texts on sampling theory see ~\cite{GARCIA200263,SamplingTutorial,UnserM2000SyaS}.

To introduce the topic I will at first restrict our attention to the one-dimensional case with uniform sample lattice before generalizing to higher dimensions and non-uniform samplings later on.

\subsection{One Dimension Uniform Sample Lattices}\label{Sec1DUniform}
Consider a generic bandlimited function, $f_\text{B}(x)$, with a bandwidth of $K$. That is, a function $f_\text{B}(x)$ such that its Fourier transform,
\begin{align}
\mathcal{F}[f_\text{B}(x)](k)\coloneqq\int_{-\infty}^\infty f_\text{B}(x) \, e^{-\ii k x} \d x,
\end{align}
has support only for wavenumbers $\vert k\vert< K$.

Suppose that we know the value of $f_\text{B}(x)$ only at the regularly spaced sample points, $x_n=n\,a+b$, with some spacing, \mbox{$0\leq a\leq a^*\coloneqq\pi/K$}, and offset, $b\in\mathbb{R}$. Let \mbox{$f_n=f_\text{B}(x_n)$} be these sample values. Having only the discrete sample data, $\{(x_n,f_n)\}_{n\in\mathbb{Z}}$, how well can we approximate the function? 

The Nyquist-Shannon sampling theorem~\cite{ShannonOriginal} tells us that from this data we can reconstruct $f_\text{B}$ exactly everywhere! That is, from this discrete data, $\{(x_n,f_n)\}_{n\in\mathbb{Z}}$, we can determine everything about the function $f_\text{B}$ everywhere. In particular, the following reconstruction is exact, 
\begin{align}\label{SincRecon}
f_\text{B}(x) 
= \sum_{n\in\mathbb{Z}} S_n\!\left(\frac{x-b}{a}\right) f_n,
\end{align}
where
\begin{align}
S(y)=\frac{\sin(\pi y)}{\pi y}, \quad\text{and}\quad
S_n(y)=S(y-n), 
\end{align}
are the normalized and shifted sinc functions. Note that $S_n(m)=\delta_{nm}$ for integers $n$ and $m$. Moreover, note that each $S_n(x)$ is $L_2$ normalized and that taken together the set $\{S_n(x)\}_{n\in\mathbb{Z}}$ forms an orthonormal basis with respects to the $L_2$ inner product. The fact that any bandlimited function can be reconstructed in this way is equivalent to the fact that this orthonormal basis spans the space of bandlimited functions with bandwidth of $K=\pi$.

\begin{figure}
\includegraphics[width=0.4\textwidth]{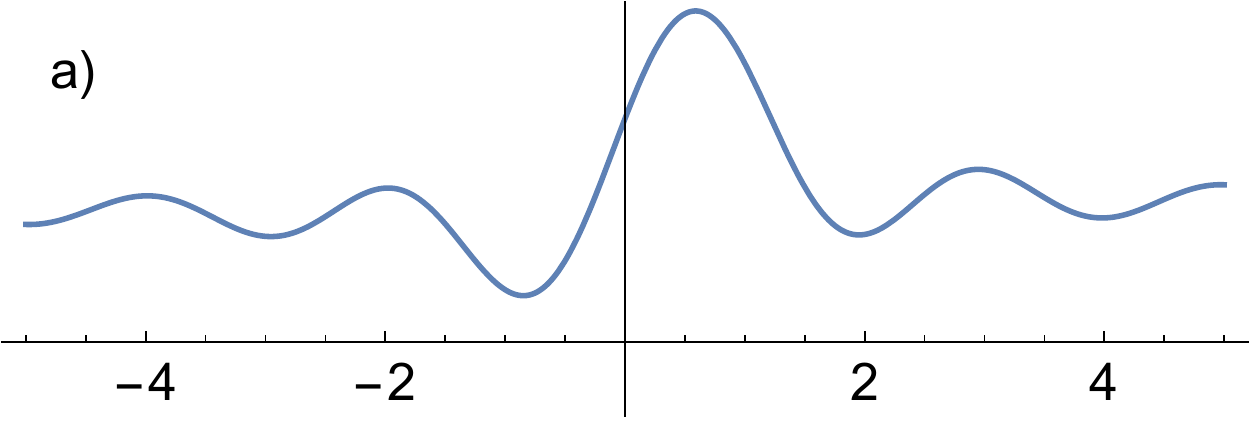}
\includegraphics[width=0.4\textwidth]{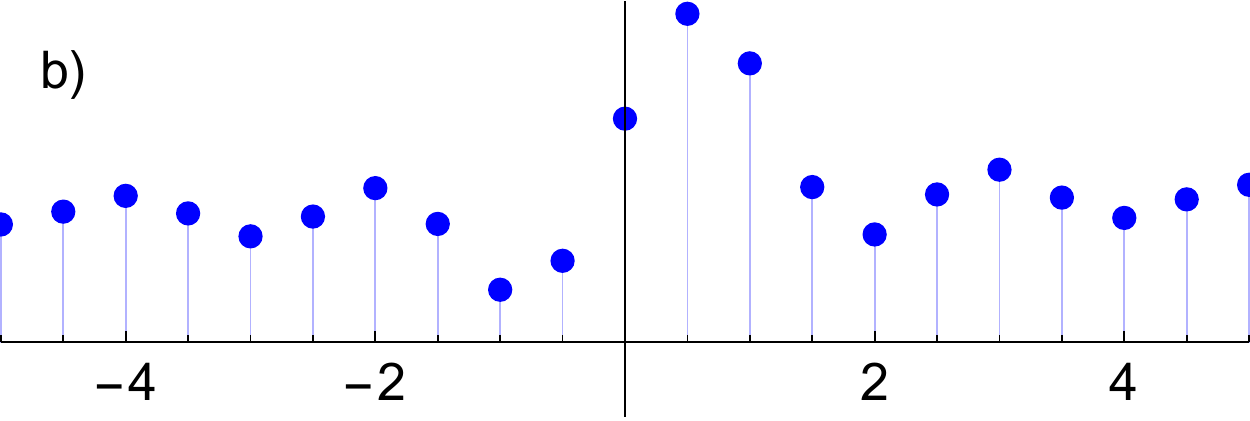}
\includegraphics[width=0.4\textwidth]{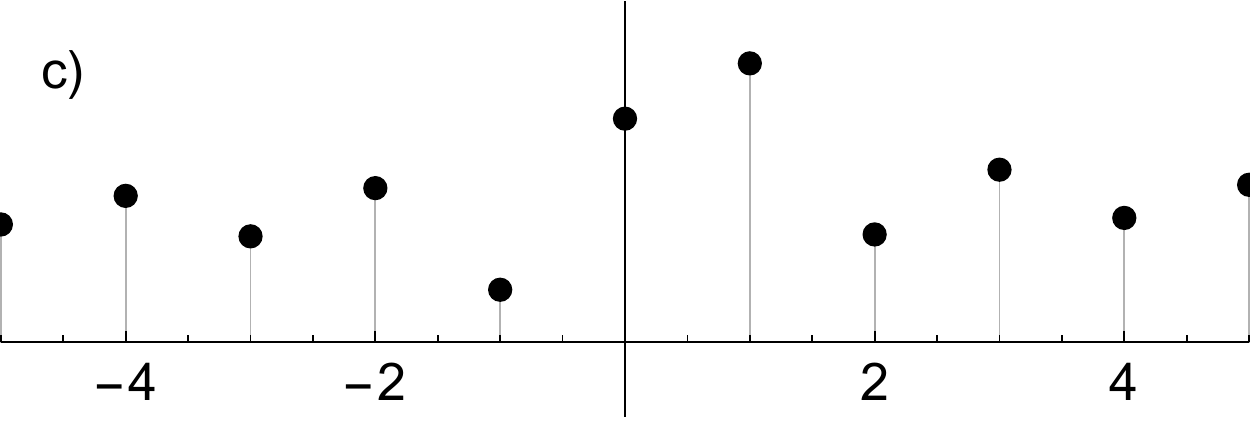}
\includegraphics[width=0.4\textwidth]{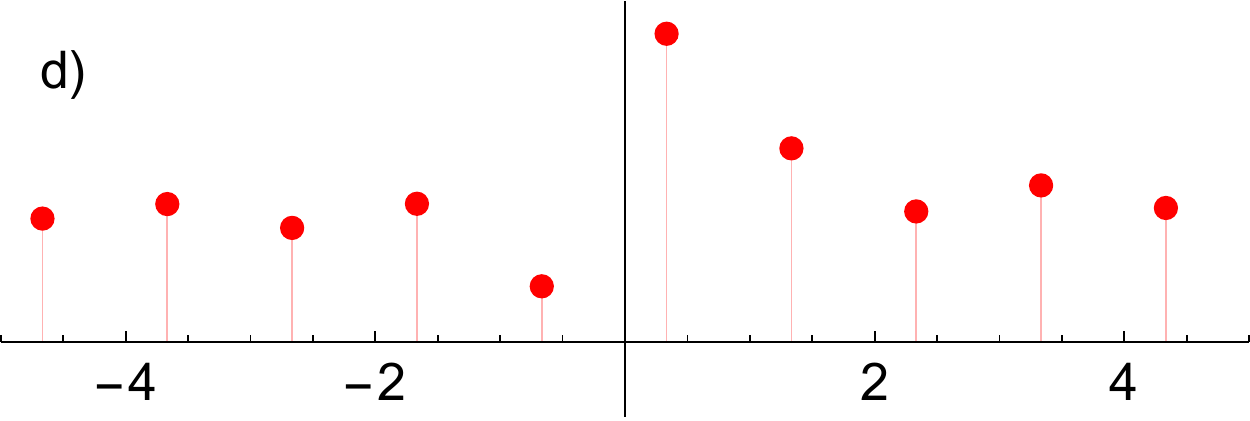}
\includegraphics[width=0.4\textwidth]{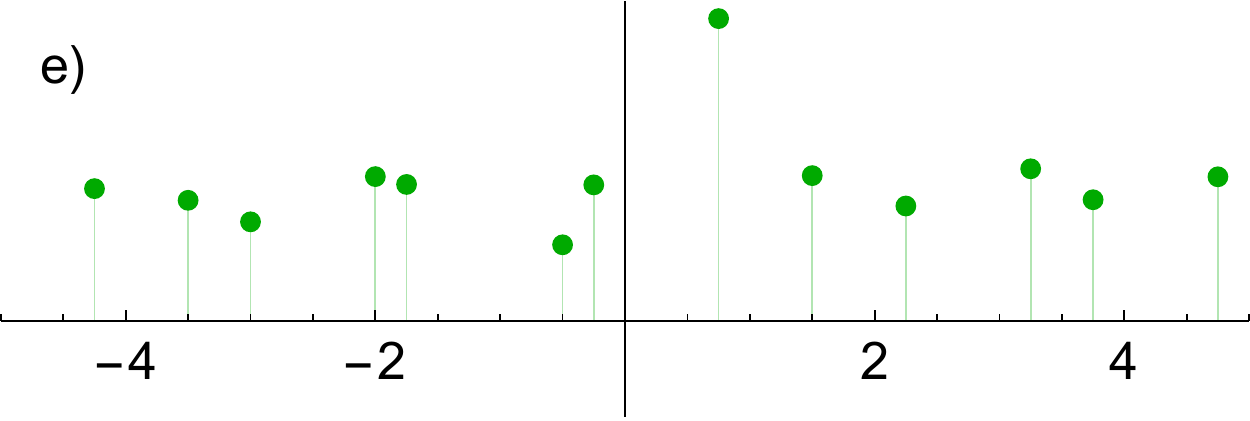}
\caption{Several different (but completely equivalent) graphical representations of the bandlimited function \mbox{$f_\text{B}(x)=1+S(x-1/2)+x\,S(x/2)^2$} with bandwidth of \mbox{$K=\pi$} and consequently a critical spacing of $a^*=\pi/K=1$. Subfigure a) shows the function values for all $x$. b) shows the values of $f_\text{B}$ at $x_n=n/2$. Since $1/2<a^*=1$ this is an instance of oversampling. c) shows the values of $f_\text{B}$ at $x_n=n$. This is an instance of critical sampling. d) shows the values of $f_\text{B}$ at $x_n=n+1/3$. This too is an instance of critical sampling. From any of these samplings we can recover the function $f_\text{B}$ exactly everywhere using Eq.~\eqref{SincRecon}. Subfigure e) shows a non-uniform sampling of $f_\text{B}$. We can recover $f_\text{B}$ exactly everywhere from here using Eq.~\eqref{GenRecon}. }\label{Fig1DSamples}
\end{figure}

As a concrete example, let us consider the function $f_\text{B}(x)=1+S(x-1/2)+x\,S(x/2)^2$, shown in Fig.~\ref{Fig1DSamples}a). This function has a bandwidth of $K=\pi$ and so has a critical sample spacing of $a^*=\pi/K=1$. Thus, we can fully reconstruct $f_\text{B}(x)$ knowing only its values at $x_n=n\,a+b$ for any spacing $a\leq a*=1$. In particular the sample values at $x_n=n/2$ are sufficient to exactly reconstruct the function, see Fig.~\ref{Fig1DSamples}b). So too are the sample values at the integers $x_n=n$ and at $x_n=n+1/3$, see Fig.~\ref{Fig1DSamples}c) and Fig.~\ref{Fig1DSamples}d). In each of these cases the reconstruction is given by Eq.~\eqref{SincRecon}.

Everything about this function can be reconstructed from any uniform sample lattice with $a\leq a^*=1$. In particular, the value of $f_\text{B}$ two third's of the way between sample point, $f_\text{B}(2/3)$, is fixed by $\{(n,f_\text{B}(n))\}_{n\in\mathbb{Z}}$ even though we have no sample at or even near $x=2/3$. The derivative of $f_\text{B}$ at zero, $f_\text{B}'(0)$, is fixed by $\{(n,f_\text{B}(n))\}_{n\in\mathbb{Z}}$ even though the only sample point we have in this neighborhood is $f_\text{B}(0)$. Moreover, the derivative at $x=2/3$, namely $f_\text{B}'(2/3)$, is fixed by $\{(n,f_\text{B}(n))\}_{n\in\mathbb{Z}}$ even we have no sample points in the neighborhood.

On first exposure this may be shocking: how can a function's behavior everywhere be fixed by its value at a discrete set of points? When $f_\text{B}$ is represented discretely, where has all of the information gone? Where is the information about the derivative at $x=2/3$ stored in the discrete representation? 

To see this, it is convenient (but not necessary) to organize our sample values $f_n=f(x_n)$ into a vector as,
\begin{align}
\bm{f}&=(\dots,f_{-1},f_0,f_1,f_2,\dots)^\intercal
= \sum_{n\in\mathbb{Z}} f_{n} \, \bm{w}_n
\end{align}
where $\bm{w}_n = (\dots,0,0,1,0,0,\dots)^\intercal\in\mathbb{R}^\mathbb{Z}$ with the 1 in the $n^\text{th}$ position. 

The values that $f_\text{B}$ takes at the sample points $x_n$ can be recovered from $\bm{f}$ as
\begin{align}
\nonumber
f_\text{B}(x_n) 
&= f_n
= \bm{w}_n^\intercal\bm{f},
\end{align}
Next notice that translating a bandlimited function preserves its bandwidth. As such both $f_\text{B}(x)$ and $f_\text{B}(x+\epsilon a)$ can be represented as Eq.~\eqref{SincRecon}. Using this fact, we can recover the values that $f_\text{B}$ takes away from the sample points (i.e., at \mbox{$x=x_n+\epsilon\,a$} for $\epsilon\in\mathbb{R}$) as
\begin{align}\label{TBdef}
f_\text{B}(x_n+\epsilon\,a) 
&= \sum_{m=-\infty}^\infty S_m(n+\epsilon) \ f_m
=\bm{w}_n^\intercal \, T_\text{B}^\epsilon \, \bm{f}
\end{align}
where the entries of the matrix $T^\epsilon_\text{B}$ are $[T^\epsilon_\text{B}]_{i,j}=S_i(j+\epsilon)$. Note $T_\text{B}^\epsilon$ acts as the translation operator for this representation of bandlimited functions. If $\bm{f}$ represents $f_\text{B}(x)$ then $T^\epsilon_\text{B}\bm{f}$ represents $f_\text{B}(x+\epsilon \,a)$.

From this translation operator we can identify the derivative operator for bandlimited functions, $D_\text{B}$, as 
\begin{align}
\nonumber
D_\text{B}\coloneqq\lim_{\epsilon\to0} \frac{T^\epsilon_\text{B}-\openone}{\epsilon}. \end{align}
It should be noted that $D_\text{B}$ and $T_\text{B}^\epsilon$ commute and moreover we have the usual relationship between derivatives and translations, $T_\text{B}^\epsilon=\exp(\epsilon\, D_\text{B})$.

From the above definition of $D_\text{B}$ one can easily work out its matrix entries as \mbox{$[D_\text{B}]_{i,j}=(-1)^{i-j}/(i-j)$} when $i\neq j$ and 0 when $i=j$. That is, 
\begin{align}\label{DBMatrix}
D_\text{B}&=\text{Toeplitz}(\dots,\!\frac{1}{4},\!\frac{-1}{3},\!\frac{1}{2},\!-1,\!0,\!1,\!\frac{-1}{2},\!\frac{1}{3},\!\frac{-1}{4},\!\dots)
\end{align}
Note that $D_\text{B}$ acts as the derivative operator for this representation of bandlimited functions. If $\bm{f}$ represents $f_\text{B}(x)$ then $\frac{1}{a}D_\text{B}\bm{f}$ represents $f_\text{B}'(x)$.

Comparing this with the $D$ operator introduced in Eq.~\eqref{BigToeplitz} we see that they are numerically identical. Indeed, $D_\text{B}=D$ and moreover $T_\text{B}^\epsilon=T^\epsilon$. If we were to extend our discussion to two-dimensional functions we could find a discrete representation of the rotation operator for bandlimited functions, $R_\text{B}^\theta$. This would come out numerically equal to the $R^\theta$ operator introduced earlier in Eq.~\eqref{RthetaDef}, namely $R_\text{B}^\theta=R^\theta$. See Appendix~\ref{AppB} for further discussion. Thus, the discrete notions of derivative, translation, and rotation that we have been using up until now are intimately connected with bandlimited functions.

It should be noted that $D=D_\text{B}$ gives us the following remarkable derivative approximation (which is exact for bandlimited functions):
\begin{align}\label{ExactDerivative}
\partial_x f(x)
&\approx2\sum_{m=1}^\infty (-1)^{m+1} \frac{f(x+m\,a)-f(x-m\,a)}{2\,m\,a}.
\end{align}
Relatedly, we have the second derivative approximation (which is exact for bandlimited functions):
\begin{align}\label{ExactDerivative2}
\partial_x^2 f(x)
&\approx2\sum_{m=1}^\infty (-1)^{m+1} \frac{f(x+m\,a)+f(x-m\,a)}{m^2\,a^2}\\
\nonumber
&-\frac{\pi^2}{3}f(x).
\end{align}
Namely, when $f$ is bandlimited with bandwidth of $K$ and $a\leq\pi/K$ then these formulas are exact. Moreover, if the Fourier transform of $f$ is mostly supported in $[-K,K]$ with thin tails (e.g, Gaussian tails) outside this region, then these are very good derivative approximations. 

Ultimately we can compute any derivative of $f_\text{B}$ anywhere from our sample data as,
\begin{align}
\nonumber
\partial_x^r\,f_\text{B}(x_n+\epsilon\,a) 
=\frac{1}{a^r}\,\bm{w}_n^\intercal \, D_\text{B}^r T_\text{B}^\epsilon \, \, \bm{f}.
\end{align}
Thus, we can recover any value or derivative of $f_\text{B}$ from its values on any sufficiently dense uniform sample lattice.

Note that each $f_\text{B}$ is representable in this way on a wide number of sample lattices with differing spacings $a<\pi/K$ and differing offsets $b$. Translating between these different descriptions of $f_\text{B}$ is equivalent to a change of basis on the vector of sample values $\bm{f}$.

\subsection{Non-Uniform Sample Lattices}\label{Sec2DSampling}
\begin{figure*}[t!]
\centering
\includegraphics[width=0.95\textwidth]{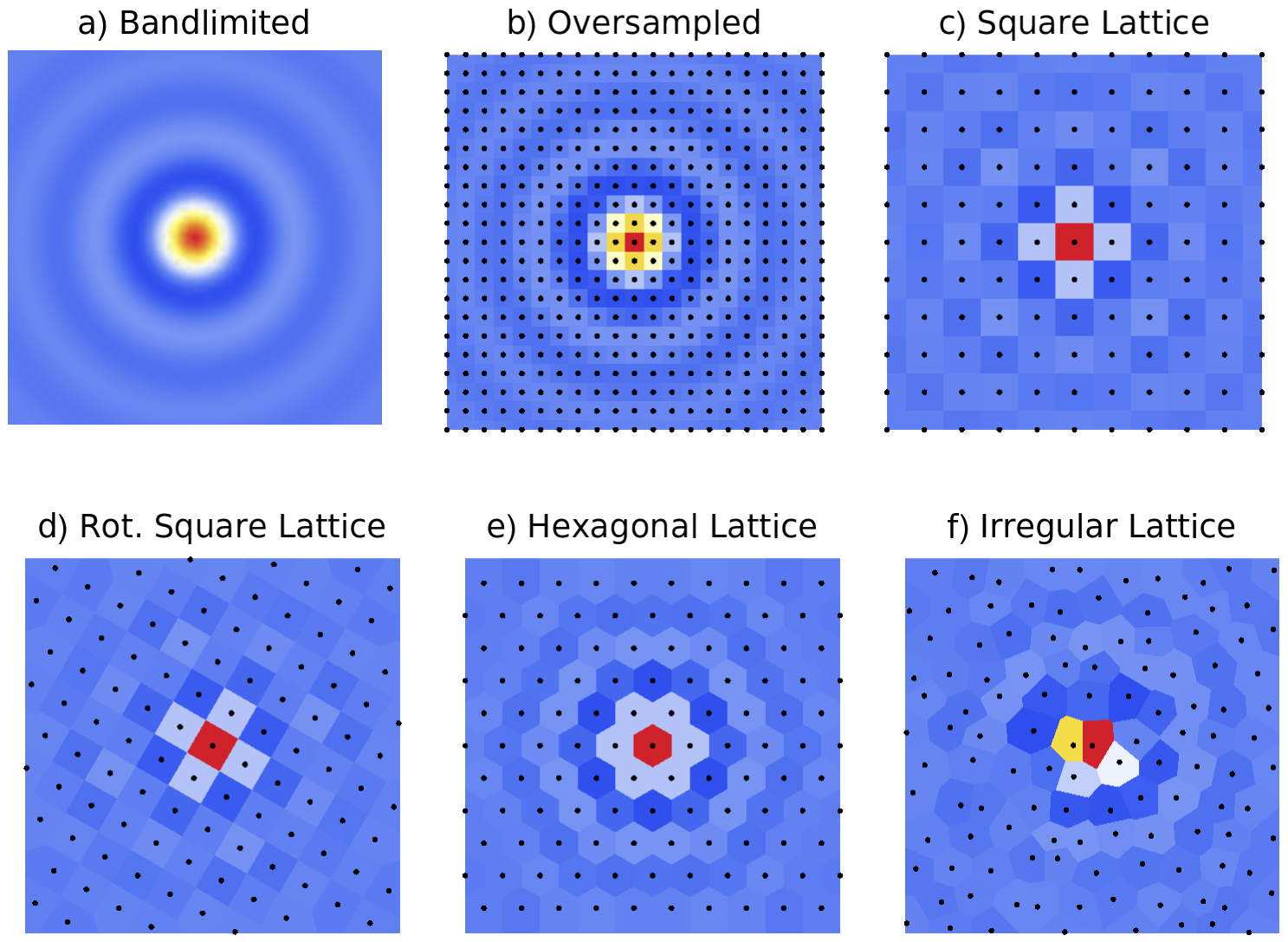}
\caption{Several different (but completely equivalent) graphical representations of the bandlimited function given by Eq.~\eqref{J1}. This function has a bandwidth of $\sqrt{k_x^2+k_y^2}<K=\pi$ and so has a critical spacing of $a^*=\pi/K=1$ in every direction. In each subfigure, the colored regions are the Voronoi cells around the sample points (black). Subfigure a) shows the function values for all $x$. b) shows $f_\text{B}$ sampled on a square lattice with $z_{n,m}=(n/2,m/2)$. Since $1/2<a^*=1$ this is an instance of oversampling. c) shows $f_\text{B}$ sampled on a square lattice with $z_{n,m}=(n,m)$. This is an instance of critical sampling since $a=a^*=1$. d) shows $f_\text{B}$ sampled on a square lattice with $z_{n,m}=(n+m,n-m)/\sqrt{2}$. e) $f_\text{B}$ sampled on a hexagonal lattice of with \mbox{$z_{n,m}=(n+m/2,\sqrt{3}m/2)\in\mathbb{R}^2$}. f) shows $f_\text{B}$ sampled on an irregular lattice.}
\label{Fig2DSamples}
\end{figure*}
The previous subsection showed how any value or derivative of $f_\text{B}$ can be recovered from its values on any sufficiently dense uniform sample lattice. Moreover, it showed how changing between representing $f_\text{B}$ with different uniform sample lattices is ultimately just a change of basis on $\bm{f}$. 

As I will now discuss, we can also represent $\bm{f}$ on any sufficiently dense non-uniform lattice. To motivate this, consider first an oversampling of $f_\text{B}$. For example, figure Fig.~\ref{Fig1DSamples}b) shows $f_\text{B}$ sampled at twice the necessary frequency. This is a representation of $\bm{f}$ in an overcomplete basis. Imagine oversampling by a factor of ten with a spacing of $a=a^*/10$. Intuitively, this sample lattice has ten times the information needed to recover the function exactly. If we were to delete all but every tenth data point we would still be able to recover the function. But what if we just half of the sample points, but did so randomly? This would result in a non-uniform sample lattice. See for instance Fig.~\ref{Fig1DSamples}e). Hopefully, the reader has some intuition that at least some non-uniform sample lattices are sufficient to exactly reconstruct $f_\text{B}$.

The scope of such non-uniform samplings is established by various non-uniform sampling theorems~\cite{GARCIA200263,SamplingTutorial}. The details of these theorems are not important here; They can all be summarized as saying that reconstruction is possible when our non-uniform sample points are ``sufficiently dense'' in some technical sense. The sampling shown in Fig. \ref{Fig1DSamples}e) is sufficiently dense. The reconstruction in the non-uniform case is significantly more complicated than it is in the uniform case. Rather than Eq.~\eqref{SincRecon}, in the non-uniform case our reconstruction is of the form,
\begin{align}\label{GenRecon}
f_\text{B}(x)=\sum_{m=-\infty}^\infty G_m(x;\{z_n\}_{n\in\mathbb{Z}}) \, f_\text{B}(z_m) 
\end{align}
for some reconstruction functions, $G_m$, which depend in a complicated way on the location of all of the other sample points, $\{z_n\}_{n\in\mathbb{Z}}$.

\subsection{Higher Dimensional Sampling}
The same story about bandlimited functions is largely true in higher dimensions as well. A two-dimensional function $f_\text{B}(x,y)$ is bandlimited if is Fourier transform $\mathcal{F}[f_\text{B}(x,y)](k_x,k_y)$ is compactly supported in the \mbox{($k_x$, $k_y$)-plane}. Specifying the value of the bandwidth is less straightforward in the high dimensional case as the Fourier transform's support may have different extents in different directions. However, any compact region can be bounded in a square. We can thus always imagine $f_\text{B}(x,y)$ as being bandlimited with \mbox{$k_x,k_y\in[-K,K]$} for some $K>0$. As such, we can represent $f_\text{B}(x,y)$ with a (sufficiently dense) uniform sample lattice in both the $x$ and $y$ directions. That is, we can represent any bandlimited $f_\text{B}(x,y)$ in terms of its sample values on a sufficiently dense square lattice. Once we have such a uniform sampling, the reasoning carried out above applies unchanged. We can represent $f_\text{B}(x,y)$ on any sufficiently dense non-uniform lattice.

For a concrete example consider the bandlimited function shown shown in Fig.~\ref{Fig2DSamples}a), namely,
\begin{align}\label{J1}
f_\text{B}(x,y)=J_1(\pi\,r)/(\pi\,r) 
\end{align}
where $J_1$ is the first Bessel function and $r=\sqrt{x^2+y^2}$. This function is bandlimited with $\sqrt{k_x^2+k_y^2}<K=\pi$ and hence critical spacing $a^*=\pi/K=1$. Moreover, note that this function is rotation invariant.

Given this function's bandwidth of $K=\pi$, we can represent it via its sample values taken on a square lattice with spacing $a=1/2\leq a^*=1$, see Fig. \ref{Fig2DSamples}b). We can also use a coarser square lattice with a spacing of $a=a^*=1$, see Fig. \ref{Fig2DSamples}c). We could also use a rotated square lattice, see Fig. \ref{Fig2DSamples}d). Sampling the function on a hexagonal lattice also works, see Fig. \ref{Fig2DSamples}e). Finally we can use a non-uniform lattice of sample points, see Fig. \ref{Fig2DSamples}f). From each of these discrete representations, we could recover the original bandlimited function everywhere exactly via some generalization of Eq.~\eqref{SincRecon} in the uniform cases and Eq.~\eqref{GenRecon} in the irregular case.

Thus, there is no conceptual barrier to representing a rotationally invariant bandlimited function on a square lattice. Indeed, there is no issue with representing such a function on any sufficiently dense lattice. In light of the analogy proposed in this paper, we can see this as analogous to the unsurprising fact that there is no conceptual barrier to representing rotationally invariant functions in Cartesian coordinates. There is no requirement that our representation (be it a choice of coordinates or a choice of sample points) latches onto the symmetries of what is being represented. 

Thus we have a non-uniform sampling theory for higher dimensions. But what about a sampling theory on curved spaces? While such things are not relevant for the aims of this paper, recently notable progress has been made on developing a sampling theory for curved manifolds~\cite{CurvedSampling,Martin2008}.

\section{Externalizing H1-H7 - Part 2}\label{SecExtPart2}
Having now reviewed the sampling property of bandlimited functions, let's return to the task of embedding $\bm{\Phi}(t)$ into our spacetime manifold $\mathcal{M}$. As discussed in Sec.~\ref{SecExtPart1}, for H4-H7 our translation-matching constraint forces the new field \mbox{$\phi:\mathcal{M}\to\mathbb{R}$}  to be bandlimited in the coordinate system $d_\text{trans.}$ with bandwidth $k_1,k_2\in[-K,K]$ on each time-slice. Similarly for H1-H3. In either case let's note this by giving the field a subscript $\text{B}$ as \mbox{$\phi_\text{B}:\mathcal{M}\to\mathbb{R}$}. 

Before continuing on, it should be noted that the above discussed translation-matching constraint guarantees that $\phi_\text{B}\circ d_\text{trans.}$ will have the sampling property discussed in the previous section on each time-slice.

\subsection{A Principled Choice of Profile}
What freedom remains for our choice of embedding? Focusing on H4-H7, in coordinate $d_\text{trans.}$ our remaining freedom is a (potentially time-dependent) choice of profile \mbox{$f(t;x,y)\in F^\pi$} appearing in Eq.~\eqref{Embedf}. A translated and rescaled copy of this profile is associated with each basis vector $\bm{e}_n\otimes\bm{e}_m$. As noted following Eq.~\eqref{Eftilde}, up to a local recaling of Fourier space (which is a symmetry of the dynamics) our embedding is fixed. However, in order to completely fix this embedding, we need to pick a profile $f(t;x,y)$.

The discussion in the previous section offers us a promising candidate for this profile namely \mbox{$f(t;x,y)=S_0(x)S_0(y)$}. This profile is notable as it is the unique one in $F^\pi$ with the following property\footnote{To give the technical details, this $f$ is the unique function in $F^\pi$ which evaluates to zero at all integer arguments except $(0,0)$ where it returns $1$. That is, \mbox{$f(t;n,m)=\delta_{n0}\delta_{m0}$} for $n,m\in\mathbb{Z}$. This uniqueness is a direct consequence of the Nyquist-Shannon sampling theorem.}: it makes $\phi_\text{B}(t,x,y)$ evaluated at \mbox{$z_{n,m}(t)=(t,n\,a,m\,a)$} take on the value of $\phi_{n,m}(t)$. That is, it makes our original discrete variables $\phi_{n,m}(t)$ the sample values of $\phi_\text{B}(t)$ at the sample points $z_{n,m}(t)$. Let us call this choice of profiles the lattice-as-sample-points constraint.

To be clear, what hinges on our choice of profile $f$ is not whether or not $\phi_\text{B}(t)$ has the sampling property discussed in the previous section; This property is guaranteed independent of our choice of $f$. Rather, what is at stake is just whether our original discrete variables $\phi_{n,m}(t)$ \mbox{\textit{themselves}} are sample values.

It follows from the lattice-as-sample-points constraint that $E(t)$ acts on the planewave basis as
\begin{align}\label{Embed0}
E(t):\bm{\Phi}(k_1,k_2)\mapsto\phi(x,y;k_1/a,k_2/a)
\end{align}
independent of time. That is, discrete planewaves are mapped onto continuous planewaves (rescaled by $a$). Moreover, this constraint forces $E(t)$ to act on the basis $\bm{e}_n\otimes\bm{e}_m$ as
\begin{align}\label{Embed}
E(t): \bm{e}_n\otimes\bm{e}_m\mapsto
S_{n}(x/a) \, S_{m}(y/a)
\end{align}
independent of time. In particular, applying this to Eq.~\eqref{PhiVec2} gives us
\begin{align}\label{PhiSincRecon1}
&\text{H1-H3:}\\
\nonumber
&\quad\phi_\text{B}(t,x)
=\sum_{n\in\mathbb{Z}} S_{n}(x/a) \ \phi_{n}(t)\\
\label{PhiSincRecon2}
&\text{H4-H7:}\\
\nonumber
&\quad\phi_\text{B}(t,x,y)
=\sum_{n,m\in\mathbb{Z}} S_{n}(x/a) \, S_{m}(x/a) \, \phi_{n,m}(t)
\end{align}
where analogous reasoning applies for H1-H3.

Thus, from these translation-matching and lattice-as-sample-points constraints it follows that our new field must be the result of a certain bandlimited reconstruction. Namely, for H4-H7 it must come from a reconstruction using sample values $\phi_{n,m}(t)$ at sample points $z_{n,m}(t)$. Similarly for H1-H3 it must come from using sample values $\phi_{n}(t)$ at sample points \mbox{$z_{n}(t)=(t,n\,a)$}.

\begin{figure}[t]
\includegraphics[width=0.45\textwidth]{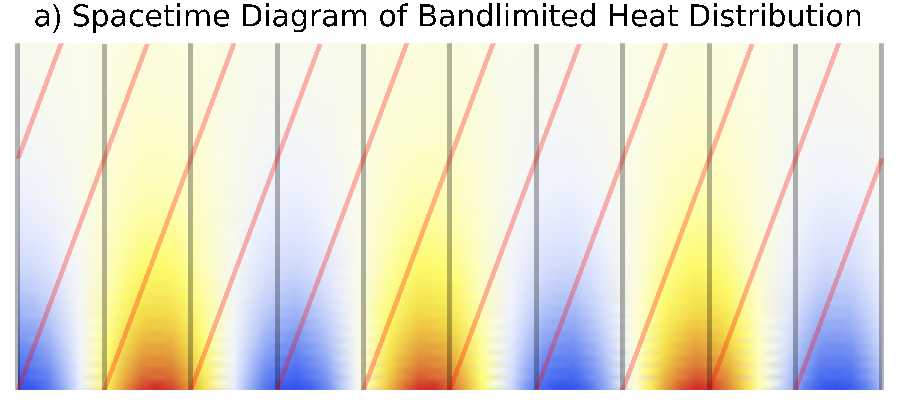}
\includegraphics[width=0.45\textwidth]{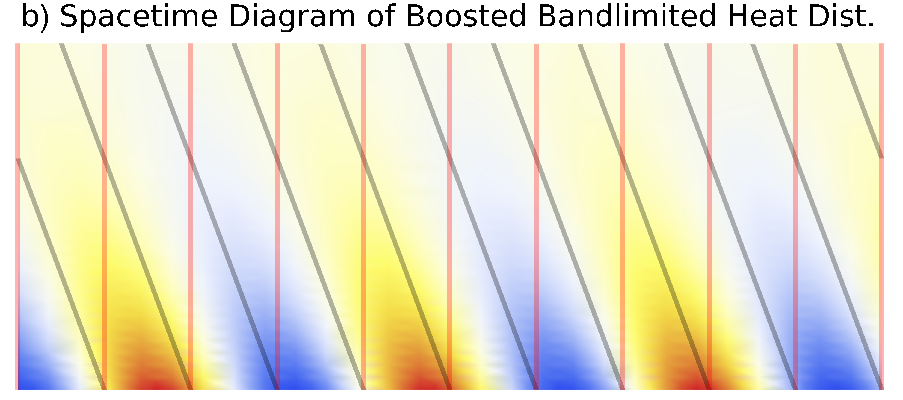}
\caption{Subfigure a) shows how the sample points $z_n(t)=(t,n\,a)$ lie on the manifold $\mathcal{M}$ in the coordinates given by $d_\text{trans.}$ as vertical black lines. These field $\phi_\text{B}$ takes on values at these sample points $\phi_n(t)=\phi_\text{B}(z_n(t))$ which obey the discrete dynamics Eq.~\eqref{H1Long}. From these sample points, we can uniquely reconstruct the bandlimited-in-space field $\phi_\text{B}$ shown in the background using Eq.~\eqref{PhiSincRecon1}. This reconstructed field obeys the continuum dynamics Eq.~\eqref{DH1bandlimited}. Once embedded, these original sample points no longer play a special role. We are free to redescribe the state and its dynamics by resampling it on any sufficiently dense collection of sample points. For instance, we might describe $\phi_\text{B}$ using the boosted sample point (red diagonal lines). These new sample values obey Eq.~\eqref{DH1boosted}. Subfigure b) shows the above situation after a boost has been applied to both the field and the sample points. The black and red sample values still obey Eq.~\eqref{H1Long} and Eq.~\eqref{DH1boosted} respectively, but the field $\phi_\text{B}$ now obeys Eq.~\eqref{DH1bandlimitedboosted}.}\label{FigHeatSpaceTime}
\end{figure}

Fig.~\ref{FigHeatSpaceTime}a) shows for H1-H3 these sample points \mbox{$z_{n}(t)=(t,n\,a)$} as they lie on the spacetime manifold in the coordinates $d_\text{trans.}$ as vertical black lines. Fig.~\ref{FigHeatSpaceTime}a) also shows for H1 what the reconstructed bandlimited function $\phi_\text{B}:\mathcal{M}\to\mathbb{R}$ might look like in these coordinates. One can imagine an analogous figure for H4-H7 with the sample points forming a square lattice extended through time in the coordinates $d_\text{trans.}$. 

One may worry that we are here embedding onto a square 2D lattice for each of H4-H7 whereas for H5 and H7 we naturally ought to embed onto a hexagonal 2D lattice. This was after all, the lattice structure picked out by our symmetry analysis in the first interpretation. However, one need not worry for two reasons: Firstly, there is no real sense in which the sample points $z_{n,m}(t)=(t,n\,a,m\,a)$ form a square lattice. This sample lattice only appears square in this particular coordinate system; in other coordinate systems they won't. In other coordinates, a square lattice may look hexagonal, see Fig.~\ref{FigSkew}. Indeed, the coordinates we are using here may ultimately lie on the spacetime manifold $\mathcal{M}$ in a curvilinear way. As such one ought not to think of these sample points as being arranged in a square lattice on the spacetime manifold. Indeed, at this point the manifold still does not have a metric.

Secondly, after we have used the sample points $z_{n}(t)$ and $z_{n,m}(t)$ to build $\phi_\text{B}$ they no longer play any special role in the theory. As discussed in the previous section, we are always free to resample a bandlimited function on any sufficiently dense sample lattice. Recall Fig.~\ref{Fig2DSamples}. Thus even staying in this coordinate system (where our initial sample points do form a square lattice) we are free to resample $\phi_\text{B}$ on a hexagonal lattice. Moreover, are also free to resample $\phi_\text{B}$ on a boosted lattice, e.g., the red lines shown in Fig.~\ref{FigHeatSpaceTime}a). The effect that such resamplings have on the theory's dynamical equations will be discussed in Appendix~\ref{SecResample}.

Thus, in a sense the lattice sites $L$ from our first interpretation still exist, they are embedded onto our new spacetime manifold $\ell\in L\mapsto z_\ell(t) \in\mathcal{M}(t)$. However, once embedded, these lattice sites no longer play any special role in the theory; We can always resample.

\vspace{0.25cm}

To review: In Sec.~\ref{SecExtPart1} we saw how the translation-matching constraint forced us to embed $\bm{\Phi}(t)$ onto $\mathcal{M}$ as a some bandlimited function, $\phi_\text{B}:\mathcal{M}\to\mathbb{R}$ (i.e., bandlimited in coordinates $d_\text{trans.}$ with $k_1,k_2\in[-K,K]$ on each time-slice.). In this section, we further demanded the lattice-as-sample-points constraint. This allowed us to think of our initial discrete variables $\phi_\ell(t)$ as being the values that $\phi_\text{B}$ takes at some sample points $z_\ell(t)\in\mathcal{M}$. Taken together these two constraints completely fix how $\phi_\text{B}:\mathcal{M}\to\mathbb{R}$ is built from $\bm{\Phi}(t)$ (or equivalently $\phi_\ell(t)$). In particular, in the global coordinate system given by $d_\text{trans.}$ we must have Eq.~\eqref{PhiSincRecon1} and Eq.~\eqref{PhiSincRecon2}.

In the next subsection, I will rewrite the dynamics of each of H1-H7 in terms of this new field $\phi_\text{B}$ in the coordinates $d_\text{trans.}$. Following this, in Secs.~\ref{SecHeat3} and \ref{SecHeat3Extra}, I will begin interpreting this theory independently of how they were constructed. In particular, in Sec.~\ref{SecFullGenCov} I will give a coordinate-independent formulation of these theories.

\subsection{Bandlimited Dynamics}\label{SecHeatBandDyn}
The previous subsections has given us a principled way to embed $\phi_\ell(t)$ onto a time-slice of our new spacetime manifold $\mathcal{M}(t)$ as a certain bandlimited function. Namely, using Eq.~\eqref{PhiSincRecon1} and Eq.~\eqref{PhiSincRecon2}. Equivalently we can think of embedding $\bm{\Phi}(t)$ onto a time-slice of our new spacetime manifold $\mathcal{M}(t)$ as a certain bandlimited function. Namely, using Eq.~\eqref{Embed}. Either of these perspectives allow us to translate the kinematics of H1-H7 into this new continuous setting. This subsection will additionally translate over the dynamics into this new continuous setting. 

Let's begin with H1. This translation is aided by the fact that the derivative is the generator of translations, i.e., $h(x+a)=\text{exp}(a\, \partial_x) h(x)$. For H1 we have,
\begin{align}\label{DH1bandlimited}
\frac{\partial}{\partial t}\phi_\text{B}
&=\sum_{n\in\mathbb{Z}} S_{n}(x/a) \ \frac{\d}{\d t}\phi_{n}(t)\\
\nonumber
&=\alpha\!\sum_{n\in\mathbb{Z}}\! S_{n}(x/a) \big[\phi_{n+1}(t)-2\phi_{n}(t)+\phi_{n-1}(t)\big]\\
\nonumber
&=\alpha\,
[\phi_\text{B}(t,x - a) - 2\phi_\text{B}(t,x) + \phi_\text{B}(t,x + a)]\\
\nonumber
&=\alpha \, [\exp(-a\,\partial_x)-2+\exp(a\,\partial_x)] \phi_\text{B}(t,x)\\
\nonumber
&=\alpha \, [2\text{cosh}(a\,\partial_x)-2] \ \phi_\text{B}(t,x).
\end{align}
Similarly for the other six theories we have:
\begin{align}\label{DH2bandlimited}
\text{H2}:&\ \partial_t\phi_\text{B}
\!=\!\frac{\alpha}{6} [-\text{cosh}(2a\partial_x)\!+\!16\text{cosh}(a\partial_x)\!-\!15]\phi_\text{B}\\
\label{DH3bandlimited}
\text{H3}:&\ \partial_t\phi_\text{B}
=\alpha_0\,\partial_x^2 \, \phi_\text{B}\\
\label{DH4bandlimited}
\text{H4}:&\ \partial_t\phi_\text{B}
=\alpha\,[\text{cosh}(a\,\partial_x)+\text{cosh}(a\,\partial_y)\!-\!2]\phi_\text{B}\\
\label{DH5bandlimited}
\text{H5}:&\ \partial_t\phi_\text{B}
=\alpha\,[\text{cosh}(a\,\partial_x)+\text{cosh}(a\,\partial_y)\\
\nonumber
&\qquad\qquad \ \ \ +\text{cosh}(a\,(\partial_y-\partial_x))\!-\!3]
\phi_\text{B},\\
\label{DH6bandlimited}
\text{H6}:&\ 
\partial_t\phi_\text{B}
=\frac{\alpha_0}{2} \, (\partial_x^2+\partial_y^2) \, \phi_\text{B}\\
\label{DH7bandlimited}
\text{H7}:&\ 
\partial_t\phi_\text{B}
=\frac{\alpha_0}{3}\left(\partial_x^2+\partial_y^2+(\partial_x-\partial_y)^2\right) \, \phi_\text{B}
\end{align}
where $\alpha_0\coloneqq a^2\alpha$. 

\begin{figure}[t]
\includegraphics[width=0.4\textwidth]{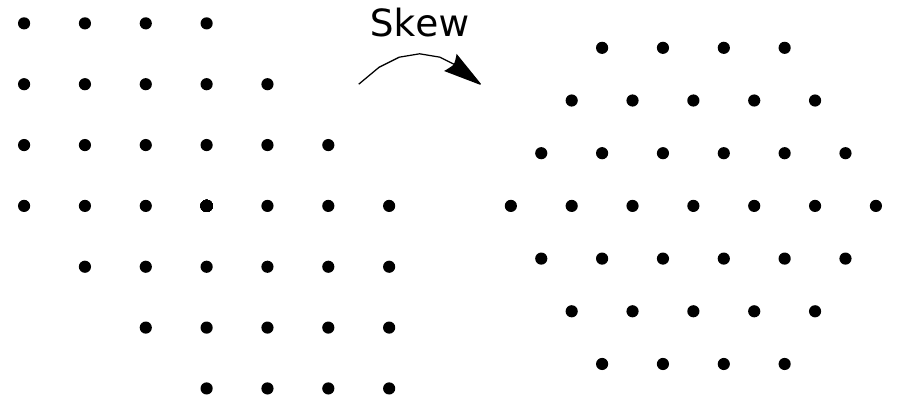}
\caption{As this figure shows, a certain linear transformation of coordinates (namely Eq.~\eqref{CoorH7H6}) maps a square lattice to a hexagonal one.}\label{FigSkew}
\end{figure}

Note that the dynamics of H3 and H6 are exactly the same as those of the continuum heat equations H00 and H0 discussed in Sec.~\ref{SecSevenHeat} with a dispersion rate $\bar{\alpha}=\alpha_0$. Moreover, after a coordinate transformation, 
\begin{align}\label{CoorH7H6}
t\mapsto t,\quad
x\mapsto x + \frac{1}{2} y,\quad
y\mapsto \frac{\sqrt{3}}{2}y
\end{align}
the dynamics of H7 exactly maps onto the dynamics of H6 (and consequently, H0). Note that this change of coordinates is equivalent to Eq.~\eqref{SkewH7H6} applied in continuum Fourier space. Moreover, note that Eq.~\eqref{CoorH7H6} maps a square lattice to a hexagonal one, see Fig.~\ref{FigSkew}. As I will discuss, later H6 and H7 can be seen as describing the same bandlimited theory, just using different sample points. 

Applying this coordinate transformation to H5's bandlimited dynamics gives us
\begin{align}\label{DH5bandlimitedSkew}
\text{H5:}\ \partial_t \phi_\text{B}
=\frac{\alpha}{3}\,\big[ &2\text{cosh}(a\,\partial_x)-2\\
\nonumber
+&2\text{cosh}(a(\sqrt{3} \partial_y+\partial_x)/2)-2\\
\nonumber
+&2\text{cosh}(a(\sqrt{3} \partial_y-\partial_x)/2)-2\big]\phi_\text{B}.
\end{align}
Note that this dynamics now manifestly has a one-sixth rotational symmetry.

Sampling the above dynamics on a hexagonal lattice gives us our original dynamics for H5. Sampling the above dynamics on a square lattice gives us the following. Given Eq.~\eqref{ExactDerivative} sampling on a square lattice effectively means taking $a\,\partial_x\to D_\text{n}$ and $a\,\partial_y\to D_\text{m}$ and replacing $\phi_\text{B}(t)$ with $\bm{\Phi}(t)$. For the above dynamics of H5 this gives:
\begin{align}\label{DH5Skew}
\text{H5:}\ \partial_t \bm{\Phi}(t)
=\frac{\alpha}{3}\,\big[ &2\text{cosh}(D_\text{n})-2\\
\nonumber
+&2\text{cosh}((\sqrt{3} D_\text{m}+D_\text{n})/2)-2\\
\nonumber
+&2\text{cosh}((\sqrt{3} D_\text{m}-D_\text{n})/2)-2\big]\bm{\Phi}(t).
\end{align}
Note that before this resampling H5 was local in the intuitive sense nearest-neighbor couplings only, see Eq.~\eqref{H5Long}. After resampling it is infinite-range. Thus, the intuitive notion of locality in terms of nearest neighbors is unstable under resampling. This will be discussed further in Sec.~\ref{SecBandlimitedLocality} and Appendix~\ref{SecResample}.

We can easily solve each of these dynamical equations. Just as in Sec.~\ref{SecSevenHeat} these dynamics admit a complete basis of planewave solutions, given by Eq.~\eqref{PlaneWaveCont}. A few key differences should be noted between these continuum planewaves and the discrete planewaves considered earlier, namely $\bm{\Phi}(k)$ and $\bm{\Phi}(k_1,k_2)$. Firstly, there is a difference of scale: note how our embedding Eq.~\eqref{Embed0} rescales the planewaves by a factor of $1/a$. Secondly, the discrete planewaves $\bm{\Phi}(k_1,k_2)$ repeated themselves cyclically with period $2\pi$ outside of the region \mbox{$k_1,k_2\in[-\pi,\pi]$}. The continuum planewaves do not do this. Now each planewave with $k_1,k_2\in\mathbb{R}$ is distinct. Here, however, by its construction $\phi_\text{B}(t)$ only has support over planewaves with \mbox{$k_1,k_2\in[-K,K]$} with $K=\pi/a$.

Each of these planewaves are solutions when multiplied by $\text{exp}(-\Gamma \, t)$ with the following wavenumber dependent decay rates: 
\begin{align}
\text{H1}:& \ \Gamma= \alpha \, (2-2\,\text{cos}(k\,a))\\
\nonumber
\text{H2}:& \ \Gamma= \frac{\alpha}{6}\,(\text{cos}(2\,k\,a)-16\,\text{cos}(k\,a)+15)\\
\nonumber
\text{H3}:& \ \Gamma= \alpha_0 \, k^2\\
\nonumber
\text{H4}:& \ \Gamma= \alpha \left(2\!-\!\text{cos}(k_1\,a)\!-\!\text{cos}(k_2\,a)\right)\\
\nonumber
\nonumber
\text{H5}:& \ \Gamma= \frac{2\alpha}{3} \left(3\!-\!\text{cos}(k_1\,a)\!-\!\text{cos}(k_2\,a)\!-\!\text{cos}(k_2\,a-k_1\,a)\right)\\
\nonumber
\text{H6}:& \ \Gamma= \frac{\alpha_0}{2} \, (k_1^2+k_2^2)\\
\nonumber
\text{H7}:& \ \Gamma=
\frac{\alpha_0}{3}\left(k_1^2+k_2^2+(k_2-k_1)^2\right).
\end{align}
Note that the decay rates for H3 and H6 are exactly the same as those of the continuum heat equations H00 and H0 discussed with a dispersion rate of $\bar{\alpha}=\alpha_0$. Moreover, after the coordinate transformation, Eq.~\eqref{SkewH6H7}, 
the decay rate for H7 exactly maps onto that of H6 (and consequently, H0). More will be said about this later.

We are now ready to make a third attempt at interpreting these theories. 

\section{A Third Attempt at Interpreting H1-H7 - Part 1}\label{SecHeat3}
The previous three sections have (after much effort) given us a new formulation of H1-H7 to interpret. In particular, this new formulation has aimed to externalize many of the internal symmetries which we discovered in our second interpretation in Sec.~\ref{SecHeat2}.

On this third interpretation I will be taking the formulations of H1-H7 in terms of $\phi_\text{B}$ seriously: namely Eq.~\eqref{DH1bandlimited}-\eqref{DH7bandlimited}. Taken literally as written, what are these theories about? 
Intuitively these theories are about a field $\phi_\text{B}$ which maps points on a manifold ($p\in\mathcal{M}$) onto temperatures ($\phi_\text{B}(p)\in\mathbb{R}$). That is a field $\phi_\text{B}:\mathcal{M}\to \mathcal{V}$ with a manifold $\mathcal{M}$ and value space $\mathcal{V}=\mathbb{R}$. Thus, taking $\phi_\text{B}:\mathcal{M}\to \mathcal{V}$ seriously as a fundamental field leads us to thinking of $\mathcal{M}$ as the theory's underlying manifold and $\mathcal{V}=\mathbb{R}$ as the theory's value space. Indeed, on this interpretation
H1-H7 are continuum spacetime theories of the sort we
are used to interpreting (albeit ones which consider only bandlimited fields).

Note that here we have $\mathcal{M}\cong\mathbb{R}^n$ with $n=2$ for H1-H3 and $n=3$ for H4-H7. As such, in either case we have access to global coordinate systems. The fields $\phi_\text{B}$ considered by this interpretation are bandlimited in the following sense. There exists a fixed special diffeomorphism $d_\text{coor.}:\mathbb{R}^n\to\mathcal{M}$ which gives us a fixed special global coordinate system for $\mathcal{M}$. In these special coordinates the field is bandlimited on each time-slice. That is $\phi_\text{B}\circ d_\text{trans.}\in F^K$ on each time-slice. It is in these special coordinates that the fields $\phi_\text{B}$ obey the dynamical equations Eq.~\eqref{DH1bandlimited}-Eq.~\eqref{DH7bandlimited}.

Of course, one ought to be suspicious of any ``special coordinates'' appearing in a supposedly fundamental formulation of a theory. Where did it come from? In our construction of $\phi_\text{B}$ in Sec.~\ref{SecExtPart1}, this special coordinate system is the one which is associated to the translation operations $T^\epsilon$ by our translation-matching condition, namely $d_\text{coor.}=d_\text{trans.}$. Here however, we are trying to interpret H1-H7 as formulated above independent of how we arrived here. Thus, presently, this special coordinate system $d_\text{coor.}$ is just something promised to us by God. A coordinate-free view of these theories will be given in Sec.~\ref{SecFullGenCov}.

As I will now discuss, taking $\mathcal{M}$ to be these theories' underlying manifold has substantial consequences for these theories' locality properties and symmetries. To preview: this third interpretation either fixes all of our issues with the first and second interpretations. 

In our first interpretation there was some tension between our theories' differences in locality and their differences in convergence rate in the continuum limit. The second interpretation addressed this tension in a hamfistedly way: denying that there are even differences in locality in the first place. This move also had the unfortunate consequence of undercutting our ability to use the lattice sites to reason about locality. This interpretation improves on this by instead bringing the locality of these theories into harmony with their convergence rates in the continuum limit.

In our first interpretation, H6 and H7 were seen as distinct theories with radically different symmetries despite there being a nice one-to-one correspondence between a substantial portion of their solutions. In our second interpretation, H6 and H7 were more satisfyingly seen to be substantially equivalent, differing by only a change of basis in the value space. Here too H6 and H7 will turn out to be substantially equivalent, differing only by a change of coordinates. Moreover, as I will discuss, we can here see H6 and H7 in their entirety as parts of a larger unified theory viewed in two ways with differently limited representational capacities.

Our first interpretation outright denied the possibility that H1-H7 could have continuous symmetries. Our second interpretation fixed this oversight by revealing H1-H7's hidden continuous translation and rotation symmetries. However, these were unintuitively categorized as internal symmetries. This interpretation maintains all of these hidden symmetries, but more satisfyingly categorizes them as external symmetries.

As I discussed in Sec.~\ref{SecHeat2}, the improvements that the second interpretation made over the first one all centered around the following realization. The lattice structures appearing in H1-H7 are merely representational artifacts and as such should not be taken seriously as a substantive part of the theory. As I will discuss, this realization is deepened and clarified on this third interpretation.

Let's begin by discussing locality in this bandlimited setting.

\subsection{Bandlimited Locality}\label{SecBandlimitedLocality}
To begin, let's develop a sense of comparative locality for H1-H7 according to this third interpretation. First, however, a brief review of locality on our first two interpretations. Recall that on our first interpretation we found an intuitive notion of locality in terms of the number of lattice sites instantaneously coupled together. We there found \mbox{$\text{H1}>\text{H2}>\text{H3}$} and \mbox{$\text{H4},\text{H5}>\text{H6},\text{H7}$} with higher rated theories being more local. Recall that these locality ratings were in tension with the rates at which these theories converge in the continuum limit: \mbox{$\text{H3}>\text{H2}>\text{H1}$} and \mbox{$\text{H6},\text{H7}>\text{H4},\text{H5}$} with higher rated theories converging faster in the continuum limit.

Our second interpretation resolved this tension by denying that these theories have any differences in terms of locality at all. In particular, our second interpretation internalized the lattice sites and thereby undercut any possibility of using them to reason about locality. On this view, there is no longer a tension between differences in locality and differences in convergence rates because there are no longer differences in locality. This dissolves the tension albeit in a less than satisfying way.

Here, however, the lattice sites are not internalized but rather embedded into a continuum manifold. It is natural that along with this substantial change in the theories' underlying manifolds, come substantial changes in the relevant notion of locality. Here, as I will discuss, rather than being undercut, all of our initial intuitive locality ratings are reversed. This does more than just dissolve the tension with the convergence rates. Rather, it resolves the tension by bringing our assessments of locality and convergence rates into harmony.

Let's begin by assessing the locality of the differential equations governing H1-H7 in terms of $\phi_\text{B}$. In general, differential equations are considered local when they only involve derivatives up to a fixed finite order. Each of these derivatives is a local operation and there is no way to build from a finite set of them something non-local. However, when one is allowed an infinite number of derivatives one can create non-local dynamics. Recall that $h(x+a)=\text{exp}(a\, \partial_x) h(x)$. Indeed, this is exactly what is going on in the dynamical equations of H1, H2, H4 and H5. 

From a bandlimited perspective, these ``nearest-neighbor'' theories are highly non-local theories despite them being rated the most local by our first interpretation. Concretely, in H1, H2, H4, and H5 the bandlimited field $\phi_\text{B}(t)$ is instantaneously coupled to itself a distance of $a$ or even $2\,a$ away. By contrast, H3, H6, and H7 are perfectly local from the bandlimited perspective. On this new bandlimited notion of locality we have the ratings $\text{H3}>\text{H2}>\text{H1}$ and $\text{H6},\text{H7}>\text{H4},\text{H5}$. These are essentially the reverse  judgments of what we had on our first interpretation. Moreover, they match with the rates at which these theories converge in the continuum limit.

But which of these two notions of locality should we care about? Well, this depends on what we take the theory's underlying spacetime manifold to be. Indeed, it's not surprising that changes in the underlying manifold as drastic as $Q\to\mathcal{M}$ will have drastic consequences for the relevant notion of locality. The question is then which of these two potentially underlying manifolds should we care about?

As I will discuss in the next subsection, we have reason to prefer $\mathcal{M}$ over $Q$ on grounds of symmetry. As I have already discussed, taking $Q$ to be the manifold underlying H1-H7 systematically underpredicts the symmetries these theories can and do have. As I will soon discuss, $\mathcal{M}$ does not have this issue. This alone is already a substantial reason to prefer $\mathcal{M}$ over $Q$. This in addition to the above-discussed resolution of the tension between locality and convergence rate.

Thus, we have two different notions of locality associated with two different framings of these theories. Moreover, we have substantial reason to prefer one of these framings over the other and hence substantial reason to prefer one of these notions of locality over the other. When properly understood, it turns out that nearest-neighbor lattice theories are ultimately tremendously non-local.

One may still be puzzled, however. Suppose we stick with viewing $\mathcal{M}$ as the underlying manifold. One may reason (poorly) as follows: Sure, when we view the dynamics of H3 (or H6 or H7) in terms of $\phi_\text{B}$ it is perfectly local. However, when we view the very same dynamics in terms of its sample points (which are after all extremely local objects: samples at a point) we find dynamics like Eq.~\eqref{DH3} which couples the sample points to each other at an infinite range via the operator $D$. This dynamics seems very non-local. In summary: describing local dynamics in terms of local degrees of freedom (sample values at various points) seems to give us non-local dynamics. What gives?

The oversight in the above line-of-thought is thinking that the sample points $\phi_\text{B}(z)$ correspond to localized degrees of freedom of the bandlimited field $\phi_\text{B}$. Surprisingly, they do not. Yes, the sample value of a bandlimited function $f_\text{B}$ at some point $x_0$ can be understood as
\begin{align}\label{DeltaSample}
f_\text{B}(x_0) = \int \d x \, f_\text{B}(x) \, \delta(x-x_0).
\end{align}
However, it is also true for every bandlimited function $f_\text{B}$ with bandwidth $k\in[-\pi/a^*,\pi/a^*]$ and for every point $x_0$ that,
\begin{align}\label{WeightedAvg}
f_\text{B}(x_0) = \frac{1}{a}\int \d x \, f_\text{B}(x) \, S\left(\frac{x-x_0}{a}\right).
\end{align}
for every $a<a^*$. This is because $\delta(x)$ and $S(x/a)$ have identical Fourier transforms for $k\in[-\pi/a^*,\pi/a^*]$. For bandlimited functions like $f_\text{B}$, these two kernels are identical. We thus have two mathematical representations for what it means to evaluate a bandlimited function at a point.

These two understandings of what it means to evaluate a bandlimited function at a point correspond closely with the two ways we considered embedding our dynamics onto a manifold. On the first view, we have $\phi_\text{B}$ constructed from $\phi_\ell(t)$ as follows. We associated each lattice site $\ell\in L$ at each times $t\in\mathbb{R}$ with a point on $\mathcal{M}(t)$ via an embedding $z_\ell(t)\in\mathcal{M}(t)$. By this means each lattice site at each time ends up being associated with a point on the manifold $\mathcal{M}(t)$. By the translation-matching and lattice-as-sample-points constraints, $\phi_\text{B}$ takes a value of $\phi_\ell(t)$ at $z_\ell(t)$. This parallels the point-like way that Eq.~\eqref{DeltaSample} extracts the sample value $f_\text{B}(x_0)$ from $f_\text{B}$.

By contrast, we can also see on the second view we have $\phi_\text{B}$ constructed from $\bm{\Phi}(t)$ as via some vector space isomorphism, $E(t)$. In particular, the translation-matching constraint led us to associate a bandlimited profile (not a point) to each lattice site, see Eq.~\eqref{Embedf}. This parallels the smeared-out way that Eq.~\eqref{WeightedAvg} extracts the sample value $f_\text{B}(x_0)$ from $f_\text{B}$.

As I will now discuss, the second representation of ``evaluating a bandlimited function at a point'' is more in line with the nature of bandlimited functions than the first is. Mathematically, this is because its kernel (the sinc function) is bandlimited whereas the other's (the Dirac delta) is not. Said differently, in Eq.~\eqref{DeltaSample}, we project the bandlimited function onto a kernel outside of the bandlimited universe, whereas in Eq.~\eqref{WeightedAvg} we stay within the bandlimited universe. 

It is standard to think of functions as something which maps some ``point'' in an input space to some ``point'' in an output space. Thus, standardly, at the core of being a function is the notion of ``evaluating a function at a point''. This is often taken as a primitive unanalyzable operation: it's just what functions do. However, pretend for a moment that we meet an alien species who have never thought of functions in this way. Rather, they take as an primitive unanalyzable operation: integrating two functions against each other. (More generally, suppose they hold this attitude towards integrating a function against a distribution.) When they ask us what we mean by ``evaluating a function at a point'', we would likely answer them by pointing to an equation like Eq.~\eqref{DeltaSample}. 

To this they may ask, how do we know that such distributions as the Dirac delta exist when and where we need them? In a bandlimited context they do not. Thus, bizarrely, for bandlimited functions the supposedly basic notion of "evaluating a function at a point" breaks down. This likely has consequences for how we think of the spacetime manifold (if points aren't the sort of things we can evaluate functions at, what are they?) but this is a question for another paper.

There is a physical story which runs parallel to this mathematics regarding the localization of degrees of freedom and counterfactuals. My claim is that if we restrict our attention to bandlimited functions, then bandlimited functions have no localized degrees of freedom. I am here understanding degrees of freedom as things which can vary independently from each other. This is a counterfactual context-sensitive notion in that it depends on both what the other candidate degrees of freedom are and how we are allowed to vary them. One cannot (while keeping $f_\text{B}$ bandlimited) change the value of $f_\text{B}$ only at one point or even only in a compact region. Suppose you could. The difference between the function before and after the change would itself have to be bandlimited (the set of bandlimited functions is closed under subtraction). But this is impossible since no bandlimited function can be compactly supported. Every compactly supported function has non-zero support over all wavenumbers. Thus, any bandlimited counter-instance must be globally different\footnote{While bandlimited theories have a sort of counterfactual non-locality among their allowed states (being bandlimited is a non-trivial global constraint), they can remain perfectly local in terms of their dynamics, see H3 H6 and H7. This is reminiscent of quantum mechanics where the non-locality associated with entanglement is paired with perfectly local dynamics. These are both tame sorts of non-locality which do not lead to faster-than-light signalling.}.

To be clear: whether or not the sample value $f_\text{B}(z_n)$ is a local degree of freedom of $f_\text{B}$ depends on context. Suppose we fix $f_\text{B}$ by giving its values at some (potentially non-uniform) sample lattice $z_n$. In one sense, all of these sample values are degrees of freedom because we can vary them all independently. Changing each of these would change $f_\text{B}$ almost everywhere, but its values at all the other sample points would remain the same. One cannot however, vary one of these sample values while only changing the function only locally. 

Thus, for both physical and mathematical reasons it is improper to associate the sample values of a bandlimited function with the sample point (besides as a mere label). Nor can one associate the sample values with the weighted average of the function over some compact region\footnote{In this sense Fig.~\ref{Fig2DSamples} and Figs.~\ref{FigEvolutionH4}, \ref{FigEvolutionH5}, and \ref{FigEvolutionH6} are misleading in the following sense. These figures associate each sample value with the sample point it was taken at. As discussed above, this is slightly misleading but ultimately understandable. However, Fig.~\ref{Fig2DSamples} and Fig.~\ref{FigEvolutionH4}, \ref{FigEvolutionH5} and \ref{FigEvolutionH6} also casually associates each sample value with the Voronoi cells surrounding its sample point. One must resist any temptation to associate the sample value with any sort of weighted average taken within these cells.}. If the sample value is to be associated with some weighted average it must be over the whole domain, e.g., as in Eq.~\eqref{WeightedAvg}. 

Ultimately, for H3 the apparent tension in between the locality of the dynamics for $\phi_\text{B}(t,x)$ and the non-locality of the dynamics for its sample values, $\phi_n(t)=\phi_\text{B}(z_n(t))$ is resolved as thus. The sample values themselves are to be understood as non-local objects. Hence, it is unsurprising that non-local things obey non-local dynamics.

We thus have good reason to favor the bandlimited notion of locality introduced here over the intuitive one introduced in Sec.~\ref{SecIntuitiveLocality}. Another such reason is visible in Eq.~\eqref{DH5Skew} and will discussed further in Appendix~\ref{SecResample}: unlike the bandlimited notion of locality, the intuitive notion of locality discussed above is fragile and not preserved under resampling.

\subsection{Bandlimited Symmetries}
How does this externalization move affect our theory's capacity for symmetry? Now that we have a continuous spacetime manifold $\mathcal{M}$ underlying these theories, we have a greatly increased capacity for external symmetries, namely \mbox{$\text{Auto}(\mathcal{M})=\text{Diff}(\mathcal{M})$}. By construction all of the symmetries which we wished to externalize $G^\text{dym}_\text{to-be-ext}$ fit inside of $\text{Diff}(\mathcal{M})$. 

In addition to these external symmetries, we also have some capacity for internal symmetries, namely $\text{Auto}(\mathcal{V})$. Note that our (potentially off-shell) bandlimited fields are themselves vectors $\phi_\text{B}(t,x,y)\in F^K$. Namely, they are closed under addition and scalar multiplication and hence form a vector space. This addition and scalar multiplication is carried out in parallel at each spacetime point. Thus, the field's value space $\mathcal{V}=\mathbb{R}$ is also structured like a vector space. Therefore, we have $\text{Auto}(\mathcal{V})=\text{Aff}(\mathbb{R})$ such that our internal symmetries are linear-affine rescalings of $\phi_\text{B}$. We might also find gauge symmetries by allowing these to vary smoothly across the manifold. Thus, in total the possibly symmetries for our theories under this interpretation are,
\begin{align}\label{BandlimitedSymmetries}
s:\quad\phi_\text{B}\mapsto C \, \phi_\text{B}\circ d +c
\end{align}
where $C:\mathcal{M}\to\mathbb{R}$ and $c:\mathcal{M}\to\mathbb{R}$ are some real scalar functions and $d:\mathcal{M}\to\mathcal{M}$ is some diffeomorphism \mbox{$d\in\text{Diff}(\mathcal{M})$}. 

Are these more or less possible symmetries than we had on our second interpretation? $\text{Diff}(\mathcal{M})$ is a tremendously large group and is in fact bigger than $\text{Aff}(\mathbb{R}^L)$. One might expect that we have strictly more possible symmetries here. However, looks can be deceiving. 

We can translate our old potential symmetries into the new continuum setting as follows. Note that because $\mathbb{R}^L \cong F^K$ we have $\text{Aff}(\mathbb{R}^L)\cong \text{Aff}(F^K)$. Using $d_\text{coor.}$ we can think of $\text{Aff}(F^K)$ varying smoothly over time $\text{Diff}(\mathbb{R})$ as a subgroup of $\text{Aff}(F(\mathcal{M}))$. Note that we can also see the above discussed transformations (i.e., \mbox{$\text{Aff}(\mathbb{R})$} varying smoothly over \mbox{$\text{Diff}(\mathcal{M})$}) as a subgroup of $\text{Aff}(F(\mathcal{M}))$. 

Viewed this way, neither is strictly larger. Indeed, $\text{Aff}(F^K)$ includes transformations in Fourier space which are inaccessible from $\text{Aff}(\mathbb{R})$ and $\text{Diff}(\mathcal{M})$. In particular, the local Fourier rescaling symmetry from Sec.~\ref{SecHeat2} is no longer available to us. On the other hand, our new symmetries contain transformations which are not closed over $F^K$ at each timeslice. However, this comparison seems unfair since our fields are in $F^K$, what if we restrict our attention to diffeomorphisms closed over $F^K$? With this restriction, our new range of potential symmetries is a strict subset of our old one.

Thus, in moving from the second to the third interpretation, we have actually lost some capacity for symmetry. The source of this reduction in capacity is us imposing structure on the theory by our choice of manifold and value space.

As I will now discuss, besides the local Fourier rescaling symmetry, all of the symmetries discussed in Sec.~\ref{SecHeat2} are maintained on this third interpretation. Moreover, these symmetries are reformulated and recategorized. As I will now show, on this interpretation the continuous translation and rotation symmetries identified earlier are now honest-to-goodness manifold symmetries, represented by diffeomorphisms $d\in\text{Diff}(\mathcal{M})$ on the manifold.

\subsubsection*{Symmetries of H1-H7: Third Attempt}
Which transformations of the form Eq.~\eqref{BandlimitedSymmetries} are symmetries for H1-H7? A technical investigation of the symmetries of H1-H7 on this interpretation is carried out in Appendix~\ref{AppB}, but the results are the following. For H1-H3 the dynamical symmetries of the form Eq.~\eqref{BandlimitedSymmetries} are:
\begin{flushleft}
\begin{enumerate}
 \item[1)] continuous translation which maps \mbox{$\phi_\text{B}(t,x)\mapsto \phi_\text{B}(t,x-x_0)$} for $x_0\in\mathbb{R}$,
 \item[2)] a negation symmetry which maps \mbox{$\phi_\text{B}(t,x)\mapsto \phi_\text{B}(t,-x)$},
 \item[3)] constant time shifts which map \mbox{$\phi_\text{B}(t,x)\mapsto \phi_\text{B}(t-t_0,x)$} for $t_0\in\mathbb{R}$,
 \item[4)] global linear rescaling which maps \mbox{$\phi_\text{B}(t,x)\mapsto c_1\,\phi_\text{B}(t,x)$} for some $c_1\in\mathbb{R}$,
 \item[5)] local affine rescaling which maps \mbox{$\phi_\text{B}(t,x)\mapsto c_1\,\phi_\text{B}(t,x)+c_2(t,x)$} for some $c_2(t,x)$ which is also a solution of the dynamics.
\end{enumerate}
\end{flushleft}
These are exactly the same symmetries that we found on the previous interpretation (sans local Fourier rescaling) just reformulated and recategorized: the translation and reflection symmetries are here now external symmetries. Translation here in space and time are generated by \mbox{$\exp(-t_0\partial_t)$} and \mbox{$\exp(-x_0\partial_x)$} respectively for some $t_0,x_0\in\mathbb{R}$. Notice that translating a bandlimited function by any amount preserves its bandwidth. Thus $\phi_\text{B}(t,x)$ remains in $F^K$ upon translation.

Thus, the first big lesson from Sec.~\ref{SecHeat2} is repeated here: despite the fact that our discrete theories H1-H3 can be represented on a lattice, they nonetheless have a continuous translation symmetry. This continuous translation symmetry was hidden on our first interpretation because we there took the lattice to be a fundamental part of the manifold. Here, we do not take the lattice structure so seriously. We have embedded it onto the manifold where it then disappears from view as just one of many possible sample lattices.

Let's next consider H4. For H4 the dynamical symmetries of the form Eq.~\eqref{BandlimitedSymmetries} are:
\begin{flushleft}\begin{enumerate}
 \item[1)] continuous translation which maps \mbox{$\phi_\text{B}(t,x,y)\mapsto \phi_\text{B}(t,x-x_0,y-y_0)$} for some $x_0,y_0\in\mathbb{R}$,
 \item[2)] two negation symmetries which map
 \mbox{$\phi_\text{B}(t,x,y)\mapsto \phi_\text{B}(t,-x,y)$} and \mbox{$\phi_\text{B}(t,x,y)\mapsto \phi_\text{B}(t,x,-y)$} respectively,
 \item[3)] a 4-fold symmetry which maps \mbox{$\phi_\text{B}(t,x,y)\mapsto \phi_\text{B}(t,y,-x)$},
 \item[4)] constant time shifts which map \mbox{$\phi_\text{B}(t,x)\mapsto \phi_\text{B}(t-t_0,x)$} for $t_0\in\mathbb{R}$,
 \item[5)] global linear rescaling which maps \mbox{$\phi_\text{B}(t,x)\mapsto c_1\,\phi_\text{B}(t,x)$} for some $c_1\in\mathbb{R}$,
 \item[6)] local affine rescaling which maps \mbox{$\phi_\text{B}(t,x)\mapsto c_1\,\phi_\text{B}(t,x)+c_2(t,x)$} for some $c_2(t,x)$ which is also a solution of the dynamics.
\end{enumerate}\end{flushleft}
These are exactly the same symmetries that we found on the previous interpretation (sans local Fourier rescaling) just reformulated and recategorized: the translation, reflection, and 4-fold symmetries are here now external symmetries. Thus, H4 has two continuous translation symmetries despite initially being represented on a lattice, Eq.~\eqref{H4Long}.

Let's next consider H5. For H5 the dynamical symmetries of the form Eq.~\eqref{BandlimitedSymmetries} are:
\begin{flushleft}\begin{enumerate}
 \item[1)] continuous translation which maps \mbox{$\phi_\text{B}(t,x,y)\mapsto \phi_\text{B}(t,x-x_0,y-y_0)$} for some $x_0,y_0\in\mathbb{R}$,
 \item[2)] an exchange symmetry which maps \mbox{$\phi_\text{B}(t,x,y)\mapsto \phi_\text{B}(t,y,x)$},
 \item[3)] a 6-fold symmetry which maps \mbox{$\phi_\text{B}(t,x,y)\mapsto \phi_\text{B}(t,-y,x+y)$}. (Roughly, this permutes the three terms in Eq.~\eqref{DH5bandlimited}),
 \item[4)] constant time shifts which map \mbox{$\phi_\text{B}(t,x)\mapsto \phi_\text{B}(t-t_0,x)$} for $t_0\in\mathbb{R}$,
 \item[5)] global linear rescaling which maps \mbox{$\phi_\text{B}(t,x)\mapsto c_1\,\phi_\text{B}(t,x)$} for some $c_1\in\mathbb{R}$,
 \item[6)] local affine rescaling which maps \mbox{$\phi_\text{B}(t,x)\mapsto c_1\,\phi_\text{B}(t,x)+c_2(t,x)$} for some $c_2(t,x)$ which is also a solution of the dynamics.
\end{enumerate}\end{flushleft}
These are exactly the same symmetries that we found on the previous interpretation (sans local Fourier rescaling) just reformulated and recategorized: the translation, reflection, and 6-fold symmetries are here external symmetries. Thus, H5 has two continuous translation symmetries despite initially being represented on a lattice, Eq.~\eqref{H4Long}. Note that after a coordinate transformation Eq.~\eqref{CoorH7H6} the above noted 6-fold symmetry is realized straight-forwardly as one-sixth rotations, see Eq.~\ref{DH5bandlimitedSkew}.

Before moving on to analyze the symmetries of H6 and H7, let's first see what this interpretation has to say about them being equivalent to one another. As discussed in Sec.~\ref{SecHeat2} H6 and H7 have a solution preserving vector space isomorphism between a substantial portion of their solutions. Since our second and third interpretations are related by a solution-preserving vector space isomorphism, \mbox{$E:\mathbb{R}^L\to F^K$}, the same is broadly true here. However, some of the details change.

As I noted in Sec.~\ref{SecHeat2} each of H4-H7 are equivalent to each other in a weak sense: approximately in the continuum limit regime, $\vert k_1\vert,\vert k_2\vert\ll \pi$. Here H4-H7 are approximately equivalent in the $\vert k_1\vert,\vert k_2\vert\ll K=\pi/a$ regime. Note however that this is not the continuum limit regime, we are already in the continuum. Rather this is the regime which is significantly below the bandwidth $K$. While each of H4-H7 are equivalent in this weak sense, H6 and H7 are equivalent in a much stronger sense: H6 and H7 is in \textit{exact} one-to-one correspondence over \textit{the whole of} $\sqrt{k_1^2+k_2^2} < K$ (and indeed more, but ultimately not all of $F^K$).

This exact correspondence is mediated by Eq.~\eqref{SkewH7H6} and Eq.~\eqref{SkewH6H7} in Fourier space or equivalently the coordinate transformations Eq.~\eqref{CoorH7H6} and its inverse. On our second interpretation, we ran into technical trouble here due to the $2\pi$ periodic identification of the discrete planewaves $\bm{\Phi}(k_1,k_2)$ stemming from Euler's identity. This caused our maps between H6 and H7 not to be each other's inverses over the whole of $\mathbb{R}^L\cong\mathbb{R}^\mathbb{Z}\otimes\mathbb{R}^\mathbb{Z}$. We were led to consider instead the subspaces $\mathbb{R}^L_\text{H6}$ and $\mathbb{R}^L_\text{H7}$ where these transformations were each other's inverse.

In the present interpretation, something similar happens but with some key differences. Unlike before, here our maps between H6 and H7 given wide scope are invertible and indeed each other's inverses. Wide scope here meaning seen as acting on $F(\mathbb{R}^3)$ the set of all functions $f(t,x,y)$. However, as before, seen as acting on the relevant vector space $F^K$ we have issues. Namely, Eq.~\eqref{SkewH7H6} and Eq.~\eqref{SkewH6H7}, understood as acting on $F^K$ are closed: a function supported over $k_1,k_2\in[-K,K]$ may have support outside of here after these transformations.

In Sec.~\ref{SecHeat2} we fixed the non-invertible issue by focusing on the subspaces $\mathbb{R}^L_\text{H6}$ and $\mathbb{R}^L_\text{H7}$ where these transformations are each other's inverses. We might overcome our current not-closed-over-$F^K$ issue here in the same way defining 
\begin{align}
F^K_\text{H7}\coloneqq&\text{span}(\phi(x,y;k_1,k_2)\vert\,\text{in }F^K\\
\nonumber
&\text{ before and after applying Eq.~\eqref{SkewH7H6}})\\
F^K_\text{H6}\coloneqq&\text{span}(\phi(x,y;k_1,k_2)\vert\,\text{in }F^K\\
\nonumber
&\text{ before and after applying Eq.~\eqref{SkewH6H7}})\\
\label{FKrotinv}
F^K_\text{rot.inv.}\coloneqq&\text{span}(\phi(x,y;k_1,k_2)\vert \sqrt{k_1^2+k_2^2} < K).
\end{align}
Note that under our embedding $E(t)$ these subspaces are isomorphic to $\mathbb{R}^L_\text{H6}$, $\mathbb{R}^L_\text{H7}$ and $\mathbb{R}^L_\text{rot.inv.}$.

Restricted to $F^K_\text{H6}$ and $F^K_\text{H7}$ these transformations are invertible and indeed are each other's inverses. We thus have a solution-preserving vector-space isomorphism between a substantial portion of H6 and H7's solutions just as before. The fact that these transformations (namely, Eq.~\eqref{CoorH7H6} and its inverse) are of the form Eq.~\eqref{BandlimitedSymmetries} means that for any symmetry transformation for H6 (restricted to $F^K_\text{H6}$) there is a corresponding symmetry transformation for H7 (restricted to $F^K_\text{H7}$) and vice versa.

Thus our second big lesson from Sec.~\ref{SecHeat2} is repeated here: despite H6 and H7 being initially presented to us with very different lattice structures (i.e., a square lattice versus a hexagonal lattice) they have nonetheless turned out to be substantially equivalent to one another. This substantial equivalence was hidden from us on our first interpretation because we there took the lattice too seriously. As I discuss in Sec.~\ref{SecHeat2}, this reduced their continuous rotation symmetries down to quarter rotations and one-sixth rotations respectively and thereby made them inequivalent. Here, we do not take the lattice structure so seriously. We have embedded it onto the manifold where it then disappears from view as just one of many possible sample lattices.

One substantial difference from our second interpretation should be noted here. We are not forced to shrink $F^K$ down to $F^K_\text{H6}$ and $F^K_\text{H7}$ to see H6 and H7 as equivalent; we have another option. Namely, rather than shrinking $F^K$ we could also expand it. Consider a theory just like H6 on this interpretation, but with a bandwidth of $2\,K$ instead of $K$. Clearly, H6 is a subtheory of this expanded theory. Note that the coordinate transformation which maps H7 onto H6, namely Eq.~\eqref{SkewH7H6}, skews its support in Fourier space outside of $k_1,k_2\in[-K,K]$. However, its support remains inside of $k_1,k_2\in[-2K,2K]$. Thus, H7 is also a subtheory of our expanded theory. We can thus see H6 and H7 both as a part of some larger theory. 
As I will discuss in Sec.\ref{PerfectRotation}, H6 and H7 are the parts of this extended theory which is visible to us when we restrict ourselves to only certain sets of representational tools. This line of thought will lead us to a perfectly rotation invariant lattice theory.

For the rest of this subsection I will only discuss the symmetries H6; analogous conclusions are true for H7 after applying Eq.~\eqref{CoorH7H6}. For H6 the dynamical symmetries of the form Eq.~\eqref{BandlimitedSymmetries} are:
\begin{flushleft}\begin{enumerate}
 \item[1)] continuous translation which maps \mbox{$\phi_\text{B}(t,x,y)\mapsto \phi_\text{B}(t,x-x_0,y-y_0)$} for some $x_0,y_0\in\mathbb{R}$,
 \item[2)] two negation symmetries which map
 \mbox{$\phi_\text{B}(t,x,y)\mapsto \phi_\text{B}(t,-x,y)$} and \mbox{$\phi_\text{B}(t,x,y)\mapsto \phi_\text{B}(t,x,-y)$} respectively,
 \item[3)] continuous rotation which maps \mbox{$\phi_\text{B}(t,x,y)$} to \mbox{$\phi_\text{B}(t,x\cos(\theta)-y\sin(\theta),x\sin{\theta}+y\cos(\theta))$} for some $\theta\in\mathbb{R}$. (This being a symmetry requires some qualification as I will discuss below.),
 \item[4)] constant time shifts which map \mbox{$\phi_\text{B}(t,x)\mapsto \phi_\text{B}(t-t_0,x)$} for $t_0\in\mathbb{R}$,
 \item[5)] global linear rescaling which maps \mbox{$\phi_\text{B}(t,x)\mapsto c_1\,\phi_\text{B}(t,x)$} for some $c_1\in\mathbb{R}$,
 \item[6)] local affine rescaling which maps \mbox{$\phi_\text{B}(t,x)\mapsto c_1\,\phi_\text{B}(t,x)+c_2(t,x)$} for some $c_2(t,x)$ which is also a solution of the dynamics.
\end{enumerate}\end{flushleft}
These are exactly the same symmetries that we found on the previous interpretation (sans local Fourier rescaling) just reformulated and recategorized: the translation, reflection, and rotation symmetries are here now external symmetries. Thus, H6 (and H7) each have two continuous translation symmetries despite initially being represented on a lattice.

In addition to this, H6 has a (qualified) continuous rotation symmetry. But in what sense is this only a qualified symmetry? Much of the discussion from the previous section following Eq.~\eqref{RthetaDef} applies here as well with some key differences. 

As I noted in Sec.~\ref{SecHeat2}, each of H4-H7 are rotation invariant in a weak sense: approximately in the continuum limit regime, $\vert k_1\vert,\vert k_2\vert\ll \pi$. Here H4-H7 are approximately rotation invariant in the $\vert k_1\vert,\vert k_2\vert\ll K=\pi/a$ regime. Note however that this is not the continuum limit regime, we are already in the continuum. Rather, this is the regime which is significantly below the bandwidth $K$. While each of H4-H7 are rotation invariant in this weak sense, H6 is rotation invariant in a much stronger sense: H6 is \textit{exactly} rotation over \textit{the whole of} $\sqrt{k_1^2+k_2^2} < K$.

Rotation is here generated by \mbox{$\exp(-\theta (x \partial_y - y \partial_x))$} for some $\theta\in\mathbb{R}$. Acting on the continuum planewave basis this transformation merely rotates their wavenumbers in Fourier space. On our second interpretation, we ran into technical trouble here due to the $2\pi$ periodic identification of the discrete planewaves $\bm{\Phi}(k_1,k_2)$ stemming from Euler's identity. This caused $R^\theta$ not to be invertible over the whole of $\mathbb{R}^L\cong\mathbb{R}^\mathbb{Z}\otimes\mathbb{R}^\mathbb{Z}$. We were led to consider only the rotation invariant subspace $\mathbb{R}^L_\text{rot.inv.}$.

In the present interpretation, something similar happens but with some key differences. Unlike before, here given wide scope rotations are always invertible. Wide scope here meaning seen as acting on $F(\mathbb{R}^3)$ the set of all functions $f(t,x,y)$. However, as before, seen as acting on the relevant vector space $F^K$ we have issues. Namely, understood as acting on $F^K$ rotation is not closed: some $f$ supported over $k_1,k_2\in[-K,K]$ may have support outside of here after being rotated.

In Sec.~\ref{SecHeat2} we fixed the non-invertible issue by focusing on the rotation invariant subspace $\mathbb{R}^L_\text{rot.inv.}\subset \mathbb{R}^L$. We might overcome our current not-closed-over-$F^K$ issue here in the same way. While rotation is not a closed over $F^K$  it is closed and invertible over $F^K_\text{rot.inv.}$ defined in Eq.~\eqref{FKrotinv}. Note that under our embedding $E(t)$ this subspace is isomorphic to the vector space $\mathbb{R}^L_\text{rot.inv.}$ defined in Eq.~\eqref{RLrotinv}. Within $F^K_\text{rot.inv.}$ the above discussed rotation transformation is of the form Eq.~\eqref{BandlimitedSymmetries} and maps solutions to H6 onto solution to H6 in an invertible way, and is hence a symmetry. H7 also has a rotation symmetry in the same sense after applying Eq.~\eqref{CoorH7H6}.

This adds to our first big lesson from Sec.~\ref{SecHeat2}: despite the fact that H6 and H7 can be represented on a square and hexagonal lattice respectively, they nonetheless both have a continuous rotation symmetry. This, in addition to their continuous translation symmetries. These continuous translation and rotations symmetries were hidden from us on our first interpretation because we there took the lattice representations too seriously. Here, we do not take the lattice structure so seriously. We have embedded it onto the manifold where it then disappears from view as just one of many possible sample lattices. 

One substantial difference from our second interpretation should be noted here. We are not forced to shrink $F^K$ down to $F^K_\text{rot.inv.}$ to make H6 rotation invariant; we have another option. Rather than shrinking $F^K$ down to its rotation invariant core, we could also expand it by adding in every state reachable from $F^K$ by rotation yielding $F^{\sqrt{2}K}_\text{rot.inv.}$. Consider a theory just like H6 on this interpretation, but defined over $F^{\sqrt{2}K}_\text{rot.inv.}$ instead of $F^K$. Clearly, H6 is a subtheory of this expanded theory. 

As I will discuss in Sec.\ref{PerfectRotation}, H6 is the part of this extended theory which is visible to us when we restrict ourselves to only a certain set of representational tools. This line of thought will lead us to a perfectly rotation invariant lattice theory.

\vspace{0.25cm}

To summarize: this third attempt at interpreting H1-H7 has fixed all of the issues with our first and second interpretations. Firstly, the tension between our theories' differences in locality and their differences in convergence rate in the continuum limit has been harmoniously resolved. This improves upon the hamfisted dissolution of tension given by our second interpretation. Secondly, like on the second interpretation, here H6 and H7 are seen to be substantially equivalent. Here however, we can moreover see H6 and H7 in their entirety as parts of a larger unified theory. Finally, like on the second interpretation, we have here exposed H1-H7's hidden continuous translation and rotation symmetries. Here however, these are more satisfyingly categorized as external symmetries.

Like the previous interpretation, this interpretation by and large invalidates all of the first intuitions laid out in Sec.~\ref{SecCentralClaims}. Indeed here the lattice seems to play a merely representational role in the theory: it does not restrict our symmetries. Moreover, theories initially appearing with different lattices may nonetheless turn out to be substantially equivalent. The process for switching between lattice structures is here done by Nyquist-Shannon resampling (more on this in Appendix~\ref{SecResample}). Indeed, the third big lesson from Sec.~\ref{SecHeat2} is repeated here: there is no sense in which these lattice structures are essentially ``baked-into'' these theories; our bandlimited theories make no reference to any lattice structure. 

As I will discuss in Sec.~\ref{SecDisGenCov} these three lessons lay the foundation for a rich analogy between the lattice structures which appear in our discrete spacetime theories and the coordinate systems which appear in our continuum spacetime theories. This ultimately gives rise to a discrete analog of general covariance.

Before that however, a bit more must be said about this third interpretation.

\section{A Third Attempt at Interpreting H1-H7 - Part 2}\label{SecHeat3Extra}
The previous section presented a third interpretation of our seven discrete theories H1-H7. This section will flesh out this interpretation in two ways. Firstly, I will provide some more explicit demonstrations of these theories' symmetries. In particular, I will explicitly demonstrate the fact that the symmetries of these theories are independent of which sample lattice we use to represent them. Secondly, I will write our above formulations of H1-H7 in terms of $\phi_\text{B}$ in the coordinate-independent language of differential geometry. This will help clarify their symmetries and assumed background structures.

\subsection{Demonstration of Bandlimited Symmetries}
\begin{figure*}[t]
\centering
\includegraphics[width=0.9\textwidth]{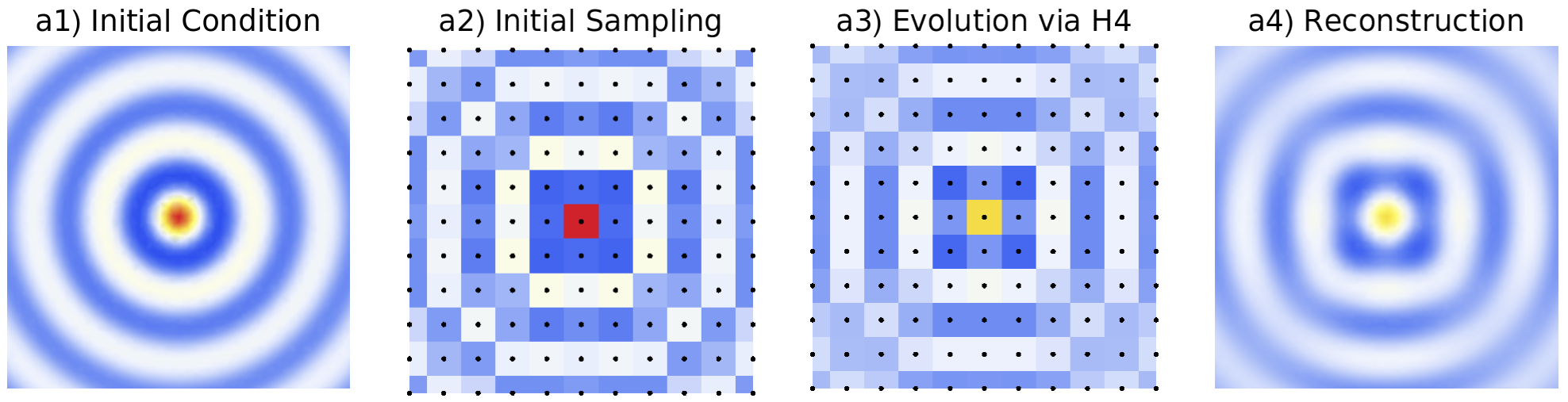}
\centering
\includegraphics[width=0.9\textwidth]{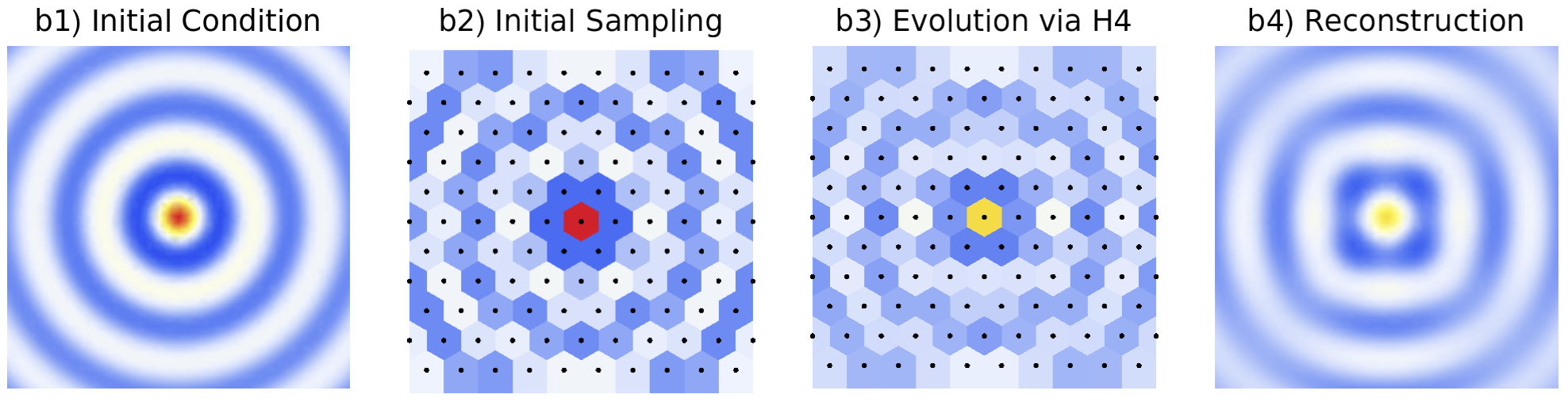}
\centering
\includegraphics[width=0.9\textwidth]{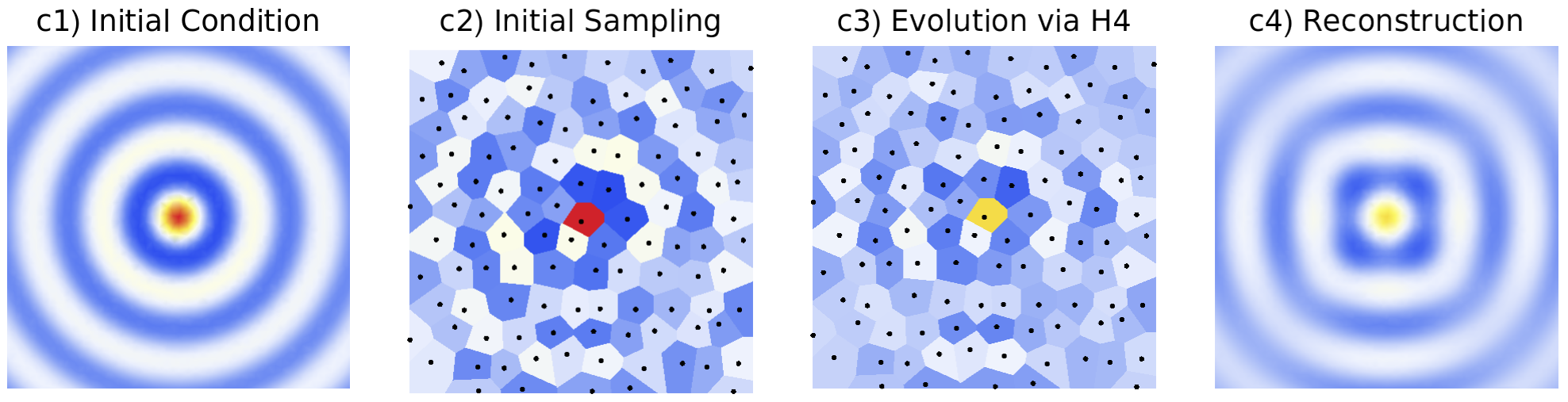}
\caption{The dynamics of H4 is here shown being carried out in a variety of lattice representations. In the left most column the initial condition is shown in its bandlimited representation, given by Eq.~\eqref{InCond}. In the rightmost column the final evolved state is shown in its bandlimited representation. Here the evolution time is $t=0.8$ and the diffusion rate is $\alpha=1$. This state can be found in four different ways. Firstly by applying the bandlimited dynamics Eq.~\eqref{DH4bandlimited} to the bandlimited initial state. The other three ways are shown in the three rows of this figure. The first row shows the initial condition being sampled onto a square lattice. This is then evolved forward in time via the discrete dynamics Eq.~\eqref{H4Long}. The bandlimited representation of the final state is then recovered through the methods discussed in Sec.~\ref{SecSamplingTheory}. The second and third rows show the same process carried out on a hexagonal lattice and an irregular lattice. Notice that the final state has a 4-fold symmetry regardless of how the dynamics is represented. Notice that the final state is the same regardless of how the dynamics is represented. We can represent any bandlimited dynamics on any (sufficiently dense) lattice.}\label{FigEvolutionH4}
\end{figure*}

\begin{figure*}[t]
\centering
\includegraphics[width=0.9\textwidth]{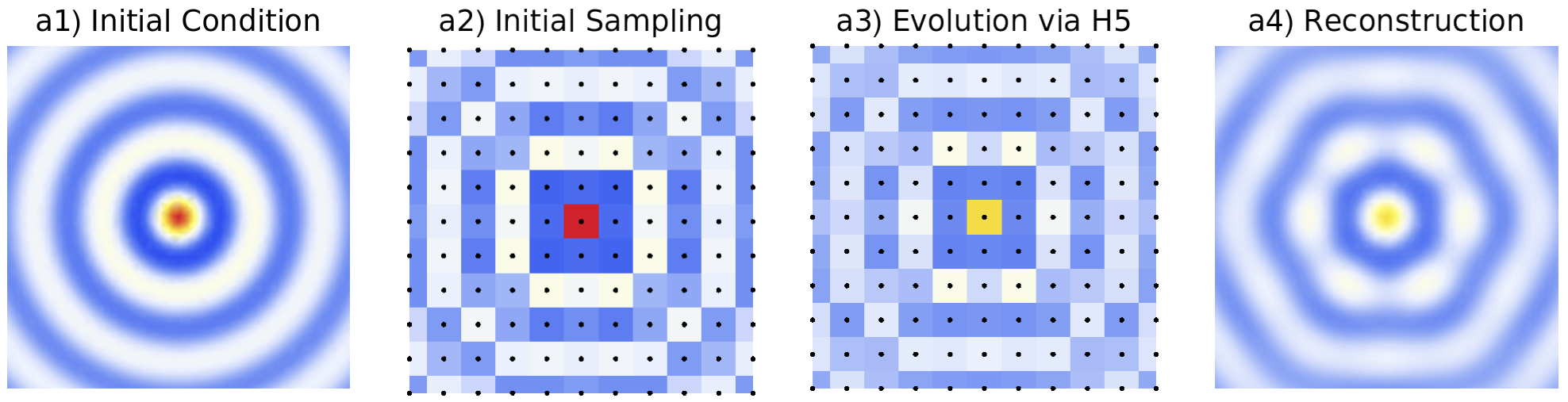}
\centering
\includegraphics[width=0.9\textwidth]{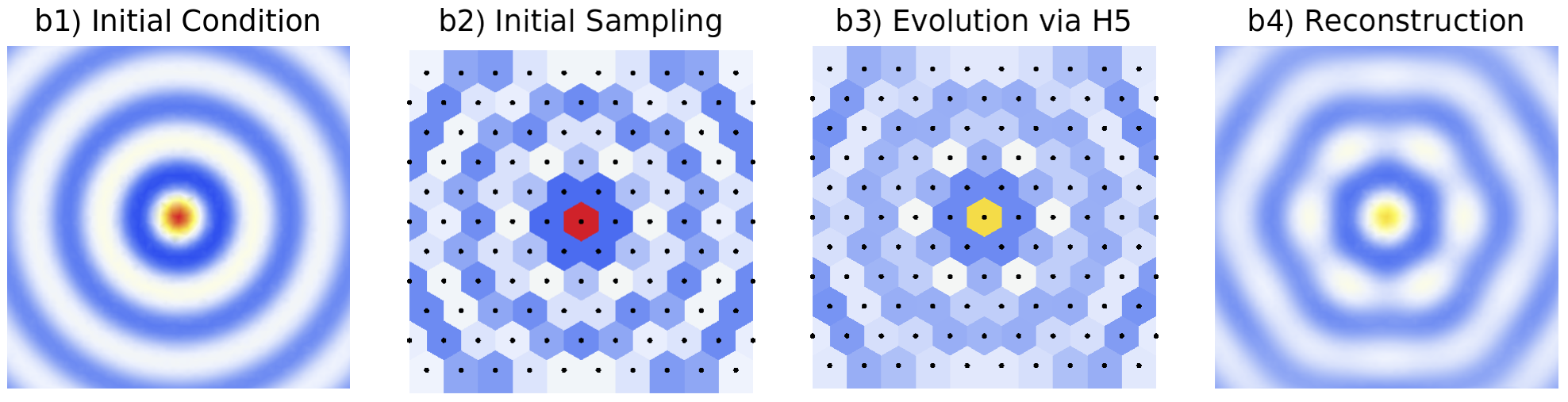}
\centering
\includegraphics[width=0.9\textwidth]{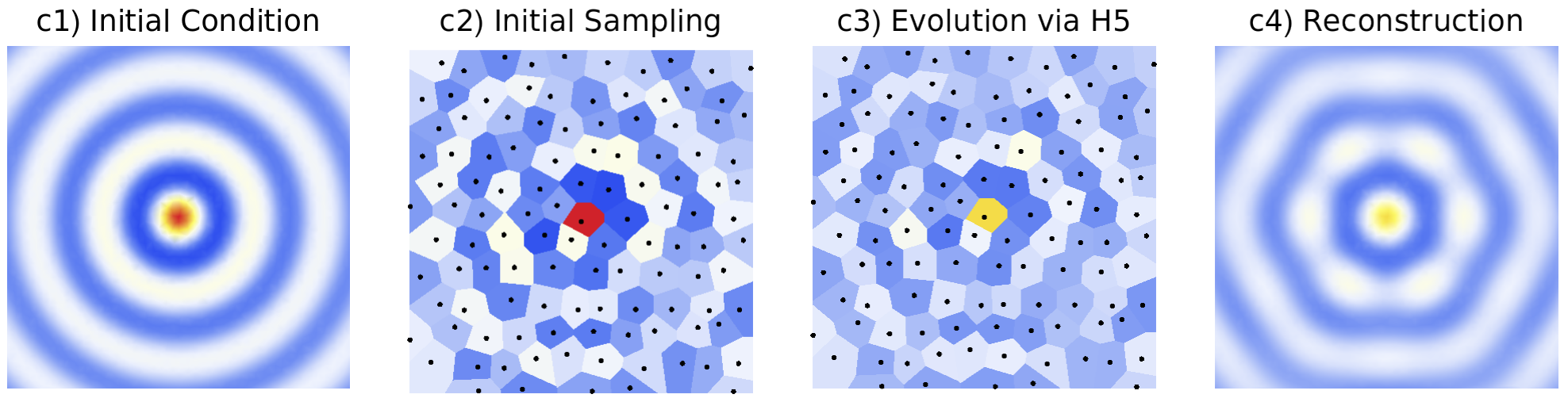}
\caption{The dynamics of H5 is here shown being carried out in a variety of lattice representations. In the left most column the initial condition is shown in its bandlimited representation, given by Eq.~\eqref{InCond}. In the rightmost column the final evolved state is shown in its bandlimited representation. Here the evolution time is $t=8.3$ and the diffusion rate is $\alpha=1$. This state can be found in four different ways. Firstly by applying the bandlimited dynamics Eq.~\eqref{DH5bandlimitedSkew} to the bandlimited initial condition. The other three ways are shown in the three rows of this figure. The second row shows the initial condition being sampled onto a hexagonal lattice. This is then evolved forward in time via the discrete dynamics Eq.~\eqref{H5Long}. The bandlimited representation of the final state is then recovered through the methods discussed in Sec.~\ref{SecSamplingTheory}. The second and third rows show the same process carried out on a square lattice and an irregular lattice. In the first row, the sample points obey Eq.~\ref{DH5Skew}. Notice that the final state has a 6-fold symmetry regardless of how the dynamics is represented. Notice that the final state is the same regardless of how the dynamics is represented. We can represent any bandlimited dynamics on any (sufficiently dense) lattice.}\label{FigEvolutionH5}
\end{figure*}

\begin{figure*}[t]
\centering
\includegraphics[width=0.9\textwidth]{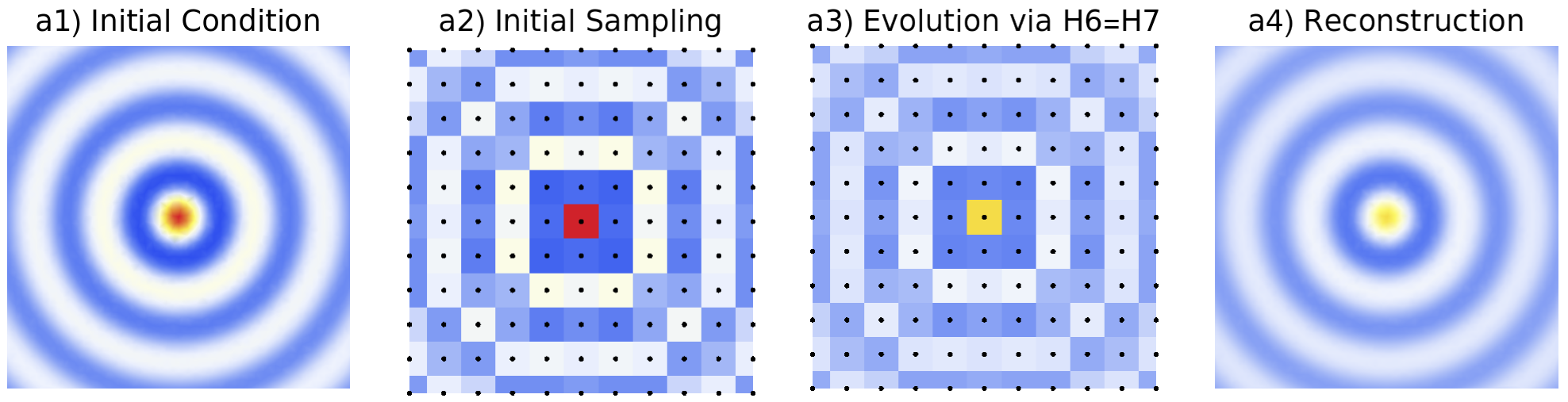}
\centering
\includegraphics[width=0.9\textwidth]{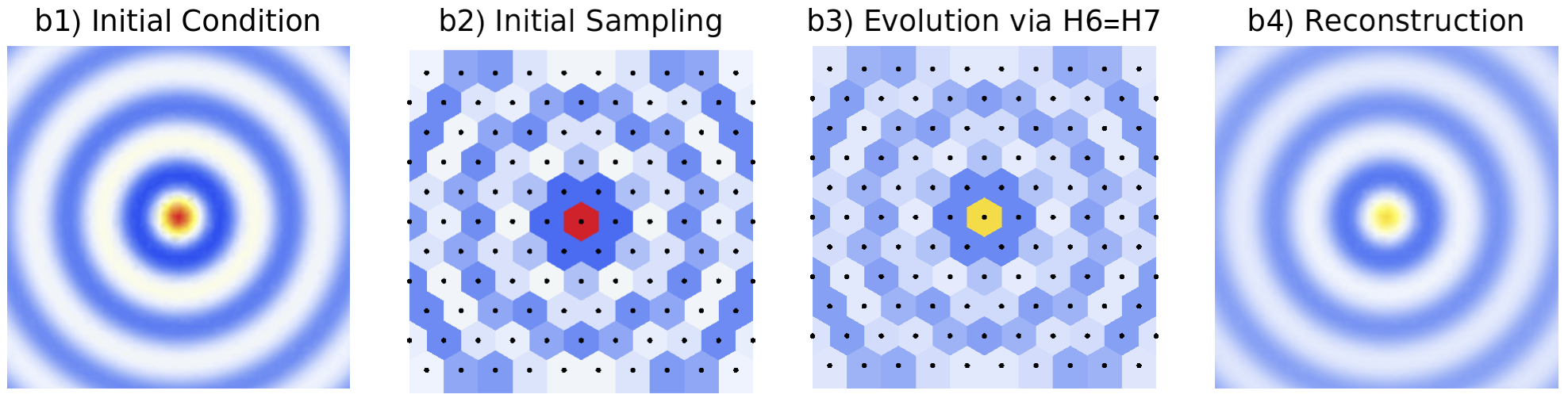}
\centering
\includegraphics[width=0.9\textwidth]{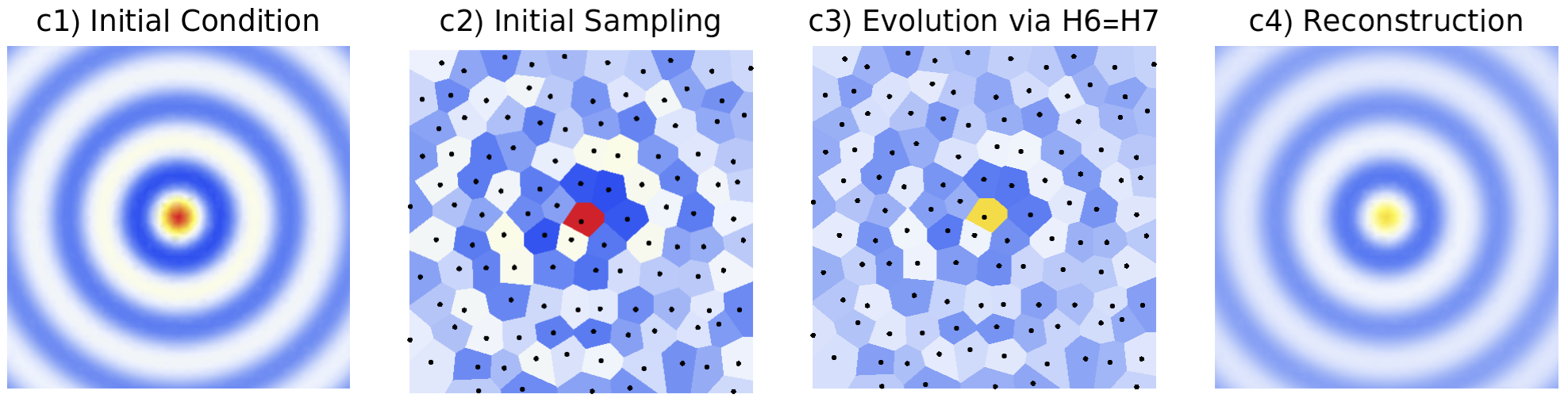}
\caption{The dynamics of H6 is here shown being carried out in a variety of lattice representations. In the left most column the initial condition is shown in its bandlimited representation, given by Eq.~\eqref{InCond}. In the rightmost column the final evolved state is shown in its bandlimited representation. Here the evolution time is $t=1$ and the diffusion rate is $\alpha=1$. This state can be found in four different ways. Firstly by applying the bandlimited dynamics Eq.~\eqref{DH6bandlimited} to bandlimited initial condition. The other three ways are shown in the three rows of this figure. The first row shows the initial condition being sampled onto a square lattice. This is then evolved forward in time via the discrete dynamics Eq.~\eqref{DH6}. The bandlimited representation of the final state is then recovered through the methods discussed in Sec.~\ref{SecSamplingTheory}. The second and third rows show the same process carried out on a hexagonal lattice and an irregular lattice. In the second row, the sample points obey Eq.~\ref{DH7}. Notice that the final state is rotation invariant regardless of how the dynamics is represented. Notice that the final state is the same regardless of how the dynamics is represented. We can represent any bandlimited dynamics on any (sufficiently dense) lattice.}\label{FigEvolutionH6}
\end{figure*}

In this subsection I will provide some more explicit demonstrations of these theories' symmetries. In particular, I will explicitly demonstrate the fact that the symmetries of some theory's dynamics has nothing to do with the symmetries of the lattice it is represented on. Just as we can represent any bandlimited state on any sufficiently dense lattice, so too can we represent any bandlimited dynamics on any sufficiently dense lattice. To see this consider Figs. \ref{FigEvolutionH4}, \ref{FigEvolutionH5} and \ref{FigEvolutionH6}. 

In each of these figures I begin from some initial heat distribution with a bandlimited representation,
\begin{align}\label{InCond}
\phi_\text{B}(0,x,y)=\frac{J_1(\pi r)}{\pi r}
+\frac{J_0(\pi r)-J_2(\pi r)}{2}
\end{align}
where $J_n(r)$ is the $n^\text{th}$ Bessel function and $r=\sqrt{x^2+y^2}$. This function is shown in the first columns of Figs. \ref{FigEvolutionH4}, \ref{FigEvolutionH5} and \ref{FigEvolutionH6}. Note that this state is rotationally invariant. 

This function is bandlimited with bandwidth of $K=\pi$. We can therefore represent this function by sampling it on a square lattice with $a=1$. We could equivalently represent this function by sampling it on a hexagonal lattice or even on an irregular lattice. For each of H4, H5 and H6=H7 such representations are shown in the second column of Figs. \ref{FigEvolutionH4}, \ref{FigEvolutionH5} and \ref{FigEvolutionH6} respectively. 

For each of these theories, we then have a choice of which representation to carry out the dynamics in. I here consider four options: as a bandlimited function, as samples on a square lattice, as samples on a hexagonal lattice or as samples on an irregular lattice. The various options for H4, H5 and H6=H7 are shown in Figs. \ref{FigEvolutionH4}, \ref{FigEvolutionH5} and \ref{FigEvolutionH6} respectively. 

Let's begin with the dynamics of H5 represented on a hexagonal lattice. This is shown in the middle row of Fig.~\ref{FigEvolutionH5}. The bandlimited representation of the initial heat distribution is shown Fig.~\ref{FigEvolutionH5}b1). The initial sample points on the hexagonal lattice are shown Fig.~\ref{FigEvolutionH5}b2). These can be evolved forward in time using Eq.~\eqref{H5Long}. The resulting time-evolved sample points are shown in Fig.~\ref{FigEvolutionH5}b3). From these we can reconstruct a bandlimited representation for the state using the techniques discussed in Sec.~\ref{SecSamplingTheory}. The resulting reconstruction is shown in Fig.~\ref{FigEvolutionH5}b4).

Alternatively, we could have carried out this evolution with no lattice representation at all. That is, we could have skipped from Fig.~\ref{FigEvolutionH5}b1) directly to Fig.~\ref{FigEvolutionH5}b4). We could do this by applying the dynamics Eq.~\eqref{DH5bandlimitedSkew} directly to the bandlimited initial condition Eq.~\eqref{InCond}. It is in this sense that the bandlimited and discrete representations of our dynamics are equivalent.

The first and third rows of Fig.~\ref{FigEvolutionH5} show the exact same evolution via H5 represented on different lattices, namely a square lattice and an irregular lattice. In the first row the evolution is carried out by a square resampled version of Eq.~\eqref{H5Long} namely Eq.~\eqref{DH5Skew}. In the third row the evolution is carried out by whatever resampling of Eq.~\eqref{H5Long} corresponds to this irregular lattice. 

Notice that the final state has a 6-fold symmetry regardless of how the dynamics is represented. Moreover, notice that the final state is the same regardless of how the dynamics is represented. Just as we can represent any bandlimited state on any lattice, so too can we represent any bandlimited dynamics on any lattice.

Fig.~\ref{FigEvolutionH4} makes the same demonstration for H4. Notice that the final state has a 4-fold symmetry regardless of how the dynamics is represented. Notice that the final state is the same regardless of how the dynamics is represented.

Likewise, Fig.~\ref{FigEvolutionH6} makes the same demonstration for H6. Notice that the final state is rotation invariant regardless of how the dynamics is represented. Notice that the final state is the same regardless of how the dynamics is represented. Representing dynamics on a lattice and carrying out the dynamical evolution on that lattice does not need to introduce ``lattice artifacts''.

These figures demonstrate clear as can be that the lattice structure we use to describe a theory has nothing to do with its dynamical symmetries. We can represent any bandlimited dynamics on any sufficiently dense lattice.

\subsection{Bandlimited General Covariance}\label{SecFullGenCov}
As the discussion throughout Sec.~\ref{SecHeat3} has shown, giving our discrete theory a bandlimited representation has had many of the same benefits one expects from a generally covariant formulation. Namely, we have exposed certain parts of our theory as merely representational artifacts and in the process we have come to a better understanding of our theory's symmetries and background structures. This is the work of the titular discrete analog of general covariance. This analogy will be spelled out in detail in Sec.~\ref{SecDisGenCov}. 

Before that, however, I will show how to combine this discrete analog with our usual continuum notion of general covariance. Note that on this interpretation H1-H7 are ultimately continuum spacetime theories of the sort we are used to interpreting (albeit ones which consider only bandlimited fields). Hence we can apply to these theories the standard techniques of general covariance which I review in Appendix~\ref{SecGenCov}.

As I discussed at the beginning of Sec.~\ref{SecHeat3} on this interpretation each of H1-H7 are about a bandlimited field \mbox{$\phi_\text{B}:\mathcal{M}\to \mathcal{V}$} with a value space $\mathcal{V}=\mathbb{R}$ and a manifold $\mathcal{M}\cong\mathbb{R}^n$ with $n=2$ for H1-H3 and $n=3$ for H4-H7. As such, in either case we have access to global coordinate systems for $\mathcal{M}$. The fields $\phi_\text{B}$ considered by this interpretation are bandlimited in the following sense. There exists a fixed special diffeomorphism $d_\text{coor.}:\mathbb{R}^n\to\mathcal{M}$ which gives us a fixed special global coordinate system for $\mathcal{M}$. In these special coordinates the field is bandlimited on each time-slice. That is, $\phi_\text{B}\circ d_\text{trans.}\in F^K$ on each time-slice. It is in these special coordinates that the fields $\phi_\text{B}$ obey the dynamical equations Eq.~\eqref{DH1bandlimited}-Eq.~\eqref{DH7bandlimited}.

Of course, we know from the lessons of continuum general covariance that one ought to be suspicious of any ``special coordinates'' appearing in a supposedly fundamental formulation of a theory. This section will remove any reference to these (or any) coordinates from our bandlimited formulation of H1-H7. As is expected of such a generally covariant reformulation, this will reveal these theories' underlying geometric background structures.

For simplicity, however, I will just consider H4 and H6 here. To start let us first write the continuum theory H0 in the coordinate-free language of differential geometry. After substantial work (see Appendix~\ref{SecGenCov}) one can rewrite Eq.~\eqref{HeatEq0} in the coordinate-free language of differential geometry as follows:
\begin{align}\label{H0GenCov}
\text{H0}:\qquad\text{KPMs:}\quad&\langle \mathcal{M}, t_\text{ab}, h^\text{ab}, \nabla_\text{a},T^\text{a},\psi\rangle\\
\nonumber
\text{DPMs:}\quad&
T^\text{a}\,\nabla_\text{a}\psi
=\frac{\bar{\alpha}}{2} \, h^\text{bc} \nabla_\text{b}\nabla_\text{c}\psi.
\end{align}
The geometric objects used in this formulation are as follows. $\mathcal{M}$ is a 2+1 dimensional differentiable manifold. $h^\text{ab}$ and $t_\text{ab}$ are space and time metric fields. They are each symmetric with signatures $(0,1,1)$ and $(1,0,0)$ respectively. Moreover, these metrics are orthogonal to each other (i.e., with $h^\text{ab} t_\text{bc}=0$). These metrics allow us to compute lengths and durations. $\nabla_\text{a}$ is a covariant derivative operator which is compatible with these metrics (i.e., $\nabla_\text{a}h^\text{bc}=0$ and $\nabla_\text{a}t_\text{bc}=0$). Note that $\nabla_\text{a}$ is not uniquely determined by the metric in this case because neither $h^\text{ab}$ nor $t_\text{ab}$ are invertible. $\nabla_\text{a}$ is flat, $R^\text{a}{}_\text{bcd}=0$. The covariant derivative operator $\nabla_\text{a}$ allows us to do parallel transport and compute derivatives of non-scalar fields.

The quadruple, $\langle \mathcal{M}, t_\text{ab}, h^\text{ab}, \nabla_\text{a}\rangle$, picked out by the above discussed objects is a Galilean spacetime~\cite{ReadThesis}. However, in addition to these structures, our spacetime also has a constant unit time-like vector field $T^\text{a}$ with \mbox{$\nabla_\text{a}T^\text{b}=0$} and \mbox{$t_\text{ab}T^\text{a}T^\text{b}=1$}. This vector field picks out a standardized way of moving forward in time (i.e, translation generated by $T^\text{a}\nabla_\text{a}$). That is, $T^\text{a}$ provides a rest frame.

Each of the above discussed objects is considered fixed in this formulation: they do not vary from model to model and do not evolve dynamically. The only dynamical field in this theory is the real scalar field $\psi:\mathcal{M}\to\mathbb{R}$ representing the temperature field. Mathematical structures satisfying the above conditions (independent of whether it satisfies the dynamics) are the theory's kinematically possible models (KPMs). This theory's dynamically possible models (DPMs) are the subset of the KPMs which additionally satisfy the theory's dynamics. The dynamical equation in Eq.~\eqref{H0GenCov} says that the derivative of the temperature field in the $T^\text{a}$ direction is proportional to its second derivative in space.

Next let's consider H4. Rewritten in the coordinate-free language of differential geometry, H4 becomes:
\begin{align}\label{DH4GenCov}
\text{H4:}\ \ \ \text{KPMs:}\ &\langle \mathcal{M}, t_\text{ab}, h^\text{ab}, \nabla_\text{a},T^\text{a},X^\text{a},Y^\text{a},\phi_\text{B}\rangle\\
\nonumber
\text{DPMs:}\ &
T^\text{a}\,\nabla_\text{a}\phi_\text{B}
=\frac{\alpha}{2} \left(F(X^\text{b}\nabla_\text{b})+F(Y^\text{c}\nabla_\text{c})\right)\phi_\text{B}.
\end{align}
where
\begin{align}
F(z)=2\text{cosh}(a\,z)-2.
\end{align}
At the level of dynamics, H4 and H0 are radically different. Note that the dynamics of H0 is local, only involving finite order of derivative. By contrast, the dynamics of H4 is highly non-local.

These theories are also substantially different at the level of KPMs. The three key differences between H4 and the continuum theory H0 are as follows. Firstly, H4 resticts us to manifolds $\mathcal{M}\cong\mathbb{R}^3$ whereas H0 does not. Secondly, the field $\phi_\text{B}$ in H4 is bandlimited with bandwidth $k_1,k_2\in[-K,K]$ on each time-slice whereas the field $\psi$ in H0  is not. I owe the reader a geometric explanation of what this means.

Finally, H4 has two extra pieces of spacetime structure which H0 lacks. Namely, $X^\text{a}$ and $Y^\text{a}$ are a pair of fixed constant space-like unit vectors which are orthogonal to each other. That is, 
\begin{align}
\nabla_\text{a}X^\text{b}&=0 &
\nabla_\text{a}Y^\text{b}&=0\\
\nonumber
t_\text{ab} X^\text{a}&=0 &
t_\text{ab} Y^\text{a}&=0\\
\nonumber
h_\text{ab} X^\text{a}X^\text{b}&=1 &
h_\text{ab} Y^\text{a}Y^\text{b}&=1\\
\nonumber
h_\text{ab} X^\text{a}Y^\text{b}&=0
\end{align}
Note that the inverse space metric $h_\text{ab}$ is only well defined for spacelike vectors, see \cite{ReadThesis}. Roughly, $X^\text{a}$ and $Y^\text{a}$ here serve to pick out the directions for the rotational anomalies appearing in Fig.~\ref{FigEvolutionH4}.

In terms of these spacetime structures, what does it mean to say that $\phi_\text{B}$ is  ``bandlimited with bandwidth $k_1,k_2\in[-K,K]$ on each time-slice''. Let's begin with ``time-slice''. Like H0 this theory makes a clear distinction between space and time. A time-slice is any surface $H\in\mathcal{M}$ such that all of its tangent vectors $x^\text{a}$ have $t_\text{ab}\,x^\text{a}=0$. 

What does it mean to be bandlimited on a time-slice? Consider the following eigen-problem for functions \mbox{$f:H\to\mathbb{R}$} defined on a time-slice, 
\begin{align}
X^\text{a}X^\text{b}\nabla_\text{a}\nabla_\text{b} f = -k_1^2 f\\
\nonumber
Y^\text{a}Y^\text{b}\nabla_\text{a}\nabla_\text{b} f = -k_2^2 f
\end{align}
A function $h:H\to\mathbb{R}$ is bandlimited if and only if it is the sum of these eigensolutions over a compact range of Fourier space. The extent of this range is its bandwidth. In these terms we can state the requirement that $\phi_\text{B}$ is bandlimited with bandwidth $k_1,k_2\in[-K,K]$ on each time-slice.

Note that spelling out what it means for $\phi_\text{B}$ to be bandlimited did not require talking about $T^\text{a}$. Thus, this geometric definition of being bandlimited can be applied in Galilean spacetimes as well.

Let's move on to H6. Rewritten in the coordinate-free language of differential geometry, H6 becomes:
\begin{align}\label{DH6GenCov}
\text{H6:}\ \ \ \text{KPMs:}\quad&\langle \mathcal{M}, t_\text{ab}, h^\text{ab}, \nabla_\text{a},T^\text{a},X^\text{a},Y^\text{a},\phi_\text{B}\rangle\\
\nonumber
\text{DPMs:}\quad&
T^\text{a}\,\nabla_\text{a}\psi
=\frac{\alpha_0}{2} \, h^\text{bc} \nabla_\text{b}\nabla_\text{c}\phi_\text{B}.
\end{align}
At the level of dynamics, H6 and H0 are nearly identical. The dynamics for H0 has a diffusion rate of $\bar{\alpha}$ whereas H6 has $\alpha_0=\alpha\,a^2$. This can be disregarded by setting $\alpha_0=\bar{\alpha}$.

However, at the level of KPMs, H6 and H0 are substantially different. The three key differences between H6 and the continuum theory H0 are as follows. 
Firstly, H6 resticts us to manifolds $\mathcal{M}\cong\mathbb{R}^3$ whereas H0 does not. Secondly, the field $\phi_\text{B}$ in H6 is bandlimited with bandwidth $k_1,k_2\in[-K,K]$ on each time-slice in the sense discussed above whereas the field $\psi$ in H0  is not. Secondly, H6 has the same extra spacetime structures $X^a$ and $Y^a$ that H4 does. Indeed, at the level of KPMs H6 and H4 are identical.

One might notice that these extra pieces of spacetime structure don't play any role in the dynamics of H6. Why then are they there? $X^a$ and $Y^a$ are needed to spell out what it means for $\phi_\text{B}$ to be bandlimited with bandwidth $k_1,k_2\in[-K,K]$ on each time-slice. This region in Fourier space is not rotation invariant, we need some extra structure to point out in which directions the corners go.

This is a very strange sort of spacetime structure. It doesn't participate in the dynamics, all it does is allows for us to articulate a certain restriction on the space of KPMs. As I will discuss in Sec.\ref{PerfectRotation}, it is best to think of H6 as being the part of an extended theory which is visible to us when we restrict our set of representational tools. On this view, while $X^a$ and $Y^a$ are real spacetime structures for H4, for H6 they are merely representational artifacts. 

Before that, however, allow me to spell out in detail the discrete analogs of general covariance which have been developed throughout this paper.

\section{Two Discrete Analogs of General Covariance}\label{SecDisGenCov}
Three lessons have been repeated throughout this paper. Each of these lessons is visible in both our second and third attempts at interpreting H1-H7. Combined these lessons give us a rich analogy between lattice structures and coordinate systems: Lattice structure is rather less like a fixed background structure and rather more like a coordinate system, i.e., merely a representational artifact.

These three lessons run counter to the three first intuitions one is likely to have regarding lattice structure discussed in Sec.~\ref{SecCentralClaims}. Namely, that lattices and lattice structure: restrict our symmetries, distinguish our theories, and are fundamentally ``baked-into'' the theory. As we have seen, they do not restrict our symmetries, they do not distinguish our theories and they are merely representational not fundamental. In particular, we have learned the following three lessons.

Our first lesson was that taking the lattice structure seriously as a fixed background structure or as a time-slice of the underlying manifold systematically under predicts the symmetries that discrete spacetime theories can and do have. Indeed, discrete theories can have significantly more symmetries than our first intuitions might allow for. As Sec.~\ref{SecHeat2} and Sec.~\ref{SecHeat3} have shown each of H1-H7 has a continuous translation symmetry despite being introduced with explicit lattice structures. Moreover, H6 and H7 even have a continuous rotational symmetry. The fact that a lattice structure was used in the initial statement of these theory's dynamics does not in any way restrict their symmetries. There is no conceptual barrier to having a theory with continuous symmetries represented on a discrete lattice. (In \cite{DiscreteGenCovPart2} I discuss how discrete theories can even have Lorentzian boost symmetries.)

In light of the proposed analogy between lattice structure and coordinate systems this first lesson is not mysterious. Coordinate systems are neither background structure nor a fundamental part of the manifold. The use of a certain coordinate system does not in any way restrict a theory's symmetries. Indeed, it is a familiar fact that there is no conceptual barrier to having a rotationally invariant theory formulated on a Cartesian coordinate system.

Our second lesson was that discrete theories which are initially presented to us with very different lattice structures may nonetheless turn out to be completely equivalent theories. Indeed, as we have seen, two of our discrete theories (H6 and H7) have a nice one-to-one correspondence between a substantial portion of their solutions. This despite the fact that these theories were initially presented to us with different lattice structures (i.e., a square lattice and a hexagonal lattice respectively). However, despite this nice correspondence, when in Sec.~\ref{SecHeat1} we took these lattice structures seriously as a fixed background structure, we found that H6 and H7 were inequivalent; namely, they were here judged to have different symmetries. 

Only in Sec.~\ref{SecHeat2} and Sec.~\ref{SecHeat3} when stopped take the lattice structure so seriously did we ultimately see H6 and H7 as having the same symmetries. Indeed, in these later two interpretations H6 and H7 were seen to be substantially equivalent. They are simply re-descriptions of a single theory with slightly different limitations in representational capacities. In Sec.~\ref{SecHeat2} this re-description is a change of basis in the theory's value space, whereas in Sec.~\ref{SecHeat3} this re-description is merely a change of coordinates. What we thought were distinct lattice theories are really just different sample points being used to describe one-and-the-same bandlimited field theory.

In light of the proposed analogy between lattice structure and coordinate systems this second lesson is not mysterious. Unsurprisingly, continuum theories presented to us in different coordinate systems may turn out to be equivalent. Moreover, we can always re-describe any continuum theory in any coordinates we wish.

Our third lesson was that, in addition to being able to switch between lattice structures, we can also reformulate any discrete theory in such a way that it has no lattice structure whatsoever. I have shown two ways of doing this. In Sec.~\ref{SecHeat2} this was done by internalizing the lattice structure into the theory's value space. In Sec.~\ref{SecHeat3} this was done by embedding the discrete theory onto a continuous manifold using bandlimited functions. Adopting a lattice structure and switching between them was then handled using Nyquist-Shannon sampling theory discussed in Sec.~\ref{SecSamplingTheory}.

In light of the proposed analogy between lattice structure and coordinate systems this third lesson is not mysterious. This is analogous to the familiar fact, discussed in Sec.~\ref{SecIntro}, that any continuum theory can be written in a generally covariant (i.e., coordinate-free) way. Thus, the two above-discussed ways of reformulating a discrete theory to be lattice-free are each analogous to reformulating a continuum theory to be coordinate-free (i.e., a generally covariant reformulation). Thus we have not one but two discrete analogs of general covariance. See Fig.~\ref{FigTwoAnalogies}.

\begin{figure}[t]
\begin{flushleft}
\text{{\bf Internal Discrete General Covariance:}}
\end{flushleft}
$\begin{array}{rcl}
\text{Coordinate Systems} & \!\!\leftrightarrow\!\! & \text{Lattice Structure}\\
\text{Changing Coordinates} & \!\!\leftrightarrow\!\! & \text{Changing Lattice Structures}\\
\ & \ & \text{by changing basis in value space}\\
\text{Gen. Cov. Formulation} & \!\!\leftrightarrow\!\! & \text{Basis-Free Formulation}\\
\text{(i.e., coordinate-free)} & \ & \text{(i.e., lattice-free)}\\
\end{array}$\\
\begin{flushleft}
\text{{\bf External Discrete General Covariance:}}
\end{flushleft}
$\begin{array}{rcl}
\text{Coordinate Systems} & \!\!\leftrightarrow\!\! & \text{Lattice Structure}\\
\text{Changing Coordinates} & \!\!\leftrightarrow\!\! & \text{Changing Lattice Structures}\\
\ & \ & \text{by Nyquist-Shannon resampling}\\
\text{Gen. Cov. Formulation} & \!\!\leftrightarrow\!\! & \text{Bandlimited Formulation}\\
\text{(i.e., coordinate-free)} & \ & \text{(i.e., lattice-free)}\\
\end{array}$
\caption{A schematic of the two notions of discrete general covariance introduced in this paper. The internal strategy is applied to H1-H7 in Sec.~\ref{SecHeat2} whereas the external strategy is applied to H1-H7 in Sec.~\ref{SecHeat3}. These are compared in Sec.~\ref{SecDisGenCov}.}
\label{FigTwoAnalogies}
\end{figure}

Before contrasting these two analogies, let's recap what they agree on. In either case, as one would hope, our discrete analog helps us to disentangle a discrete theory's substantive content from its merely representational artifacts. In particular, in both cases, lattice structure is revealed to be non-substantive and merely representational as is the lattice itself. Lattice structure is no more attached or baked-into to our discrete spacetime theories than coordinate systems are to our continuum theory. In either case, getting clear about this has helped us to expose our discrete theory's hidden continuous symmetries.

What distinguishes these two notions of discrete general covariance is how they treat the lattice structure after it has been revealed as being coordinate-like and so merely representational. The approach in Sec.~\ref{SecHeat2} was to internalize the lattice structure into the theory's value space. By contrast, the approach in Sec.~\ref{SecHeat3} was to keep the lattice structure external, but to flesh it out into a continuous manifold such that it is no longer fundamental. Let us therefore call these two notions of discrete general covariance internal and external respectively.

These two approaches pick out very different underlying manifolds for our discretely-representable spacetime theories. As a consequence, they license very different conclusions about locality and symmetries. In particular, while for H1-H7 these internal and external approaches have more-or-less agreed as to what symmetries there are, they have disagreed about how they are to be classified.

In each of these differences I find reason to favor the external approach. To briefly overview my feelings: It is more natural for the continuous translation and rotation symmetries of H1-H7 to be classified as external. Moreover, keeping the lattice structure external as a part of the manifold, allows us to draw intuitions about locality from it. In particular, the external approach has a better way  of resolving some puzzles about locality and convergence in the continuum limit. However, neither of these reasons are decisive and I think either the internal or external approach is likely to be fruitful for further investigation/use.

\section{A Perfectly Rotation Invariant Lattice Theory}\label{PerfectRotation}
As I will now discuss, we can leverage this analogy between lattice structures and coordinate systems to construct a perfectly rotation invariant lattice theory. 

As discussed in Sec.~\ref{SecFullGenCov} H6 is not rotation invariant in an unqualified way due to its KPMs not being closed under rotations. Stating the condition that $\phi_\text{B}$ be bandlimited with $k_1,k_2\in[-K,K]$ on each time-slice requires the use of two space-like fields $X^a$ and $Y^a$ which break rotation invariance.

However, as discussed in Sec.~\ref{SecHeat3}, there is a substantial portion of H6 which is rotation invariant, namely $F^K_\text{rot.inv.}$ with $\sqrt{k_1^2+k_2^2}<K$ on each time-slice. Moreover, we can see H6 as a part of an extended theory defined over $F^{\sqrt{2}K}_\text{rot.inv.}$ with $\sqrt{k_1^2+k_2^2}<\sqrt{2}K$ on each time-slice. Let's now define being in $F^K_\text{rot.inv.}$ on a time-slice geometrically in terms of the metrics $t_\text{ab}$ and $h^\text{ab}$.

This definition comes in two parts. First let's define the space $F_\text{B}$ of bandlimited function (regardless of their bandwidth). Consider the eigen-problem for functions on a time-slice $f:H\to\mathbb{R}$
\begin{align}\label{GeometricBandwidth}
h^\text{bc} \nabla_\text{b}\nabla_\text{c} f= -k^2 f
\end{align}
A function $g:H\to\mathbb{R}$ is in $F_\text{B}$ if and only if it is the sum of these eigensolutions over a compact range $\mathcal{K}(g)$ of Fourier space. A function is further within $F^K_\text{rot.inv.}$ if and only if this compact region $\mathcal{K}(g)$ itself is contained within the portion of Fourier space satisfying $\sqrt{k_1^2+k_2^2}<K$.

We can thus define the following perfectly rotation invariant lattice theory:
\begin{align}\label{HKrotinvGenCov}
\text{H}_\text{rot.inv.}^K\text{:}\ \ \ \text{KPMs:}\quad&\langle \mathcal{M}, t_\text{ab}, h^\text{ab}, \nabla_\text{a},T^\text{a},\phi_\text{B}\rangle\\
\nonumber
\text{DPMs:}\quad&
T^\text{a}\,\nabla_\text{a}\psi
=\frac{\bar{\alpha}}{2} \, h^\text{bc} \nabla_\text{b}\nabla_\text{c}\phi_\text{B}.
\end{align}
with spacetime structures as defined following Eq.~\eqref{H0GenCov} and $\phi_\text{B}$ bandlimited on each time-slice with \mbox{$\sqrt{k_1^2+k_2^2}<K$} in the sense just defined.

Note that $\text{H}_\text{rot.inv.}^K$ is a subtheory of H6 which is itself a subtheory of $\text{H}_\text{rot.inv.}^{\sqrt{2}K}$. More will be said about the relationship H6 has to these theories in a moment. But first, how does $\text{H}_\text{rot.inv.}^K$ differ from H0? 

Their only difference is that we are here limited to manifolds $\mathcal{M}\cong\mathbb{R}^3$ and that
the field $\phi_\text{B}$ in $\text{H}_\text{rot.inv.}^K$ is bandlimited with bandwidth $\sqrt{k_1^2+k_2^2}<K$ on each time-slice, whereas the field $\psi$ in H0  is not. In fact, for either theory the dynamics guarantees that if the temperature field starts off bandlimited it will stay bandlimited. Thus this restriction of the allowed dynamical fields is really just a restriction on the allowed initial conditions. Thus, ultimately the only difference between $\text{H}_\text{rot.inv.}^K$ and H0 is a restriction on the initial conditions.

As innocent as this restriction on initial conditions may seem, I believe it has serious implications for the nature of the underlying manifold. As I will discuss later, $\text{H}_\text{rot.inv.}^K$ has a nice sampling property and therefore has access to discrete descriptions which make no reference to the manifold. By contrast, the manifold in H0 seems descriptively essential. Moreover, this restriction on initial conditions has serious implications for counterfactual reasoning and locality. As discussed following Eq.~\eqref{WeightedAvg}, when restricted to bandlimited functions we can no longer ask ``What would have happened, if things had been different only in this compact region?'' Any bandlimited counter-instance must be globally different.

Viewing H6 as a part of $\text{H}_\text{rot.inv.}^{\sqrt{2}K}$ sheds some light on its failure to be rotation invariant in an unqualified way. To see how, it is instructive to restate this in terms of sample lattices. H6 is a theory about all of the bandlimited functions which are representable on a fixed square sample lattice with spacing $a$. By contrast, $\text{H}_\text{rot.inv.}^{\sqrt{2}K}$ is a theory about\footnote{Technically, $\text{H}_\text{rot.inv.}^{\sqrt{2}K}$ is about this collection of bandlimited functions closed under addition.} all of the bandlimited functions representable on a fixed square sample lattices with spacing $a$ \textit{or any rotation thereof.}

Thus, H6 is the part of $\text{H}_\text{rot.inv.}^{\sqrt{2}K}$ which is visible to us when we restrict ourselves to only a certain set of representational tools. We can thus see H6's failure to be rotation invariant in an unqualified way as a problem of representational capacity and not of physics. Sure, if we limit ourselves to functions which are representable on a fixed square lattice, we lose rotation symmetry. But what physical reason do we have to limit our representational tools in this way?

Consider an analogous situation involving coordinate systems. Almost all manifolds cannot be covered with a single global coordinate system. For instance, the 2-sphere needs at least two coordinate systems to cover it. Consider this sphere under arbitrary rotations. Note that no coordinate system is closed under these transformations. If we limit ourselves to functions which are supported entirely within the scope of a single fixed coordinate system, we lose rotation symmetry. Would it be right to say that no theory which is set on a sphere can be rotation invariant? Of course not, this just reflects the fact that each of our coordinate systems individually have a limited representational capacity. We can regain rotation invariance by using multiple representational tools, i.e., multiple coordinate systems. 

Applying this lesson to H6, we ought to view it as being able to represent a rotation non-invariant part of a wider rotation invariant theory. As discussed above, making use of multiple sample lattices (i.e., all rotated square lattices with spacing $a$) we can describe a rotation invariant theory $\text{H}_\text{rot.inv.}^{\sqrt{2}K}$ over $F^{\sqrt{2}K}_\text{rot.inv.}$.

The substantial equivalence between H6 and H7 can also be seen this way. Note that the coordinate transformation which maps H7 onto H6, namely Eq.~\eqref{CoorH7H6}, maps a square lattice to a hexagonal lattice maintaining the lattice spacing $a$, see Fig.~\ref{FigSkew}. Thus, H7 is a theory about all bandlimited functions representable on a hexagonal sample lattice with spacing $a$. The subsets of H6 and H7 which are equivalent to each other ($F^K_\text{H6}\cong F^K_\text{H7}$) are exactly those parts which are representable on both a square and hexagonal lattice. We can here see H6 and H7 as parts of a larger unified theory viewed in two ways with differently limited representational capacities.

Note that even the extended theory $\text{KG}_\text{Poin.inv.}^{K}$ has an artificially limited representational capacity. What space do we find if we use all of our representational tools (i.e., every possible sample lattice)? The answer is the space $F_\text{B}$ defined above.

Contrast $F_\text{B}$ with $F(\mathbb{R}^3)$ the space of all functions $f(t,x,y)$. The Gaussian distribution is in $F(\mathbb{R}^3)$ but not $F_\text{B}$. One might gloss this difference as follows: $F(\mathbb{R}^3)$ contains actually infinite frequencies, whereas $F_\text{B}$ contains arbitrarily large but always finite frequencies. That is, $F_\text{B}$ contains only a potential infinity of frequencies. Note that while $F_\text{B}$ is closed under finite sums it  is not closed under infinite sums. 

Using $F_\text{B}$ we can thus define following perfectly rotation invariant lattice theory:
\begin{align}\label{HBGenCov}
\text{H}_\text{B}\text{:}\ \ \ \text{KPMs:}\quad&\langle \mathcal{M}, t_\text{ab}, h^\text{ab}, \nabla_\text{a},T^\text{a},\phi_\text{B}\rangle\\
\nonumber
\text{DPMs:}\quad&
T^\text{a}\,\nabla_\text{a}\psi
=\frac{\bar{\alpha}}{2} \, h^\text{bc} \nabla_\text{b}\nabla_\text{c}\phi_\text{B}.
\end{align}
with spacetime structures as defined following Eq.~\eqref{H0GenCov} and $\phi_\text{B}\in F_\text{B}$ on each time-slice in the sense defined above.

Note that on every time-slice $\phi_\text{B}\in F_\text{B}$ can be represented via its sample values on some fine-enough square lattice (square here defined by the metric $h^\text{ab}$). Doing so one would find its sample values on this lattice obey our original discrete equation for H6 , Eq.~\eqref{DH6}, with some rescaling of $\alpha_0$. Hence, as each state in this theory is representable on some lattice, this is a perfectly rotation invariant lattice theory.

One may still complain that this should not be called a lattice theory. For each state here we need a different lattice to represent it, there is no single ontologically significant lattice on which we can represent all of our states. As I will soon discuss, there is such an all-representing lattice (albeit a non-fundamental one), but first let me argue against the spirit of this complaint.

Given the deflationary attitude taken in this paper towards lattice structures, the lack of an all-representing lattice wouldn't bother me. As I have argued throughout this paper, lattice structures are merely representational artifacts playing no substantive role in our theories: they do not limit our symmetries, they do not distinguish our theories, and they are not fundamentally ``baked-into'' our theories. Indeed, as I have argued, there is a rich analogy between the lattice structures appearing in our discrete spacetime theories and the coordinate systems appearing in our continuum spacetime theories. We wouldn't refer to our continuum spacetime theories as ``coordinate theories'', they are rather coordinate-representable theories. Analogously, I claim that properly understood, there are no such things as lattice-fundamental theories, rather there are only lattice-representable theories. Hence, the above theory is a lattice theory in the strongest sense available. 

Accepting this, one might still feel there ought to be an all-representing lattice (albeit a non-fundamental one). As I will discuss, there is one but if there weren't this wouldn't be an issue. Note that we are often forced to represent our continuum theories with multiple overlapping complementary coordinate systems. That is, we often lack an all-representing coordinate system. Given the central analogy of this paper, why should the lack of an all-representing lattice bother us?

Returning to answering the complaint directly however, there is such an all-representing lattice. Consider a square lattice with some fixed spacing (square here defined by the metric $h^\text{ab}$). Consider another square lattice with twice the resolution, offset from the first lattice by some rational amount. Consider an infinite sequence of lattices each doubling in resolution with some rational offset. At some point any function $f\in F_\text{B}$ will be representable on one of these lattices. Consider the \textit{deep lattice} which results from taking the infinite union of all of these lattices (but not their closure under the limit). Every function $f\in F_\text{B}$ is representable on this deep lattice\footnote{Indeed any $f\in F_\text{B}$ can be represented on some finite length initial segment on this.}. Moreover, note that the deep lattice has a countable number of lattice sites: by construction it is a subset of the rationals $\mathbb{Q}^2$.

One final comment on $F_\text{B}$. Note that this space is closed under (finite) addition and scalar multiplication and is hence a vector space. As the deep lattice shows, this is a vector space with a countably infinite dimension. Thus, given its rotation and translation invariance $F_\text{B}$ supports a countably infinite dimensional representation of the 2D Euclidean group plus time shifts. Every function $f\in F_\text{B}$ is representable on some fine-enough square lattice. Doing so one would find its sample values on this lattice obey Eq.~\eqref{DH6} with some rescaling of $\alpha$.

As I see it, the proper view of H6 (even beginning from $\phi_\ell(t)$) is the one given by Eq.~\eqref{DH6GenCov} understood as a part of the maximally extended theory $\text{H}_\text{B}$. In particular, H6 is the part of $\text{H}_\text{B}$ which is visible to us if we restrict ourselves to only certain representational tools.

How does $\text{H}_\text{B}$ differ from the continuum theory H0? H0 assents to the existence of actually infinite frequencies whereas $\text{H}_\text{B}$ does not. To me $\text{H}_\text{B}$ appears to be on better ground empirically speaking than H0: I have never and will never see a literally infinite frequency photon. As small as the difference is between $\text{H}_\text{B}$ and H0, I expect there to be radical differences in views on spacetime, but this is a topic for another paper.

\section{Conclusion}\label{SecConclusion}
This paper has given an exhaustive study of the seven discrete heat equations presented in Sec.~\ref{SecSevenHeat}. These theories were initially formulated in terms of a real valued function over lattice sites and times, $\phi_\ell(t)\in F_L$. Next, as an infinite dimensional vector field over time, $\bm{\Phi}(t)\in \mathbb{R}^L$. And finally, as an bandlimited field $\phi_\text{B}(t)\in F^K$ over a timeslice $\mathcal{M}(t)$ of a continuous spacetime manifold $\mathcal{M}$. Each of these redescriptions was carried out by a vector space isomorphism, $F_L\cong\mathbb{R}^L\cong F^K$.

Throughout this paper, we have learned three substantial lessons about the role that lattice structures play in our discrete spacetime theories. These lessons serve to undermine the three first intuitions about lattice structure laid out in Sec.~\ref{SecCentralClaims}. As I have shown, lattice structures don't restrict symmetries, they don't distinguish our theories, and they are not fundamentally ``baked-into'' these theories. As I have discussed, these lessons lay the foundation for a rich analogy between the lattice structures which appear in our discrete spacetime theories and the coordinate systems which appear in our continuum spacetime theories. Indeed, my analysis has shown that lattice structure is rather less like a fixed background structure or a fundamental part of some underlying manifold and rather more like a coordinate system, i.e., merely a representational artifact.

Based upon this analogy this paper has introduced two discrete analogs of general covariance (see Fig.~\ref{FigTwoAnalogies}) and demonstrated their usefulness at exposing hidden symmetries and background structures in our discrete spacetime theories (i.e., lattice theories). In either case, as hoped, when applied to such theories this discrete analog helps us disentangle the theory's substantive content from its representational artifacts. 

These results are significant as they tell strongly against the intuitions laid out in Sec.~\ref{SecCentralClaims}. One might have an intuition that the world could be fundamentally set on a lattice. This lattice might be square or hexagonal and we might discover which by probing the world at the smallest possible scales looking for violations of rotational symmetry, or other lattice artifacts. Many serious efforts at quantum gravity assume that the world is set on something like a lattice at the smallest scales (although these are often substantially more complicated than the lattices assumed here). However, as this paper clearly demonstrates this just cannot be the case. 

The world cannot be ``fundamentally set on a square lattice'' (or any other lattice) any more than it could be ``fundamentally set in a certain coordinate system''. Like coordinate systems, lattice structures are just not the sort of thing that can be fundamental; they are both thoroughly merely representational. Spacetime cannot be a lattice (even when it might be representable as such). Specifically, I claim that properly understood, there are no such things as lattice-fundamental theories, rather there are only lattice-representable theories.

The primary inspiration for this paper has been the work of mathematical physicist Achim Kempf~\cite{UnsharpKempf,Kempf2000b,Kempf2003,Kempf2004a,Kempf2004b,Kempf2006,Martin2008,Kempf_2010,Kempf2013,Pye2015,Kempf2018} among others~\cite{PyeThesis,Pye2022,BEH_2020}. For an overview of Kempf's works on this topic and his closely related thesis that ``Spacetime could be simultaneously continuous and discrete, in the same way that information can be''~\cite{Kempf_2010}, see~\cite{Kempf2018}. My thesis here inserts a ``representable as'' and then analyzes which parts of these representations should be thought of as substantial. Lattice structures turn out to be coordinate-like representational artifacts.

In addition to supporting this bold claim, I expect the results of this paper to have further substantial consequences for the philosophy of space and time. For a discussion of what these results mean for quantum gravity and Lattice QFT specifically, see \cite{DiscreteGenCovPart2} where I extend my analysis on the Klein Gordon equation. My discussion here will focus on non-Lorentzian theories. However, before discussing some of the further implications of this work, let me raise a few questions regarding the nature of the argument made in this paper. I am not qualified to answer either of these questions but reserve the right to speculate. 

Firstly, essential to the argument put forward in this paper is an analogy between lattice structures and coordinate systems. But exactly what sort of analogy is this? The analogy is not between causal mechanisms, neither side plays any causal role in our theory. The analogy is not between the structural roles bits of mathematics play in the theory, neither side plays any structural role in our theory. It appears to be a thoroughly negative analogy based on the causal/structural roles each side fails to play in these theories.

Secondly, what is the modal character of my conclusion that ``Spacetime cannot be a lattice''? Does ``cannot'' here mean ``logical impossibility'', ``metaphysical impossibility'', or rather ``very unlikely if we are any good at interpreting our theories''? I have little to say about this at the moment.

Next let me discuss some of the expected implications of this work. The three interpretative strategies for discretely-representable spacetime theories discussed in this paper (i.e, naive, internal, and external) provide us with a novel examples of strong underdetermination of theory by evidence: \textit{in principle} there is no experiment which can distinguish between these interpretations. Importantly, these three interpretations are not merely notational variants of each other. Rather, these interpretations make drastically different physically-significant claims about the nature of the underlying manifold and consequently about locality and symmetry.

Next, about the above work one may ask: does it follow from my conclusion that spacetime must be continuous? No, it does not. It is true that in this paper both the internal and external approaches lead to theories with a continuous manifold. However, this is incidental to the examples considered in this paper. For instance, if we had begun from the Klein Gordon equation Eq.~\eqref{KG1Long}, the internalizing approach would have left us with no manifold whatsoever. Cases such as this are discussed in a follow-up paper \cite{DiscreteGenCovPart2}. 

People often have arguments (e.g., from quantum gravity) that spacetime must be a lattice of some sort, are such arguments in disagreement with my conclusion? This would need to be checked on a case-by-case basis, but there is room for compatibility here. Upon closer investigation, such arguments may only show that spacetime must be \textit{representable as} a lattice. In this case there is no disagreement with my conclusion. For instance, bandlimited functions can always be represented as a lattice of sample values; it's just that upon closer philosophical investigations these lattice representations cannot possibly be fundamental, as they are merely coordinate-like representational tools. However, if their arguments carry ontological force as well as representational force, then there is disagreement.

What does my conclusion mean for those who model spacetime as a lattice? (E.g., some approaches to quantum gravity) There is no issue per se with modeling spacetime as a lattice: e.g., as I just mentioned bandlimited physics is discretely representable. My conclusion only speaks to how such models should be interpreted. Indeed, those who describe their physics on a lattice for reasons of approximation (e.g. the Lattice QFT community, people studying crystalline solid state systems) are free to ignore my conclusion. However, they ought not ignore the central analogy or the three lessons which support it. 

For them my lesson is this: ``lattice artifacts'' arise under the following two conditions. They arise when 1) we represent our continuum physics on a lattice \textit{2) and then modify the dynamics to be simple in this representation}. As this paper has shown, the blame is to be placed entirely on the second step. Square-shaped lattice artifacts do not come from using a square lattice, see Fig.~\ref{FigEvolutionH4}. These artifacts come from using dynamics which are relatively simple (i.e., nearest neighbor interactions) when represented on a square lattice. One can define rotation invariant dynamics on a square lattice, Fig.~\ref{FigEvolutionH6}, but not with nearest neighbor interactions.

So how should lattice-based approaches to quantum gravity be interpreted then? I leave further commentary on this to \cite{DiscreteGenCovPart2} which extends these results to a Lorentzian context.

Another set of interesting questions arises in connection with the status of the manifold in the above discussion. Consider the following in light of the dynamical vs geometrical spacetime debate~\cite{EarmanJohn1989Weas,TwiceOver,BelotGordon2000GaM,Menon2019,BrownPooley1999,Nonentity,HuggettNick2006TRAo,StevensSyman2014Tdat,DoratoMauro2007RTbS,Norton2008,Pooley2013}. Roughly, the question there is which of dynamical and spacetime symmetries are explanatorily prior. Are spacetime structures merely codifications of the dynamical behavior of matter? Or do they have an independent existence and act to govern the dynamical behavior of matter (by for as a meta-law restricting the possible dynamics to have certain symmetries)? Moreover, consider Norton's complaint~\cite{Norton2008} that proponents of the dynamical approach must assume some prior spacetime structure (name the spacetime manifold itself) to even begin talking about the dynamics of matter let alone its dynamical symmetries. 

As I have shown in this paper, we can make sense of dynamics and dynamical symmetries without (much of) a manifold. The internalized view here got rid of the lattice sites leaving us with only time as our spacetime manifold. In \cite{DiscreteGenCovPart2} I repeat this analysis for the Klein Gordon equation. There the internalized view has no spacetime manifold whatsoever. In either case, this internalized view gives us something like a neutral staging area on which to probe a theory for its spacetime commitments to spacetime stuructures, yes, but also the spacetime manifold itself. 

Concretely, allow me to provide a schematic for the analysis done here (and in \cite{DiscreteGenCovPart2}): We are given some spacetime theory concerning fields $\varphi_\text{old}$ on a manifold $\mathcal{M}_\text{old}$ and value space $\mathcal{V}_\text{old}$. In the present cases, $\mathcal{M}_\text{old}$ was discrete, but it could be continuous as well in the schematic. One might be suspicious (for whatever reason) that this framing of the theory has gotten its spacetime structure right. One can begin to abandon this particular spacetime framing structure by first reconceptualizing the fields not as maps from $\varphi:\mathcal{M}_\text{old}\to\mathcal{V}_\text{old}$, but as vectors in some (off-shell) vector space $V_\text{old}$. 

In the present case this is done by seeing $\phi_\ell(t)$ as a vector in a vector space $F_L$. (In \cite{DiscreteGenCovPart2}, time too is internalized at this point.)

One can now abandon the old spacetime framing by taking a vector space isomorphism $V_\text{old}\cong V_\text{neutral}$, essentially forgetting about any structure within it besides those of a vector space. Crucially, we now have the old theory reformulated without explicit commitment to a spacetime manifold. 

In the present case this is done by the vector space isomorphism $F_L\cong\mathbb{R}^L$ sending $\phi_\ell(t)\in F_L$ to $\bm{\Phi}(t)\in \mathbb{R}^L$. This, in addition to us working in a basis-independent way. (In this paper, we are technically still left with a manifold of times $\mathcal{M}=\mathbb{R}$ at this point, but in \cite{DiscreteGenCovPart2} we see how it is possible to internalized time as well.)

From this neutral staging ground if $V_\text{neutral}$ we can then analyze the theories dynamics and its dynamical symmetries. All of our dynamical symmetries in the old framing are still dynamical symmetries here. Moreover, we may find a good many new symmetries according to this new spacetime neutral perspective. In particular, we may here find new translation-like transformations which are worthy of being externalized (i.e., more worthy than whatever our old translations were). 

We might then find a new spacetime manifold $\mathcal{M}_\text{new}$ and value space $\mathcal{V}_\text{new}$ for our theory which are specifically designed to fit with the theory's already-studied manifold-independent dynamics. The design of this manifold may be motivated not just by symmetry, but also locality and other consideration. In particular, we can do this taking a vector space isomorphism $V_\text{neutral}\cong V_\text{new}$ where $V_\text{new}$ is the vector space of fields $\varphi_\text{new}$ on a manifold $\mathcal{M}_\text{new}$ with value space $\mathcal{V}_\text{new}$.

In the present case, this was carried out in Sec.~\ref{SecExtPart1} where a translation-matching constraint fixed $\mathcal{M}_\text{new}\cong\mathbb{R}^d$ up to diffeomorphism. Moreover, the lattice-as-sample-points constraint gave us a unique way of embedding $\bm{\Phi}(t)$ onto our new field $\phi_\text{B}$. In particular, it fixed the isomorphism $\mathbb{R}^L\cong F^K$ sending $\bm{\Phi}(t)\in \mathbb{R}^L$ to $\phi_\text{B}(t)\in F^K$. Similarly, we can see this process in a Lorentzian setting in \cite{DiscreteGenCovPart2}.

Finally, after we have migrated from $\mathcal{M}_\text{old}$ to $\mathcal{M}_\text{new}$, some extending of the new theory may be needed. As we saw Sec.~\ref{PerfectRotation}, the new theory on $\mathcal{M}_\text{new}$ may have some artificial representational limitations built into it due to its previous representation on $\mathcal{M}_\text{old}$. In the present case, a bit of work was needed to give H6 and H7 rotation invariance in an unqualified way. Similarly, in \cite{DiscreteGenCovPart2} beginning from some discrete Klein Gordon equations I carry out the above process. A bit of work is needed at the end to see the resulting theories as representationally limited parts of a perfectly Lorentzian theory.

To summarize: recall that the dynamical approach suggests we see a theory's fixed spacetime structures (here, $t_\text{ab}$, $h^\text{ab}$ and in \cite{DiscreteGenCovPart2}, $\eta_\text{ab}$) as codifications of the dynamical behavior of matter. Recall that Norton in \cite{Norton2008} complains that proponents of the dynamical approach must assume some prior spacetime structure (name the spacetime manifold itself) to even begin talking about the dynamics of matter let alone its dynamical symmetries. I take the above schematic to satisfactorily answer Norton's complaint, especially when extended to Lorentzian settings \cite{DiscreteGenCovPart2}. However, spelling this out in detail is a task for another paper.

\begin{acknowledgments}
The authors thanks James Read,  Achim Kempf, Jason Pye, and Nick Menicucci, the Barrio RQI, and the attendees of the ILMPS 2022 conference for their helpful feedback.
\end{acknowledgments}

\bibliographystyle{apsrev4-1}
\bibliography{references}

\begin{thebibliography}{49}%
\makeatletter
\providecommand \@ifxundefined [1]{%
 \@ifx{#1\undefined}
}%
\providecommand \@ifnum [1]{%
 \ifnum #1\expandafter \@firstoftwo
 \else \expandafter \@secondoftwo
 \fi
}%
\providecommand \@ifx [1]{%
 \ifx #1\expandafter \@firstoftwo
 \else \expandafter \@secondoftwo
 \fi
}%
\providecommand \natexlab [1]{#1}%
\providecommand \enquote  [1]{``#1''}%
\providecommand \bibnamefont  [1]{#1}%
\providecommand \bibfnamefont [1]{#1}%
\providecommand \citenamefont [1]{#1}%
\providecommand \href@noop [0]{\@secondoftwo}%
\providecommand \href [0]{\begingroup \@sanitize@url \@href}%
\providecommand \@href[1]{\@@startlink{#1}\@@href}%
\providecommand \@@href[1]{\endgroup#1\@@endlink}%
\providecommand \@sanitize@url [0]{\catcode `\\12\catcode `\$12\catcode
  `\&12\catcode `\#12\catcode `\^12\catcode `\_12\catcode `\%12\relax}%
\providecommand \@@startlink[1]{}%
\providecommand \@@endlink[0]{}%
\providecommand \url  [0]{\begingroup\@sanitize@url \@url }%
\providecommand \@url [1]{\endgroup\@href {#1}{\urlprefix }}%
\providecommand \urlprefix  [0]{URL }%
\providecommand \Eprint [0]{\href }%
\providecommand \doibase [0]{http://dx.doi.org/}%
\providecommand \selectlanguage [0]{\@gobble}%
\providecommand \bibinfo  [0]{\@secondoftwo}%
\providecommand \bibfield  [0]{\@secondoftwo}%
\providecommand \translation [1]{[#1]}%
\providecommand \BibitemOpen [0]{}%
\providecommand \bibitemStop [0]{}%
\providecommand \bibitemNoStop [0]{.\EOS\space}%
\providecommand \EOS [0]{\spacefactor3000\relax}%
\providecommand \BibitemShut  [1]{\csname bibitem#1\endcsname}%
\let\auto@bib@innerbib\@empty
%</preamble>
\bibitem [{\citenamefont {Pooley}(2015)}]{Pooley2015}%
  \BibitemOpen
  \bibfield  {author} {\bibinfo {author} {\bibfnamefont {O.}~\bibnamefont
  {Pooley}},\ }\href {https://arxiv.org/abs/1506.03512} {\enquote {\bibinfo
  {title} {{Background Independence, Diffeomorphism Invariance, and the Meaning
  of Coordinates}},}\ } (\bibinfo {year} {2015}),\ \Eprint
  {http://arxiv.org/abs/1506.03512} {arXiv:1506.03512 [gr-qc]} \BibitemShut
  {NoStop}%
%%CITATION = ARXIV:1412.7827;%%
\bibitem [{\citenamefont {Norton}(1993)}]{Norton1993}%
  \BibitemOpen
  \bibfield  {author} {\bibinfo {author} {\bibfnamefont {J.~D.}\ \bibnamefont
  {Norton}},\ }\href {\doibase 10.1088/0034-4885/56/7/001} {\bibfield
  {journal} {\bibinfo  {journal} {Reports on Progress in Physics}\ }\textbf
  {\bibinfo {volume} {56}},\ \bibinfo {pages} {791} (\bibinfo {year}
  {1993})}\BibitemShut {NoStop}%
\bibitem [{\citenamefont {Friedman}(1983)}]{Friedman1983}%
  \BibitemOpen
  \bibfield  {author} {\bibinfo {author} {\bibfnamefont {M.}~\bibnamefont
  {Friedman}},\ }\enquote {\bibinfo {title} {Space-time theories},}\ in\ \href
  {http://www.jstor.org/stable/j.ctt7ztj02.5} {\emph {\bibinfo {booktitle}
  {Foundations of Space-Time Theories: Relativistic Physics and Philosophy of
  Science}}}\ (\bibinfo  {publisher} {Princeton University Press},\ \bibinfo
  {year} {1983})\ pp.\ \bibinfo {pages} {32--70}\BibitemShut {NoStop}%
\bibitem [{\citenamefont {Freidel}\ and\ \citenamefont
  {Tah}(2021)}]{TehFreidel}%
  \BibitemOpen
  \bibfield  {author} {\bibinfo {author} {\bibfnamefont {L.}~\bibnamefont
  {Freidel}}\ and\ \bibinfo {author} {\bibfnamefont {N.}~\bibnamefont {Tah}},\
  }\href {https://arxiv.org/abs/2109.08516} {\enquote {\bibinfo {title}
  {Substantive general covariance and the einstein-klein dispute: A noetherian
  approach},}\ } (\bibinfo {year} {2021}),\ \Eprint
  {http://arxiv.org/abs/2109.08516} {arXiv:2109.08516 [hist-ph]} \BibitemShut
  {NoStop}%
\bibitem [{\citenamefont {Earman}(1989)}]{EarmanJohn1989Weas}%
  \BibitemOpen
  \bibfield  {author} {\bibinfo {author} {\bibfnamefont {J.}~\bibnamefont
  {Earman}},\ }\href@noop {} {\emph {\bibinfo {title} {`World enough and
  space-time : absolute versus relational theories of space and time'}}}\
  (\bibinfo  {publisher} {MIT Press},\ \bibinfo {address} {Cambridge, Mass ;
  London},\ \bibinfo {year} {1989})\BibitemShut {NoStop}%
\bibitem [{\citenamefont {Brian~Pitts}(2006)}]{Pitts2006}%
  \BibitemOpen
  \bibfield  {author} {\bibinfo {author} {\bibfnamefont {J.}~\bibnamefont
  {Brian~Pitts}},\ }\href@noop {} {\bibfield  {journal} {\bibinfo  {journal}
  {Studies in History and Philosophy of Modern Physics}\ }\textbf {\bibinfo
  {volume} {37}},\ \bibinfo {pages} {347} (\bibinfo {year} {2006})}\BibitemShut
  {NoStop}%
\bibitem [{\citenamefont {Pooley}(2010)}]{Pooley2010}%
  \BibitemOpen
  \bibfield  {author} {\bibinfo {author} {\bibfnamefont {O.}~\bibnamefont
  {Pooley}},\ }\enquote {\bibinfo {title} {Space-time theories},}\ in\ \href
  {http://www.jstor.org/stable/j.ctt7ztj02.5} {\emph {\bibinfo {booktitle}
  {EPSA philosophical issues in the sciences}}},\ \bibinfo {editor} {edited by\
  \bibinfo {editor} {\bibfnamefont {M.}~\bibnamefont {Suárez}}, \bibinfo
  {editor} {\bibfnamefont {M.}~\bibnamefont {Dorato}}, \ and\ \bibinfo {editor}
  {\bibfnamefont {M.}~\bibnamefont {Rédei}}}\ (\bibinfo  {publisher}
  {Springer},\ \bibinfo {address} {Dordrecht ; London},\ \bibinfo {year}
  {2010})\ Chap.~\bibinfo {chapter} {19}, pp.\ \bibinfo {pages}
  {197--210}\BibitemShut {NoStop}%
\bibitem [{\citenamefont {Read}(2016)}]{ReadThesis}%
  \BibitemOpen
  \bibfield  {author} {\bibinfo {author} {\bibfnamefont {J.}~\bibnamefont
  {Read}},\ }\emph {\bibinfo {title} {Background independence in classical and
  quantum gravity}},\ \href@noop {} {Master's thesis},\ \bibinfo  {school}
  {University of Oxford} (\bibinfo {year} {2016})\BibitemShut {NoStop}%
\bibitem [{\citenamefont {Teitel}(2019)}]{Teitel2019}%
  \BibitemOpen
  \bibfield  {author} {\bibinfo {author} {\bibfnamefont {T.}~\bibnamefont
  {Teitel}},\ }\href@noop {} {\bibfield  {journal} {\bibinfo  {journal}
  {Studies In History And Philosophy Of Modern Physics}\ }\textbf {\bibinfo
  {volume} {65}},\ \bibinfo {pages} {pp41} (\bibinfo {year}
  {2019})}\BibitemShut {NoStop}%
\bibitem [{\citenamefont {Belot}(2011)}]{Belot2011}%
  \BibitemOpen
  \bibfield  {author} {\bibinfo {author} {\bibfnamefont {G.}~\bibnamefont
  {Belot}},\ }\href@noop {} {\bibfield  {journal} {\bibinfo  {journal} {General
  Relativity and Gravitation}\ }\textbf {\bibinfo {volume} {43}},\ \bibinfo
  {pages} {2865} (\bibinfo {year} {2011})}\BibitemShut {NoStop}%
\bibitem [{\citenamefont {Rovelli}(2001)}]{rovelli_2001}%
  \BibitemOpen
  \bibfield  {author} {\bibinfo {author} {\bibfnamefont {C.}~\bibnamefont
  {Rovelli}},\ }\enquote {\bibinfo {title} {Quantum spacetime: What do we
  know?}}\ in\ \href {\doibase 10.1017/CBO9780511612909.005} {\emph {\bibinfo
  {booktitle} {Physics Meets Philosophy at the Planck Scale: Contemporary
  Theories in Quantum Gravity}}},\ \bibinfo {editor} {edited by\ \bibinfo
  {editor} {\bibfnamefont {C.}~\bibnamefont {Callender}}\ and\ \bibinfo
  {editor} {\bibfnamefont {N.}~\bibnamefont {Huggett}}}\ (\bibinfo  {publisher}
  {Cambridge University Press},\ \bibinfo {year} {2001})\ p.\ \bibinfo {pages}
  {101–122}\BibitemShut {NoStop}%
\bibitem [{\citenamefont {Rovelli}(2004)}]{rovelli_2004}%
  \BibitemOpen
  \bibfield  {author} {\bibinfo {author} {\bibfnamefont {C.}~\bibnamefont
  {Rovelli}},\ }\href {\doibase 10.1017/CBO9780511755804} {\emph {\bibinfo
  {title} {Quantum Gravity}}},\ Cambridge Monographs on Mathematical Physics\
  (\bibinfo  {publisher} {Cambridge University Press},\ \bibinfo {year}
  {2004})\BibitemShut {NoStop}%
\bibitem [{\citenamefont {Smolin}(2006)}]{SmolinLee2006TCfB}%
  \BibitemOpen
  \bibfield  {author} {\bibinfo {author} {\bibfnamefont {L.}~\bibnamefont
  {Smolin}},\ }in\ \href@noop {} {\emph {\bibinfo {booktitle} {The Structural
  Foundations of Quantum Gravity}}}\ (\bibinfo  {publisher} {Oxford University
  Press},\ \bibinfo {address} {Oxford},\ \bibinfo {year} {2006})\BibitemShut
  {NoStop}%
\bibitem [{\citenamefont {Smolin}(2008)}]{SmolinLee2008Ttwp}%
  \BibitemOpen
  \bibfield  {author} {\bibinfo {author} {\bibfnamefont {L.}~\bibnamefont
  {Smolin}},\ }\href@noop {} {\emph {\bibinfo {title} {The trouble with physics
  : the rise of string theory, the fall of a science and what comes next}}}\
  (\bibinfo  {publisher} {Penguin},\ \bibinfo {address} {London},\ \bibinfo
  {year} {2008})\BibitemShut {NoStop}%
\bibitem [{\citenamefont {de~Haro}(2017)}]{deHaroSebastian2017Daeg}%
  \BibitemOpen
  \bibfield  {author} {\bibinfo {author} {\bibfnamefont {S.}~\bibnamefont
  {de~Haro}},\ }\href@noop {} {\bibfield  {journal} {\bibinfo  {journal}
  {Studies in History and Philosophy of Modern Physics}\ }\textbf {\bibinfo
  {volume} {59}},\ \bibinfo {pages} {109} (\bibinfo {year} {2017})}\BibitemShut
  {NoStop}%
\bibitem [{\citenamefont {De~Haro}(2017)}]{DeHaroSebastian2017TIoD}%
  \BibitemOpen
  \bibfield  {author} {\bibinfo {author} {\bibfnamefont {S.}~\bibnamefont
  {De~Haro}},\ }\href@noop {} {\bibfield  {journal} {\bibinfo  {journal}
  {Foundations of physics}\ }\textbf {\bibinfo {volume} {47}},\ \bibinfo
  {pages} {1464} (\bibinfo {year} {2017})}\BibitemShut {NoStop}%
\bibitem [{\citenamefont {Kempf}(2000{\natexlab{a}})}]{UnsharpKempf}%
  \BibitemOpen
  \bibfield  {author} {\bibinfo {author} {\bibfnamefont {A.}~\bibnamefont
  {Kempf}},\ }\href {\doibase 10.1103/physrevlett.85.2873} {\bibfield
  {journal} {\bibinfo  {journal} {Physical Review Letters}\ }\textbf {\bibinfo
  {volume} {85}},\ \bibinfo {pages} {2873–2876} (\bibinfo {year}
  {2000}{\natexlab{a}})}\BibitemShut {NoStop}%
\bibitem [{\citenamefont {Kempf}(2003)}]{Kempf2003}%
  \BibitemOpen
  \bibfield  {author} {\bibinfo {author} {\bibfnamefont {A.}~\bibnamefont
  {Kempf}},\ }\href {https://arxiv.org/abs/gr-qc/0306104} {\enquote {\bibinfo
  {title} {{Aspects of Information Theory in Curved Space}},}\ } (\bibinfo
  {year} {2003}),\ \Eprint {http://arxiv.org/abs/0306104} {arXiv:0306104
  [gr-qc]} \BibitemShut {NoStop}%
\bibitem [{\citenamefont {Kempf}(2004{\natexlab{a}})}]{Kempf2004a}%
  \BibitemOpen
  \bibfield  {author} {\bibinfo {author} {\bibfnamefont {A.}~\bibnamefont
  {Kempf}},\ }\href {\doibase 10.1103/PhysRevLett.92.221301} {\bibfield
  {journal} {\bibinfo  {journal} {Phys. Rev. Lett.}\ }\textbf {\bibinfo
  {volume} {92}},\ \bibinfo {pages} {221301} (\bibinfo {year}
  {2004}{\natexlab{a}})}\BibitemShut {NoStop}%
\bibitem [{\citenamefont {Kempf}(2004{\natexlab{b}})}]{Kempf2004b}%
  \BibitemOpen
  \bibfield  {author} {\bibinfo {author} {\bibfnamefont {A.}~\bibnamefont
  {Kempf}},\ }\href {\doibase 10.1103/PhysRevD.69.124014} {\bibfield  {journal}
  {\bibinfo  {journal} {Phys. Rev. D}\ }\textbf {\bibinfo {volume} {69}},\
  \bibinfo {pages} {124014} (\bibinfo {year} {2004}{\natexlab{b}})}\BibitemShut
  {NoStop}%
\bibitem [{\citenamefont {KEMPF}(2006)}]{Kempf2006}%
  \BibitemOpen
  \bibfield  {author} {\bibinfo {author} {\bibfnamefont {A.}~\bibnamefont
  {KEMPF}},\ }in\ \href {\doibase 10.1142/9789812704030_0316} {\emph {\bibinfo
  {booktitle} {The Tenth Marcel Grossmann Meeting}}}\ (\bibinfo  {publisher}
  {World Scientific Publishing Company},\ \bibinfo {year} {2006})\BibitemShut
  {NoStop}%
\bibitem [{\citenamefont {Kempf}(2000{\natexlab{b}})}]{Kempf2000b}%
  \BibitemOpen
  \bibfield  {author} {\bibinfo {author} {\bibfnamefont {A.}~\bibnamefont
  {Kempf}},\ }\href {\doibase 10.1103/PhysRevD.63.024017} {\bibfield  {journal}
  {\bibinfo  {journal} {Phys. Rev. D}\ }\textbf {\bibinfo {volume} {63}},\
  \bibinfo {pages} {024017} (\bibinfo {year} {2000}{\natexlab{b}})}\BibitemShut
  {NoStop}%
\bibitem [{\citenamefont {Martin}\ and\ \citenamefont
  {Kempf}(2008)}]{Martin2008}%
  \BibitemOpen
  \bibfield  {author} {\bibinfo {author} {\bibfnamefont {R.}~\bibnamefont
  {Martin}}\ and\ \bibinfo {author} {\bibfnamefont {A.}~\bibnamefont {Kempf}},\
  }\href {\doibase 10.1007/BF03549501} {\bibfield  {journal} {\bibinfo
  {journal} {Sampling Theory in Signal and Image Processing}\ }\textbf
  {\bibinfo {volume} {7}},\ \bibinfo {pages} {281} (\bibinfo {year}
  {2008})}\BibitemShut {NoStop}%
\bibitem [{\citenamefont {Kempf}(2010)}]{Kempf_2010}%
  \BibitemOpen
  \bibfield  {author} {\bibinfo {author} {\bibfnamefont {A.}~\bibnamefont
  {Kempf}},\ }\href {\doibase 10.1088/1367-2630/12/11/115001} {\bibfield
  {journal} {\bibinfo  {journal} {New Journal of Physics}\ }\textbf {\bibinfo
  {volume} {12}},\ \bibinfo {pages} {115001} (\bibinfo {year}
  {2010})}\BibitemShut {NoStop}%
\bibitem [{\citenamefont {Kempf}\ \emph {et~al.}(2013)\citenamefont {Kempf},
  \citenamefont {Chatwin-Davies},\ and\ \citenamefont {Martin}}]{Kempf2013}%
  \BibitemOpen
  \bibfield  {author} {\bibinfo {author} {\bibfnamefont {A.}~\bibnamefont
  {Kempf}}, \bibinfo {author} {\bibfnamefont {A.}~\bibnamefont
  {Chatwin-Davies}}, \ and\ \bibinfo {author} {\bibfnamefont {R.~T.~W.}\
  \bibnamefont {Martin}},\ }\href {\doibase 10.1063/1.4790482} {\bibfield
  {journal} {\bibinfo  {journal} {Journal of Mathematical Physics}\ }\textbf
  {\bibinfo {volume} {54}},\ \bibinfo {pages} {022301} (\bibinfo {year}
  {2013})},\ \Eprint {http://arxiv.org/abs/https://doi.org/10.1063/1.4790482}
  {https://doi.org/10.1063/1.4790482} \BibitemShut {NoStop}%
\bibitem [{\citenamefont {Pye}\ \emph {et~al.}(2015)\citenamefont {Pye},
  \citenamefont {Donnelly},\ and\ \citenamefont {Kempf}}]{Pye2015}%
  \BibitemOpen
  \bibfield  {author} {\bibinfo {author} {\bibfnamefont {J.}~\bibnamefont
  {Pye}}, \bibinfo {author} {\bibfnamefont {W.}~\bibnamefont {Donnelly}}, \
  and\ \bibinfo {author} {\bibfnamefont {A.}~\bibnamefont {Kempf}},\ }\href
  {\doibase 10.1103/physrevd.92.105022} {\bibfield  {journal} {\bibinfo
  {journal} {Physical Review D}\ }\textbf {\bibinfo {volume} {92}} (\bibinfo
  {year} {2015}),\ 10.1103/physrevd.92.105022}\BibitemShut {NoStop}%
\bibitem [{\citenamefont {Kempf}(2018)}]{Kempf2018}%
  \BibitemOpen
  \bibfield  {author} {\bibinfo {author} {\bibfnamefont {A.}~\bibnamefont
  {Kempf}},\ }\href {\doibase 10.1007/s10701-018-0163-2} {\bibfield  {journal}
  {\bibinfo  {journal} {Foundations of Physics}\ }\textbf {\bibinfo {volume}
  {48}},\ \bibinfo {pages} {1191} (\bibinfo {year} {2018})}\BibitemShut
  {NoStop}%
\bibitem [{\citenamefont {{Pye, Jason}}(2020)}]{PyeThesis}%
  \BibitemOpen
  \bibfield  {author} {\bibinfo {author} {\bibnamefont {{Pye, Jason}}},\ }\emph
  {\bibinfo {title} {On the Application of Bandlimitation and Sampling Theory
  to Quantum Field Theory}},\ \href {http://hdl.handle.net/10012/16427} {Ph.D.
  thesis},\ \bibinfo  {school} {University of Waterloo} (\bibinfo {year}
  {2020})\BibitemShut {NoStop}%
\bibitem [{\citenamefont {Pye}(2022)}]{Pye2022}%
  \BibitemOpen
  \bibfield  {author} {\bibinfo {author} {\bibfnamefont {J.}~\bibnamefont
  {Pye}},\ }\href {https://arxiv.org/abs/2202.03589} {\enquote {\bibinfo
  {title} {Lorentz-covariant sampling theory for fields},}\ } (\bibinfo {year}
  {2022}),\ \Eprint {http://arxiv.org/abs/2202.03589} {arXiv:2202.03589
  [hep-th]} \BibitemShut {NoStop}%
\bibitem [{\citenamefont {Henderson}\ and\ \citenamefont
  {Menicucci}(2020)}]{BEH_2020}%
  \BibitemOpen
  \bibfield  {author} {\bibinfo {author} {\bibfnamefont {L.~J.}\ \bibnamefont
  {Henderson}}\ and\ \bibinfo {author} {\bibfnamefont {N.~C.}\ \bibnamefont
  {Menicucci}},\ }\href {\doibase 10.1103/physrevd.102.125026} {\bibfield
  {journal} {\bibinfo  {journal} {Physical Review D}\ }\textbf {\bibinfo
  {volume} {102}} (\bibinfo {year} {2020}),\
  10.1103/physrevd.102.125026}\BibitemShut {NoStop}%
\bibitem [{\citenamefont {Kempf}(1997)}]{Kempf_1997}%
  \BibitemOpen
  \bibfield  {author} {\bibinfo {author} {\bibfnamefont {A.}~\bibnamefont
  {Kempf}},\ }\href {\doibase 10.1209/epl/i1997-00457-7} {\bibfield  {journal}
  {\bibinfo  {journal} {Europhysics Letters ({EPL})}\ }\textbf {\bibinfo
  {volume} {40}},\ \bibinfo {pages} {257} (\bibinfo {year} {1997})}\BibitemShut
  {NoStop}%
\bibitem [{\citenamefont {García}(2002)}]{GARCIA200263}%
  \BibitemOpen
  \bibfield  {author} {\bibinfo {author} {\bibfnamefont {A.~G.}\ \bibnamefont
  {García}}\ }(\bibinfo  {publisher} {Elsevier},\ \bibinfo {year} {2002})\
  pp.\ \bibinfo {pages} {63--137}\BibitemShut {NoStop}%
\bibitem [{\citenamefont {Jerri}(1977)}]{SamplingTutorial}%
  \BibitemOpen
  \bibfield  {author} {\bibinfo {author} {\bibfnamefont {A.}~\bibnamefont
  {Jerri}},\ }\href {\doibase 10.1109/PROC.1977.10771} {\bibfield  {journal}
  {\bibinfo  {journal} {Proceedings of the IEEE}\ }\textbf {\bibinfo {volume}
  {65}},\ \bibinfo {pages} {1565} (\bibinfo {year} {1977})}\BibitemShut
  {NoStop}%
\bibitem [{\citenamefont {Unser}(2000)}]{UnserM2000SyaS}%
  \BibitemOpen
  \bibfield  {author} {\bibinfo {author} {\bibfnamefont {M.}~\bibnamefont
  {Unser}},\ }\href@noop {} {\bibfield  {journal} {\bibinfo  {journal}
  {Proceedings of the IEEE}\ }\textbf {\bibinfo {volume} {88}},\ \bibinfo
  {pages} {569} (\bibinfo {year} {2000})}\BibitemShut {NoStop}%
\bibitem [{\citenamefont {Grimmer}(2022)}]{DiscreteGenCovPart2}%
  \BibitemOpen
  \bibfield  {author} {\bibinfo {author} {\bibfnamefont {D.}~\bibnamefont
  {Grimmer}},\ }\href {\doibase 10.48550/ARXIV.2205.07701} {\enquote {\bibinfo
  {title} {A discrete analog of general covariance -- part 2: Despite what
  you've heard, a perfectly lorentzian lattice theory},}\ } (\bibinfo {year}
  {2022})\BibitemShut {NoStop}%
\bibitem [{\citenamefont {Dasgupta}(2016)}]{TwiceOver}%
  \BibitemOpen
  \bibfield  {author} {\bibinfo {author} {\bibfnamefont {S.}~\bibnamefont
  {Dasgupta}},\ }\href@noop {} {\bibfield  {journal} {\bibinfo  {journal} {The
  British Journal for the Philosophy of Science}\ }\textbf {\bibinfo {volume}
  {67}},\ \bibinfo {pages} {837} (\bibinfo {year} {2016})}\BibitemShut
  {NoStop}%
\bibitem [{\citenamefont {Belot}(2000)}]{BelotGordon2000GaM}%
  \BibitemOpen
  \bibfield  {author} {\bibinfo {author} {\bibfnamefont {G.}~\bibnamefont
  {Belot}},\ }\href@noop {} {\bibfield  {journal} {\bibinfo  {journal} {The
  British Journal for the Philosophy of Science}\ }\textbf {\bibinfo {volume}
  {51}},\ \bibinfo {pages} {561} (\bibinfo {year} {2000})}\BibitemShut
  {NoStop}%
\bibitem [{\citenamefont {Menon}(2019)}]{Menon2019}%
  \BibitemOpen
  \bibfield  {author} {\bibinfo {author} {\bibfnamefont {T.}~\bibnamefont
  {Menon}},\ }\href@noop {} {\bibfield  {journal} {\bibinfo  {journal}
  {Philosophy of Science}\ }\textbf {\bibinfo {volume} {86}} (\bibinfo {year}
  {2019})}\BibitemShut {NoStop}%
\bibitem [{\citenamefont {O.~Pooley}(1999)}]{BrownPooley1999}%
  \BibitemOpen
  \bibfield  {author} {\bibinfo {author} {\bibfnamefont {H.~B.}\ \bibnamefont
  {O.~Pooley}},\ }\enquote {\bibinfo {title} {The origins of the spacetime
  metric: Bell’s lorentzian pedagogy and its significance in general
  relativity.}}\ in\ \href@noop {} {\emph {\bibinfo {booktitle} {Physics Meets
  Philosophy at the Planck Scale}}},\ \bibinfo {editor} {edited by\ \bibinfo
  {editor} {\bibfnamefont {N.}~\bibnamefont {Callender~C.}, \bibfnamefont
  {Huggett}}}\ (\bibinfo  {publisher} {Cambridge University Press},\ \bibinfo
  {address} {Cambridge},\ \bibinfo {year} {1999})\ pp.\ \bibinfo {pages}
  {256--72}\BibitemShut {NoStop}%
\bibitem [{\citenamefont {Brown}\ and\ \citenamefont
  {Pooley}(2004)}]{Nonentity}%
  \BibitemOpen
  \bibfield  {author} {\bibinfo {author} {\bibfnamefont {H.~R.}\ \bibnamefont
  {Brown}}\ and\ \bibinfo {author} {\bibfnamefont {O.}~\bibnamefont {Pooley}},\
  }\href@noop {} {\  (\bibinfo {year} {2004})}\BibitemShut {NoStop}%
\bibitem [{\citenamefont {Huggett}(2006)}]{HuggettNick2006TRAo}%
  \BibitemOpen
  \bibfield  {author} {\bibinfo {author} {\bibfnamefont {N.}~\bibnamefont
  {Huggett}},\ }\href@noop {} {\bibfield  {journal} {\bibinfo  {journal}
  {Mind}\ }\textbf {\bibinfo {volume} {115}},\ \bibinfo {pages} {41} (\bibinfo
  {year} {2006})}\BibitemShut {NoStop}%
\bibitem [{\citenamefont {Stevens}(2014)}]{StevensSyman2014Tdat}%
  \BibitemOpen
  \bibfield  {author} {\bibinfo {author} {\bibfnamefont {S.}~\bibnamefont
  {Stevens}},\ }\href {http://search.proquest.com/docview/2130069327/}
  {\enquote {\bibinfo {title} {The dynamical approach to relativity as a form
  of regularity relationalism},}\ } (\bibinfo {year} {2014})\BibitemShut
  {NoStop}%
\bibitem [{\citenamefont {Dorato}(2007)}]{DoratoMauro2007RTbS}%
  \BibitemOpen
  \bibfield  {author} {\bibinfo {author} {\bibfnamefont {M.}~\bibnamefont
  {Dorato}},\ }\href
  {http://www.tandfonline.com/doi/abs/10.1080/02698590701305891} {\bibfield
  {journal} {\bibinfo  {journal} {International studies in the philosophy of
  science}\ }\textbf {\bibinfo {volume} {21}},\ \bibinfo {pages} {95} (\bibinfo
  {year} {2007})}\BibitemShut {NoStop}%
\bibitem [{\citenamefont {Norton}(2008)}]{Norton2008}%
  \BibitemOpen
  \bibfield  {author} {\bibinfo {author} {\bibfnamefont {J.~D.}\ \bibnamefont
  {Norton}},\ }\href@noop {} {\bibfield  {journal} {\bibinfo  {journal} {The
  British Journal for the Philosophy of Science}\ }\textbf {\bibinfo {volume}
  {59}},\ \bibinfo {pages} {821} (\bibinfo {year} {2008})}\BibitemShut
  {NoStop}%
\bibitem [{\citenamefont {Batterman}\ and\ \citenamefont
  {Pooley}(2013)}]{Pooley2013}%
  \BibitemOpen
  \bibfield  {author} {\bibinfo {author} {\bibfnamefont {R.}~\bibnamefont
  {Batterman}}\ and\ \bibinfo {author} {\bibfnamefont {O.}~\bibnamefont
  {Pooley}},\ }in\ \href
  {http://oxfordhandbooks.com/view/10.1093/oxfordhb/9780195392043.001.0001/oxfordhb-9780195392043-e-16}
  {\emph {\bibinfo {booktitle} {The Oxford Handbook of Philosophy of
  Physics}}}\ (\bibinfo  {publisher} {Oxford University Press},\ \bibinfo
  {year} {2013})\BibitemShut {NoStop}%
\bibitem [{\citenamefont {Shannon}(2001)}]{ShannonOriginal}%
  \BibitemOpen
  \bibfield  {author} {\bibinfo {author} {\bibfnamefont {C.~E.}\ \bibnamefont
  {Shannon}},\ }\href {\doibase 10.1145/584091.584093} {\bibfield  {journal}
  {\bibinfo  {journal} {SIGMOBILE Mob. Comput. Commun. Rev.}\ }\textbf
  {\bibinfo {volume} {5}},\ \bibinfo {pages} {3–55} (\bibinfo {year}
  {2001})}\BibitemShut {NoStop}%
\bibitem [{\citenamefont {Feichtinger}\ \emph {et~al.}(2015)\citenamefont
  {Feichtinger}, \citenamefont {Führ},\ and\ \citenamefont
  {Pesenson}}]{CurvedSampling}%
  \BibitemOpen
  \bibfield  {author} {\bibinfo {author} {\bibfnamefont {H.~G.}\ \bibnamefont
  {Feichtinger}}, \bibinfo {author} {\bibfnamefont {H.}~\bibnamefont {Führ}},
  \ and\ \bibinfo {author} {\bibfnamefont {I.~Z.}\ \bibnamefont {Pesenson}},\
  }\href {https://arxiv.org/abs/1512.08668} {\enquote {\bibinfo {title}
  {Geometric space-frequency analysis on manifolds},}\ } (\bibinfo {year}
  {2015}),\ \Eprint {http://arxiv.org/abs/1512.08668} {arXiv:1512.08668
  [math.FA]} \BibitemShut {NoStop}%
\bibitem [{\citenamefont {Wallace}(2019)}]{WallaceAfraid}%
  \BibitemOpen
  \bibfield  {author} {\bibinfo {author} {\bibfnamefont {D.}~\bibnamefont
  {Wallace}},\ }\href {\doibase 10.1016/j.shpsb.2017.07.002} {\bibfield
  {journal} {\bibinfo  {journal} {Studies in History and Philosophy of Science
  Part B: Studies in History and Philosophy of Modern Physics}\ }\textbf
  {\bibinfo {volume} {67}},\ \bibinfo {pages} {125} (\bibinfo {year}
  {2019})}\BibitemShut {NoStop}%
\bibitem [{\citenamefont {Thorne}\ \emph {et~al.}(1973)\citenamefont {Thorne},
  \citenamefont {Lee},\ and\ \citenamefont {Lightman}}]{ThorneLeeLightman}%
  \BibitemOpen
  \bibfield  {author} {\bibinfo {author} {\bibfnamefont {K.~S.}\ \bibnamefont
  {Thorne}}, \bibinfo {author} {\bibfnamefont {D.~L.}\ \bibnamefont {Lee}}, \
  and\ \bibinfo {author} {\bibfnamefont {A.~P.}\ \bibnamefont {Lightman}},\
  }\href {\doibase 10.1103/PhysRevD.7.3563} {\bibfield  {journal} {\bibinfo
  {journal} {Phys. Rev. D}\ }\textbf {\bibinfo {volume} {7}},\ \bibinfo {pages}
  {3563} (\bibinfo {year} {1973})}\BibitemShut {NoStop}%
\end{thebibliography}%

\appendix
\section{Brief Review of General Covariance, Diffeomorphism Invariance, and Background Independence}\label{SecGenCov}
 As discussed in the introduction, a crucial step in the history of GR was Einstein's adoption of the principle of general covariance. While, ultimately, this principle is merely stylistic (with no physical content per se) it nonetheless commits us to a good and useful style of theorizing. As discussed in Sec.~\ref{SecIntro}, a generally covariant formulation of a theory disentangles its substantive content from its merely representational artifacts. In particular, reformulating in this way more clearly exposes the theory's geometric background structure, and thereby helps clarify our understanding of the theory's symmetries.

In the physics literature, three closely related concepts are often confused: general covariance, diffeomorphism invariance, and background independence (i.e., a complete lack of background structure). To demonstrate these ideas, let's go through some examples.

\subsection{Klein Gordon Equation}
Consider a real scalar field $\varphi:\mathcal{M}\to\mathbb{R}$ with mass $M$, satisfying the Klein Gordon equation,
\begin{align}\label{KG0}
\partial_{t}^2\varphi(t,x,y,z) \!=\! (\partial_x^2+\partial_y^2+\partial_z^2-\!M^2) \, \varphi(t,x,y,z).
\end{align}
This formulation is not generally covariant since when it is rewritten in different coordinates it changes form. For instance, in the coordinates $t'=t$, $x'=x$, $y'=y$ and $z'=z+\frac{1}{2}a\,t^2$, we have,
\begin{align}
\nonumber
\partial_{t'}^2\varphi(t',x',y',z') 
&= (\partial_{x'}^2+\partial_{y'}^2+\partial_{z'}^2-M^2) \, \varphi(t',x',y',z')\\
\nonumber
&- a \, \partial_{z'} \, \varphi(t',x',y',z').
\end{align}
An extra term shows up when we move into a non-inertial coordinate system. Let's fix this. Introducing a fixed Lorentzian metric field, \mbox{$\eta^\text{ab}=\text{diag}(-1,1,1,1)$} we can rewrite Eq.~\eqref{KG0} as,
\begin{align}\label{KG00}
\qquad(\eta^{\mu\nu}\partial_\mu\partial_\nu-M^2) \, \varphi = 0,
\end{align}
where $x^\mu=(t,x,y,z)$ and $\partial_\mu=(\partial_t,\partial_x,\partial_y,\partial_z)$. Unfortunately this is still not generally covariant. If we rewrite Eq.~\eqref{KG00} in arbitrary coordinates, $x'^\mu$, we find,
\begin{align}
\nonumber
\left(\eta^{\sigma\rho}\frac{\partial x'^\mu}{\partial x^\sigma}\frac{\partial x'^\nu}{\partial x^\rho}\partial_\mu\partial_\nu-M^2\right) \, \varphi +
\eta^{\sigma\rho}\frac{\partial^2 x'^\mu}{\partial x^\sigma\,\partial x^\rho}\,\partial_\mu\varphi
= 0.
\end{align}
This formulation, however, is coordinate-invariant. If we change coordinates again to $x''^\mu$ the equation keeps the same form except with $x'^\mu\to x''^\mu$.

This demonstrates an ambiguity in the usage of the term generally covariant above~\cite{WallaceAfraid}: coordinate-independent can mean coordinate-invariant (but still written in terms of coordinates) or coordinate-free (written without any reference to coordinates at all). The real benefits of general covariance come from having a coordinate-free representation. This is the notion of general covariance relevant throughout this paper. To achieve general covariance we need to reformulate Eq.~\eqref{KG00} in the coordinate-free language of differential geometry. 

Before this however, I need to introduce some terminology. Throughout this paper I will associate with any continuum spacetime theory with two spaces of models~\cite{ThorneLeeLightman,Pooley2015}: kinematically possible models (KPMs) and dynamically possible models (DPMs). Roughly, these are off-shell and on-shell models.

KPMs are all of the mathematical objects which have the right sort of structures to make sense as models of our theory (regardless of whether they satisfy the dynamics). These are represented as an ordered collection of the theory's manifold together with its geometric fields and matter fields. For our Klein Gordon example, the KPMs are given by\footnote{The name SR1 is picked to follow the notation set in~\cite{Pooley2015}.},
\begin{align}
\text{SR1:}\qquad\text{KPMs:}\quad&\langle\mathcal{M},\eta_\text{ab},\varphi\rangle
\end{align}
where $\mathcal{M}$ is a differentiable (3+1)-manifold, $\eta_\text{ab}$, is a fixed Lorentzian metric field\footnote{It's important to note the difference between $\eta_{\mu\nu}$ and $\eta_\text{ab}$. Any tensor object with greek indices (e.g., $\eta_{\mu\nu}$) is to be understood as the components of a certain tensor in a particular coordinate system. By contrast, any tensor object with roman indices (e.g., $\eta_\text{ab}$) is coordinate-free, its ``indices'' are merely there to remind us of the rank of this tensor and to help us see how it interacts with the other tensor objects.}, and $\varphi:\mathcal{M}\to\mathbb{R}$ is a dynamical real scalar field.

By contrast, a theory's dynamically possible models (DPMs), are the subset of the KPMs which obey the theory's dynamical equations. For SR1 the DPMs are picked out by,
\begin{align}
\text{SR1:}\qquad\text{KPMs:}\quad&\langle\mathcal{M},\eta_\text{ab},\varphi\rangle\\
\nonumber
\text{DPMs:}\quad&(\eta^\text{ab}\nabla_\text{a}\nabla_\text{b}-M^2)\,\varphi = 0, 
\end{align}
where $\nabla_\text{a}$ is the unique covariant derivative operator compatible with the metric, (i.e. with $\nabla_\text{c}\,\eta_\text{ab}=0$). This formulation of the Klein Gordon equation is now generally covariant (in the coordinate-free sense). 

\subsubsection*{Klein Gordon - Symmetries}
Let us now use this generally covariant formulation to help us understand this theory's symmetries. It is important to distinguish two kinds of symmetry~\cite{EarmanJohn1989Weas} (spacetime symmetries and dynamical symmetries) related to two different kinds of fields~\cite{Pooley2015} (fixed fields and dynamical fields). The latter distinction is that fixed fields are fixed by fiat to be the same in every model. By contrast, dynamical fields can vary from model to model. In SR1, $\eta_\text{ab}$ is fixed whereas $\varphi$ is dynamical.

The distinction regarding symmetries is as follows. Spacetime symmetries are those diffeomorphisms, \mbox{$d\in\text{Diff}(\mathcal{M})$}, which preserve the theory's fixed fields (regardless of the dynamical equations). Dynamical symmetries are those diffeomorphisms, \mbox{$d\in\text{Diff}(\mathcal{M})$}, which map solutions to solutions when applied to the dynamical fields of our models (leaving the fixed fields fixed). In either case, let us call these external symmetries.

For SR1 our only fixed field is $\eta_\text{ab}$. Thus SR1's spacetime symmetries are those diffeomorphisms\footnote{In this paper for simplicity I will not differentiate between the pull back and push forward of $d$. Both of these will be referred to as $d^*$ which will be called the drag along map. $d^*$ is whatever modification of $d$ is demanded by the context.} with $d^*\eta_\text{ab}=\eta_\text{ab}$. Only a small subset of $\text{Diff}(\mathcal{M})$ have this property, namely the Poincar\'e group.

For SR1, given a generic DPM, $\langle\mathcal{M},\eta_\text{ab},\varphi\rangle$ we can apply a generic diffeomorphism \mbox{$d\in\text{Diff}(\mathcal{M})$} to its dynamical fields to get some KPM, \mbox{$\langle\mathcal{M},\eta_\text{ab},d^*\varphi\rangle$}. This diffeomorphism $d$ is a dynamical symmetry when for every input DPM this output KPM is a solution to the dynamics (i.e., is also a DPM). It turns out that all and only $d$ in the Poincar\'e group maps solutions to solutions. Thus the dynamical symmetry group of SR1 is the Poincar\'e group.

In the above example, the theory's spacetime symmetries and its dynamical symmetries match. There are good reasons\footnote{If there are more dynamical symmetries than spacetime symmetries, then some of the theory's fixed fields are not being used by any of the dynamics, in which case why are they there? Conversely, if there are more spacetime symmetries than dynamical symmetries then it appears some necessary piece of spacetime structure is missing. E.g., consider a case where the dynamics are not boost invariant and so implicitly pick out a rest frame, but somehow the spacetime comes equipped with no rest frame.} for these to match in general~\cite{EarmanJohn1989Weas} but they won't always\footnote{They would not match for instance if, as a piece of fixed field, we had included a time orientation field, $\chi$, which distinguishes the future light cone from the past light cone at each event. In this case, the dynamical symmetries would still be the Poincar\'e group, but the spacetime symmetries would only be the time-orientation preserving subset of these. In general, the spacetime symmetries can be smaller than the dynamical symmetries if there is some piece of spacetime structure which is unused by the dynamics.}. In any case it is important to keep them separate conceptually. Unless otherwise specified, throughout this paper unqualified references to symmetries should be understood as meaning dynamical symmetries.

In addition to these external symmetries, our theories might also have internal symmetries and relatedly gauge symmetries. In any given theory, the dynamical fields will map\footnote{More generally, the fields might be defined as sections of a fiber bundle of $\mathcal{V}$ over $\mathcal{M}$, but let's neglect that complication here.} from the manifold $\mathcal{M}$ into some value space $\mathcal{V}$ as $\varphi:\mathcal{M}\to \mathcal{V}$. We may find additional internal symmetries of our theory within the automorphisms of this value space, $\text{Auto}(\mathcal{V})$. We might also find our theory has gauge symmetries by allowing these internal automorphisms to vary smoothly across the manifold.

For SR1, the value space of our dynamical field $\varphi$ is the real numbers $\mathcal{V}=\mathbb{R}$. Note that the space of all (potentially off-shell) fields $\varphi$ is closed under addition and scalar multiplication, and is hence a vector space. This addition and scalar multiplication is carried out at each spacetime point. Thus, the field's value space $\mathcal{V}=\mathbb{R}$ is also structured like a vector space. Therefore, \mbox{$\text{Auto}(\mathcal{V})=\text{Aff}(\mathbb{R})$} such that our internal symmetries are linear-affine rescalings of $\varphi$. Namely, $\varphi\mapsto c_1\,\varphi+c_2$ for some $c_1,c_2\in\mathbb{R}$. Of these, only global rescalings of $\varphi$ as $\varphi\mapsto c_1\,\varphi$ are symmetries of SR1.

We can find the theory's gauge symmetries by localizing its internal symmetries as $\varphi\mapsto c_1(t,x)\,\varphi+c_2(t,x)$ for some spacetime functions $c_1(t,x)$ and $c_2(t,x)$. Of these, the dynamics is only preserved when $c_1(t,x)=c_1$ is constant and $c_2(t,x)$ is itself a solution to the dynamics.

\subsubsection*{Klein Gordon - Background Structure}
Let us now use this generally covariant formulation to help us understand this theory's background structure. As mentioned in the introduction, there is ongoing debate~\cite{Norton1993,Pitts2006,Pooley2010,ReadThesis,Pooley2015,Teitel2019,Belot2011} about what exactly should and should not count as background structure. However, it is widely agreed that any fixed field ought to count as background structure. Thus, for SR1 the fixed Lorentzian metric $\eta_\text{ab}$ counts as background structure.

From this one may be tempted to reason (poorly) as follows. A theory's spacetime symmetries are just those transformations which preserve its fixed fields. Any fixed fields count as background structures. As such, minimizing background structure is the same as maximizing spacetime symmetry. Therefore, background independence is the same concept as diffeomorphism invariance. 

However, this reasoning and its conclusion are in error. There can be other kinds of background structure than fixed fields. Indeed, following~\cite{Pooley2015}, we can reformulate SR1 as, 
\begin{align}
\text{SR2:}\qquad\text{KPMs:}\quad&\langle\mathcal{M},g_\text{ab},\varphi\rangle,\\
\nonumber
\text{DPMs:}\quad&(g^\text{ab}\nabla_\text{a}\nabla_\text{b}-M^2)\,\varphi = 0\\
\nonumber
&R^\text{a}{}_\text{bcd}=0.
\end{align}
Here the fixed Lorentzian metric field, $\eta_\text{ab}$, has been replaced with a dynamical metric field, $g_\text{ab}$, with signature $(-1,1,1,1)$. The dynamical metric field varies from model to model and obeys a new dynamical equation, $R^\text{a}{}_\text{bcd}=0$, where $R^\text{a}{}_\text{bcd}$ is the Riemann tensor associated with $\nabla_\text{a}$.

Note that SR2 has the same KPMs as the Klein Gordon equation in GR would:
\begin{align}
\text{GR:}\qquad\text{KPMs:}\quad&\langle\mathcal{M},g_\text{ab},\varphi\rangle,\\
\nonumber
\text{DPMs:}\quad&(g^\text{ab}\nabla_\text{a}\nabla_\text{b}-M^2)\,\varphi = 0\\
\nonumber
&G_\text{ab}=8\pi \, T_\text{ab}.
\end{align}
Indeed, SR2 and GR only differ in their dynamical equations. One may favor SR2 over SR1 on these grounds.

What are SR2's symmetries? SR2 has no fixed fields. As such, its spacetime symmetries are the full diffeomorphism group, $\text{Diff}(\mathcal{M})$. This theory's dynamical symmetries are also the full diffeomorphism group, $\text{Diff}(\mathcal{M})$. Thus, SR2 is diffeomorphism invariant.

Ought we to conclude (using the above erroneous logic) that because SR2 is diffeomorphism invariant that it is also background independent? Clearly not. Intuitively SR1 and SR2 should have the same background structures. The difference is that in SR2 this background structure is hidden whereas SR1 is in a sense more honest, declaring its background structure up front as a fixed field. For this reason one may prefer SR1 to SR2. 

However, as SR2 clearly demonstrates, we cannot in general expect our theories to be so honest about their background structures. There can be other sorts of background structure in our theories than those which are declared upfront as fixed fields. There is a wide literature attempting to find a way of systematically identifying these hidden background structures~\cite{Norton1993,Pitts2006,Pooley2010,ReadThesis,Pooley2015,Teitel2019,Belot2011}.

To demonstrate these ideas further, let's consider one more example.

\subsection{Heat Equation}\label{HeatEqGenCov}
Let us next consider a real scalar field $\psi:\mathcal{M}\to\mathbb{R}$ satisfying the following one and two-dimensional heat equations:
\begin{align}
&\text{\bf Heat Equation 00 (H00):}\\
\nonumber
&\partial_t \psi(t,x) = \alpha_0\, \partial_x^2\,\psi(t,x)\\
&\text{\bf Heat Equation 0 (H0):}\\
\nonumber
&\partial_t \psi(t,x,y) = \frac{\alpha_0}{2}\, (\partial_x^2+\partial_y^2)\,\psi(t,x,y) 
\end{align}
with some diffusion rate $\alpha_0\geq0$. Focusing on the two-dimensional case, after substantial work, one can rewrite this equation in the coordinate-free language of differential geometry as follows:
\begin{align}\label{H0GenCovApp}
\text{H0}:\qquad\text{KPMs:}\quad&\langle \mathcal{M}, t_\text{ab}, h^\text{ab}, \nabla_\text{a},T^\text{a},\psi\rangle\\
\nonumber
\text{DPMs:}\quad&
T^\text{a}\,\nabla_\text{a}\psi
=\frac{\alpha_0}{2} \, h^\text{bc} \nabla_\text{b}\nabla_\text{c}\psi.
\end{align}
The geometric objects used in this formulation are as defined following Eq.~\eqref{H0GenCov}

What are the heat equation's symmetries? In contrast with SR1, the H0 has many more fixed fields: $t_\text{ab}$, $h^\text{ab}$, $\nabla_\text{a}$, and $T^\text{a}$ are all fixed. Ultimately, this restricts the theory's spacetime symmetries down to the two-dimensional Euclidean group (spacial translations, rotations and reflections) plus constant time shifts. Note that time inversions and Galilean boosts are not symmetries because each of these fail to preserve our rest frame $T^a$. This theory's dynamical symmetries match its spacetime symmetries. 

What are this theory's internal symmetries? The value space here is the real numbers $\mathcal{V}=\mathbb{R}$. As such, I will take $\text{Auto}(\mathcal{V})=\text{Aff}(\mathbb{R})$ such that our internal symmetries are linear-affine rescalings of $\psi$. Namely, $\psi\mapsto c_1\,\psi+c_2$ for some $c_1,c_2\in\mathbb{R}$. All such transformations preserve the dynamics. We can find this theory's gauge symmetries by localizing its internal symmetries as $\psi\mapsto c_1(t,x,y)\,\psi+c_2(t,x,y)$ for some spacetime functions $c_1(t,x,y)$ and $c_2(t,x,y)$. Of these, the dynamics is only preserved when $c_1(t,x,y)=c_1$ is constant and $c_2(t,x,y)$ is itself a solution to the dynamics.

What are the heat equation's background structures? The above discussed fixed fields will surely count as background structures. While conceivably there could be more background structures than just these, given the simplicity of this theory that seems unlikely.

\section{Analysis of Symmetries for H1-H7}\label{AppB}
This appendix will identify the dynamical symmetries for H1-H7 under the three interpretations put forward in the main text. However, it is convenient to do this in the reverse order in which these interpretations were introduced in the main text.

\subsection{Symmetries in the Third Interpretation}
Let's begin by analyzing the dynamical symmetries of H1-H7 in our third attempt at interpreting them, see Sec.~\ref{SecHeat3}.

For H6 (and relatedly H7) we can look to their generally covariant formulation Eq.~\eqref{DH6GenCov} to determine their dynamical symmetries. For H6 we find the same symmetries as we did for the continuum heat equation, H0. Namely, the two-dimensional Euclidean group (spacial translations, rotations, and reflections) plus constant time shifts, global linear rescalings and certain local affine rescalings. Technically, the space of KPMs for H6 is not closed under rotations. This has been discussed at length in Secs.\ref{SecHeat3} and \ref{SecFullGenCov}. 

Similarly, for H4 we can look to its generally covariant formulation Eq.~\eqref{DH4GenCov} to identify its symmetries. Its extra spacetime structures $X^a$ and $Y^a$ appearing in its dynamics restrict its rotation symmetries to just quarter turns and allow us only a restricted set of reflection symmetries. Applying the same analysis to H5 one would find we are restricted to one-sixth turns and only a restricted set of reflection symmetries (different from H4's).

We could also cast H1-H3, (namely Eq.~\eqref{DH1bandlimited}-\eqref{DH3bandlimited}) into a generally covariant form as well. Doing so we would find their symmetries are the one-dimensional Euclidean group (spacial translations and reflections) plus constant time shifts, global linear rescalings and certain local affine rescalings.

None of this is mysterious.

\subsection{Symmetries in the Second Interpretation}
Let us next analyze the dynamical symmetries of H1-H7 in our second attempt at interpreting them, in Sec.~\ref{SecHeat2}.

Before this, note that our redescription of H1-H7 in terms of $\phi_\ell(t)\in F_L$, $\bm{\Phi}(t)\in \mathbb{R}^L$, and $\phi_\text{B}(t)\in F^K$ were each mediated by a vector space isomorphism, \mbox{$F_L\cong\mathbb{R}^L\cong F^K$}. This gives us a nice solution-preserving one-to-one correspondence between each of our three interpretations.

Thus, given any transformation on one interpretation we can always find the equivalent transformation on the other interpretations. However, as we learned following Eq.~\eqref{GaugeVR} and Eq.~\eqref{BandlimitedSymmetries} this does not mean that these theories have the same class of possible symmetries in each interpretation. While we have a one-to-one correspondence between generic transformations in one interpretation and another, what counts as a \textit{symmetry} transformation is interpretation dependent. The scope of possible symmetry transformation varies from interpretation to interpretation.

Thus every symmetry revealed by our third interpretation gives us a candidate symmetry for our other two interpretations, but more must be done. In particular, we need to check whether these transformations are of the forms allowed in Eq.~\eqref{Permutation} and Eq.~\eqref{GaugeVR}.

All of the symmetries revealed by our third interpretation are also symmetries on the second interpretation. However, as discussed following Eq.~\eqref{BandlimitedSymmetries}, our second interpretation has a wider scope of possible symmetries than our third interpretation does. In fact, as I will soon discuss it includes local Fourier rescalings among other things.

Let's begin however, with the symmetries shared by our second and third interpretations. As revealed in Sec.~\ref{SecSamplingTheory} translation of a bandlimited function \mbox{$f_\text{B}(x)\mapsto f_\text{B}(x-\epsilon)$} is represented in terms of its vector of sample values $\bm{f}$ as $\bm{f}\mapsto T^\epsilon_\text{B}\bm{f}$, see Eq.~\eqref{TBdef}. Moreover, as discussed following Eq.~\eqref{DBMatrix}, the operator $T_\text{B}^\epsilon$ in Sec.~\ref{SecSamplingTheory} is numerically identical to the operator $T^\epsilon$ appearing in Sec.~\ref{SecHeat2}. 

Thus for H1-H3 our candidate symmetry for continuous translation is $\bm{\Phi}(t)\mapsto T^\epsilon\bm{\Phi}(t)$. Likewise for H4-H7 the candidate symmetries are $\bm{\Phi}(t)\mapsto T^\epsilon_\text{n}\bm{\Phi}(t)$ and $\bm{\Phi}(t)\mapsto T^\epsilon_\text{m}\bm{\Phi}(t)$. These are all of the form Eq.~\eqref{GaugeVR} and are thus viable symmetries under our second interpretation. Indeed, this is a symmetry of H1-H7 under our second interpretation.

Similar considerations apply for constant time shifts, reflection symmetries and for the linear and affine rescalings. It is easy to find the symmetry candidates here and check that they are of the form Eq.~\eqref{GaugeVR}. The only other non-trivial symmetry to transfer over is the continuous rotational symmetry. To see this we need the following facts. 

For functions $h(x,y)$ rotations are generated through the derivative as 
\mbox{$h(R(x,y))= \exp(\theta (x \partial_y - y \partial_x))h(x,y)$}. Suppose that $h=h_\text{B}$ is bandlimited and we sample it in two ways. Once on some square lattice, and once on another lattice identical to the first but rotated around $(n,m)=(0,0)$ by an angle $\theta$. The sample values in these two case, $\bm{h},\bm{h}'\in\mathbb{R}^\mathbb{Z}\otimes\mathbb{R}^\mathbb{Z}$ are related as $\bm{h}'=R_\text{B}^\theta \bm{h}$ where
\begin{align}
R_\text{B}^\theta \coloneqq \exp(-\theta (N_\text{n} D_\text{B,m}-N_\text{m} D_\text{B,n}))
\end{align}
with $\theta\in\mathbb{R}$ and where $N_\text{n}$ and $N_\text{m}$ are position operators which return the first and second index. Finally, since $D_\text{B}$ is numerically identical to $D$ we have $R_\text{B}^\theta=R^\theta$ as defined in Eq.~\eqref{RthetaDef}. 

This transformation is of the form Eq.~\eqref{GaugeVR} and is thus a viable symmetry under our second interpretation. Indeed, this is a symmetry of H6 under our second interpretation (at least restricted to $\mathbb{R}^L_\text{rot.inv.}$). Moreover, with a slight change of basis (namely, Eq.~\eqref{SkewH6H7}) it is for H7 as well (at least restricted to $\mathbb{R}^L_\text{rot.inv.}$ post-transformation).

But what about the local Fourier rescaling symmetry present on our second interpretation. It's straight forward to check that this is a symmetry, but how do we know there aren't other symmetries of the form Eq.~\eqref{GaugeVR} missed by our third interpretation?

As I will now discuss, there are tremendously many symmetries on our second interpretation. Let's begin the search. Firstly, note that the only time reparametrizations which preserve the dynamics are constant time shifts $t\mapsto t+\tau$. Secondly note that any non-trivial time dependence in $\Lambda(t)$ will cause it to fail to preserve the dynamics. Our possibility for affine transformations $\bm{c}(t)$ has already been accounted for above.

Thus, the only place extra symmetries could be realized is as $\bm{\Phi}(t)\mapsto\Lambda\,\bm{\Phi}(t)$ where $\Lambda$ is some time-independent invertible linear map. This will be a symmetry if and only if $\Lambda$ commutes with the relevant operator (call it $\Delta_{H1-H7}^2$) appearing in the theory's dynamical equation: namely, Eq.~\eqref{DH1}, Eq.~\eqref{DH2}, Eq.~\eqref{DH3} and Eqs.~\eqref{DH4}-\eqref{DH7}. In each case, $\Delta_{H1-H7}^2$ is diagonal in the relevant discrete Fourier basis: $\bm{\Phi}(k)$ for H1-H3 and $\bm{\Phi}(k_1,k_2)$ for H4-H7. 

From basic linear algebra, it follows that there are roughly two ways for $\Lambda$ to commute with $\Delta_{H1-H7}^2$: 1) it could either be diagonal in the discrete Fourier basis itself, or 2) it could operate within a degenerate subspaces of $\Delta_{H1-H7}^2$. The transformations $\Lambda$ which are diagonal in the discrete Fourier basis are exactly the local Fourier rescalings which we have already discussed (translations are of this form too). All other symmetries come from operating within the degenerate subspaces of $\Delta_{H1-H7}^2$.

What then are the degenerate subspaces of $\Delta_{H1-H7}^2$? The spectrum of these operators are proportional to the decay rates given namely $\lambda=-\Gamma/\alpha$. For H1-H3 these are plotted in Fig.~\ref{FigHeatDecay}. For these theories the degenerate subspaces are pairs of planewaves with $k=\pm k_0$, namely  $\bm{\Phi}(k_0)$ and $\bm{\Phi}(-k_0)$. For each wavenumber $k\in[-\pi,\pi]$ we have an $\text{GL}(\mathbb{R}^2)$ symmetry operating within this degenerate spaces. 

For H4-H7, these degenerate subspaces are much larger. For relatively small $k_1$ and $k_2$ these degenerate subspaces trace out approximate circles in Fourier space. For H6 and H7 these degenerate subspaces remain exactly circular as we increase our wavenumbers, but for H4 and H5 these become increasingly distorted. In every case, for sufficiently high wavenumbers the degenerate subspaces begin to touch the sides of the region $k_1,k_2\in[-\pi,\pi]$. Recall that we have a $2\pi$ periodic identification if planewaves on our second interpretation. 

For all but the $k_1=k_2=0$ degenerate subspaces trace out a compact figure in Fourier space. Collecting all of these degenerate Fourier modes together into a vector space we have something isomorphic to $\mathbb{R}^\mathbb{Z}$. The dynamical symmetries within each of these degenerate subspaces is $\text{GL}(\mathbb{R}^\mathbb{Z})$, i.e., tremendously large. For instance, permuting these degenerate planewaves in a discontinuous way is a symmetry here.

If this feels excessive, one can add in some structure to Fourier space to allow such discontinuous permutations. Indeed, throughout this paper I have talked about compact regions in Fourier space. We must therefore have some native topology on Fourier space. Moreover, let's give Fourier space a differentiable structure. We can then restrict our attention to symmetries generated by diffeomorphisms in Fourier space. As I mentioned above, the degenerate Fourier spaces for H4-H7 trace out closed curves in Fourier space. Our symmetries are then smooth maps which flow along these degeneracy curves. This may still seem excessive. On this view, H4 and H5 have something like rotation symmetries (smoothly rotating around their non-circular degenerate spaces). Although it should be noted that these pseudo-rotation to do not fit together with these theories translation symmetries to form a Euclidean group. 

One might further restrict the symmetries in this second interpretation by giving its Fourier space a metric structure. The only symmetries then are generated by metric preserving transformations of Fourier space (i.e., rotations). This forbids H4 and H5's pseudo-rotations but allows for H6 and H7's authentic rotations.

\subsection{Symmetries in the First Interpretation}
Let us now check whether the above discussed symmetries are still symmetries in our first interpretation. To do this we just need to check which are understandable in terms of permutations of the lattice site and time reparametrizations. That is, when are the above-discussed linear transformations $\Lambda$ permutations?

It is not hard to check when $T^\epsilon$, $T_\text{n}^\epsilon$, $T_\text{m}^\epsilon$, and $R^\theta$ reduce to permutations for H1-H6. Similarly for H7 we can check when $T_\text{n}^\epsilon$ and $T_\text{m}^\epsilon$ and $R^\theta$ after the basis change which adapts them to H7. Ultimately, this shows the symmetries under the first interpretation are just those claimed in Sec.~\ref{SecHeat1}.

\section{Nyquist-Shannon Resampling}\label{SecResample}
As discussed in Sec.~\ref{SecHeat3Extra}, any bandlimited dynamics can be represented on any sufficiently dense sample points. In this appendix I will discuss the effect that changing these sample points has on the dynamics of the sample values.

As a demonstrative example of how switching between different sets of sample points affects the form of a theories' dynamics, consider the following. Fig.~\ref{FigHeatSpaceTime}a) shows a bandlimited function which is a solution to H1's dynamics Eq.~\eqref{DH1bandlimited}. We could alternatively describe this state via the values it takes at any sufficiently dense set of sample points. For instance, we might choose the sample points given by the vertical black lines. In this case, the sample values would obey the dynamical equation Eq.~\eqref{H1Long}, that is, the 1D nearest neighbor heat equation. But what if we described the very same bandlimited dynamics on a different set of sample points?

For instance, what dynamical equation do the boosted sample points given by the diagonal red lines give us? Let 
\begin{align}
z_n^\text{Boost}(t)=(t,n\,a+v\,t)
\end{align}
be our new sample points for some speed $v$. Let 
\begin{align}
\varphi_n(t)=\phi_\text{B}(z_n^\text{Boost}(t))=\phi_\text{B}(t,n\,a+v\,t).
\end{align}
be out new sample values. What are the dynamics which these new sample values obey?

The simplest way to go about answering this question is to first rewrite the bandlimited dynamics in boosted coordinates. This boosted field $\phi_\text{B}$ is shown in Fig.~\ref{FigHeatSpaceTime}b) along with a corresponding boost applied to the sample points. Note that this boosting operation does not change the bandwidth of these functions in space. For each of H1-H3 taking 
\mbox{$\phi_\text{B}(t,x)\mapsto\phi_\text{B}(t,x+v\,t)$} the bandlimited field $\phi_\text{B}$ then obeys,
\begin{align}\label{DH1bandlimitedboosted}
\text{H1}: \partial_t\phi_\text{B}&=\alpha \, [2\text{cosh}(a\,\partial_x)-2] \ \phi_\text{B}-v\,\partial_x\phi_\text{B}\\
\nonumber
\text{H2}:\partial_t\phi_\text{B}
&=\frac{\alpha}{6} [-\text{cosh}(2a\partial_x)\!+\!16\text{cosh}(a\partial_x)\!-\!15]\phi_\text{B}\\
\nonumber
&-v\,\partial_x\phi_\text{B}\\
\nonumber
\text{H3}: \partial_t\phi_\text{B}
&=\alpha \, a^2\,\partial_x^2 \, \phi_\text{B}-v\,\partial_x\phi_\text{B}
\end{align}
Note that the appearance of this new term in the dynamics means that none of these theories are Galilean boost invariant. 

Our task is then to write these dynamics in terms of their values at the red (now stationary) sample points, $x_n=n\,a$. Recalling that $\text{exp}(a\partial_x)h(x)=h(x+a)$ we can rewrite the $\text{cosh}(a\,\partial_x)$ terms in terms of $\phi_\text{B}$ shifted by $\pm a$. Similarly for the $\text{cosh}(2a\,\partial_x)$ terms with an offset of $\pm 2a$. Writing the $\partial_x\phi_\text{B}$ is a bit more difficult but can be done using Eq.~\eqref{ExactDerivative}. Recall that Eq.~\eqref{ExactDerivative} is exact for bandlimited functions with bandwidth less than $K=\pi/a$. We can thus rewrite the dynamics of $\phi_\text{B}$ at any sample point in terms of the values it takes at the other sample points. 

Doing this and collecting these sample values into a vector \mbox{$\bm{\varphi}(t)=(\dots,\varphi_{-1}(t),\varphi_0(t),\varphi_1(t),\dots)\in\mathbb{R}^\mathbb{Z}$} as in Sec.~\ref{SecSevenHeat} we have
\begin{align}\label{DH1boosted}
\text{H1}:\quad
\frac{\d }{\d t}\bm{\varphi}(t)&=\alpha\,\Delta_{(1)}^2 \bm{\varphi}(t)-\frac{v}{a}\,D\,\bm{\varphi}(t).\\
\text{H2}:\quad
\frac{\d }{\d t}\bm{\varphi}(t)&=\alpha\,\Delta_{(2)}^2 \bm{\varphi}(t)-\frac{v}{a}\,D\,\bm{\varphi}(t)\\
\text{H3}:\quad
\frac{\d }{\d t}\bm{\varphi}(t)&=\alpha\,D^2 \bm{\varphi}(t)-\frac{v}{a}\,D\,\bm{\varphi}(t).
\end{align}
Note that the appearance of this new term in the dynamics that none of these theories are Galilean boost invariant. 

Also note how the infinite range discrete derivative operator $D$ appears in each of these equations, even when we start off with only finite range derivative approximation. Moreover, note that while before this resampling H1-H3 were local in the intuitive sense of Sec.~\ref{SecIntuitiveLocality} (i.e., nearest-neighbor couplings only), they are no longer once we have resample them. Thus, this intuitive notion of locality is uncomfortably representation dependent and hence unphysical. Another example of this loss of intuitive locality under resampling is given by Eq.~\eqref{DH5Skew}.

\end{document}